\documentclass[11pt,a4paper]{article}

\usepackage{mathptmx}
\usepackage[a4paper,text={130mm,198mm}]{geometry}

\usepackage{amsmath}
\usepackage{amsfonts}
\usepackage{amssymb}
\usepackage{amscd}
\usepackage{graphicx}
\usepackage{citesort}
\usepackage{mathrsfs}
\usepackage{bbm}
\usepackage{graphics}

\DeclareMathOperator{\tr}{tr}

\def\S{{\mathbb S}}

\def\[{\begin{equation}}
\def\]{\end{equation}}
\def\<{\begin{eqnarray}}
\def\>{\end{eqnarray}}

\def\({\left(}
\def\){\right)}

\def\AdSxS{{AdS${}_5 \times {}$S${}^5$}}

\def\S{{\mathbb S}}

\newcommand{\fGG}{{\mathbb G}}
\newcommand{\fHH}{{\mathbb H}}

\newcommand{\comm}[2]{[#1,#2]}
\newcommand{\gen}[1]{\mathfrak{#1}}
\newcommand{\cwgen}[1]{#1}

\newcommand{\Ya}[1]{\mathcal{Y}(#1)}

\newcommand{\half}{\frac{1}{2}}

\newcommand{\adjb}[2]{\text{ad}_{#1}(#2)}

\newcommand{\acomm}[2]{\{#1,#2\}}
\newcommand{\copro}{\Delta}

\newcommand{\nn}{\nonumber}
\newcommand{\genY}[1]{\widehat{\mathfrak{#1}}}

\newcommand{\struc}{f}
\newcommand{\earel}[1]{\mathrel{}&\hspace{-2\arraycolsep}#1\hspace{-2\arraycolsep}&\mathrel{}}
\newcommand{\eq}{\earel{=}}
\newcommand{\alg}[1]{\mathfrak{#1}}

\newcommand{\sls}{\alg{sl}}

\newcommand{\mathsym}[1]{{}}

\newcommand{\stateA}[1]{|#1\rangle^{\rm{I}}}
\newcommand{\stateB}[1]{|#1\rangle^{\rm{II}}}
\newcommand{\stateC}[1]{|#1\rangle^{\rm{III}}}

\renewcommand{\[}{\left[}
\renewcommand{\]}{\right]}
\renewcommand{\eqref}[1]{$\({\rm \ref{#1}}\)$}

\begin{document}

\begingroup\parindent0pt
\vspace*{2em}
\begingroup\LARGE
Yangians, S-matrices and AdS/CFT
\par\endgroup
\vspace{1.5em}
\begingroup\large
Alessandro Torrielli 
\par\endgroup
\vspace{1em}
\begingroup\itshape
Department of Mathematics, University of York, 

Heslington, York, YO10 5DD

United Kingdom
\par\endgroup
\vspace{1em}
\begingroup\ttfamily
alessandro.torrielli@york.ac.uk
\par\endgroup
\vspace{1.5em}
\endgroup

\bigskip

\paragraph{Abstract.}

This review is meant to be an account of the properties of the infinite-dimensional quantum group (specifically, Yangian) symmetry lying behind the integrability of the AdS/CFT spectral problem. In passing, the chance is taken to give a concise anthology of basic facts concerning Yangians and integrable systems, and to store a series of remarks, observations and proofs the author has collected in a five-year span of research on the subject. We hope this exercise will be useful for future attempts to study Yangians in field and string theories, with or without supersymmetry\footnote{This is an author-created, un-copyedited version of an article (invited topical review) accepted
for publication in {\it Journal of Physics A: Mathematical and Theoretical}. IOP Publishing Ltd is not
responsible for any errors or omissions in this version of the manuscript
or any version derived from it. The definitive publisher authenticated
version will be available online at [details to follow].}.

\bigskip

\bigskip


\ \ \ \ \ \ \ \ \ \ \ \ \ \ \ \  Yangians, Lie Superalgebras, Representation Theory  

\newpage

\section{Introduction}

\hfill{\it ``What makes you think that the theory will still be integrable?"}

\hfill{\it ``Unlimited optimism."}

\smallskip

\hfill{\footnotesize(M. Staudacher, replying to A. A. Migdal at the Itzykson Meeting, Paris, 2007)}

\bigskip

\bigskip

Gauge theories play a dominant role in our current understanding
of the nature of fundamental interactions at very short distances.
A prominent example of such a theory is the Standard Model of
elementary particles, which is remarkably successful in describing
the physics up to the currently available energy scale. This
description is, however, to a significant extent restricted to the
perturbative regime. The derivation of analytical results when the
coupling constant is large is an extremely challenging task. This represents an obstacle to the complete understanding
of interesting nonperturbative phenomena, like, for instance,
confinement. 

The revolutionary discovery of integrable structures
in Quantum Chromodynamics (QCD) \cite{Lipatov:1993yb}, and, more recently, in planar ${\cal{N}}=4$ Supersymmetric Yang-Mills (SYM)
theory and AdS/CFT \cite{Minahan:2002ve}, has changed this situation\footnote{According to the AdS/CFT correspondence \cite{Maldacena:1997re,Gubser:1998bc,Witten:1998qj,Aharony:1999ti,D'Hoker:2002aw}, the scaling dimension of gauge-invariant composite operators should match the energy of the corresponding closed string states. In particular, we will be focusing our attention on string states with large values of some spin or angular momentum quantum number $Q$, corresponding to composite operators containing a large (order $Q$) number of fields. The energy of these states / dimension of these operators can be expressed as $E=Q + \epsilon(Q,\lambda)$, with $\epsilon$ going to zero at weak 't Hooft coupling $\lambda\equiv g_{YM}^2 N$ ($g_{YM}$ being the Yang-Mills coupling) where the dimension reduces to the bare dimension $Q$ (see, for instance,
\cite{Tseytlin:2010jv}). The {\it anomalous} dimension $\epsilon$ is a dynamical quantity which should interpolate between the two sides of the correspondence, and which will be our main object of interest 
\cite{Beisert:2010jr}.}.
 
For a Hamiltonian system with $2n$-dimensional phase space, {\it complete integrability} stands for the existence of $n$ independent integrals of motion, written as integrals of local densities, in involution ({\it i.e.} Poisson-commuting with each other). One of these integrals of motion is the Hamiltonian itself, while the other ones are sometimes referred to as {\it higher Hamiltonians}. According to the Liouville-Arnold theorem, the equations of motion can then be solved by quadratures. This means that there exists a set of canonical coordinates (`action-angle') such that the action variables (momenta) are constants of motion, and the angles (coordinates) are linear in time and parameterize a torus. For a field theory, the number of degrees of freedom is normally infinite, and one associates {\it integrability} with the existence of an infinite number of independent local conserved charges in involution. 
In scattering theory, integrability implies pure reshuffling of momenta (`diffractionless' scattering). In general, flavour degrees of freedom can be transformed in a complicated way during the scattering.  One has `transmission' if the flavours are unchanged, `reflection' if they are exchanged. We recommend \cite{Faddeev:1987ph,Sklyanin:1980ij,Evgeny} for classical references on integrable systems (see also the excellent \cite{GlebLect}). 

A link with the Yang-Mills Millennium prize problem\footnote{For any compact gauge group $G$, one is to show that quantum Yang-Mills theory on $R^4$ exists and has a mass gap $\Delta>0$ ({\it i.e.} the lightest particle has strictly positive mass squared).} has been also advertised. The situation in AdS/CFT is quite peculiar because of conformal invariance. Moreover, 't Hooft's limit $N\to \infty$, with $\lambda=g_{YM}^2 N$ fixed, suppresses instanton contributions, according to the standard argument that the action for such configurations scales in this limit as $\frac{N}{\lambda} [ finite]$. However, one hopes that the understanding of even one single interacting four-dimensional gauge theory in this special limit will be important for progress in the Yang-Mills problem as well.  For a relatively recent report, underlying the potential role of AdS/CFT and integrability, see \cite{Douglas}.

The ${\cal{N}}=4$ theory is a quantum conformal field theory
(CFT). The information on its spectrum is encoded
in the short-distance power-law behavior of (2-point) correlators
of composite operators. In determining this behavior for all operators of the
theory one encounters a non-trivial operator-mixing, which makes
the calculations notoriously difficult. The observation of \cite{Minahan:2002ve} is that, in the planar limit, the problem translates into the equivalent
problem of finding the spectrum of certain spin-chain
Hamiltonians. This spectrum consists of spin-wave excitations and
their bound states, and the dynamics (S-matrix\footnote{We take a chance and clarify that, whenever we will be talking of S-matrices in this review, it will always be referred to the two-dimensional scattering of excitations in the integrable models effectively describing the SYM spectral problem in various regimes (spin-chain, sigma model). Never will we be talking of a spacetime SYM S-matrix (also because, in that case, conformal invariance would be an obstacle to the definition of asymptotics states).}) describing their
scattering turns out to be completely integrable \cite{Staudacher:2004tk,Beisert:2005tm}. Planarity is probably a crucial ingredient for the appearance of integrability. It would be overwhelming to give here a comprehensive list of the relevant references. They can be found in many of the available reviews (just to mention some of the most recent ones, see \cite{Arutyunov:2009ga,1751-8121-44-12-124001,1751-8121-42-25-254001,Puletti:2010ge,Beisert:2010jr}).

The result strictly
applies to infinitely long chains, which are related to gauge
theory operators composed of an infinite number of fields. When
the spin-chains are of finite length, certain corrections occur
that go under the name of `wrapping effects' \cite{Beisert:2004ry,Sieg:2005kd,Sieg:2010jt}, since the range of
the interactions exceeds the length of the spin-chain. Recently \cite{Fiamberti:2007rj,Bajnok:2008bm},
these effects have been shown to be calculable for very specific
operators and at the first few significant orders in perturbation
theory, by techniques of finite-volume integrability\footnote{These techniques involve the use of the so-called {\it L\"uscher corrections}. Such corrections do not assume integrability, but, if the theory is integrable, they are expected to complete to a set of exact integral equations for the spectrum (see also \cite{1751-8121-44-12-124003}).}. The first confirmation that one has obtained from these impressive results is that the ingredients used in the mirror theory approach \cite{Arutyunov:2007tc}, {\it i.e.} the mirror bound states, are all one needs to sum over in order to reproduce the field theory result. In other words, no excitation is missing.

The technology developed so far has been impressive, see for instance \cite{Beccaria:2009eq,Bajnok:2009vm,Fiamberti:2009jw,Lukowski:2009ce,Arutyunov:2010gb,Gromov:2010vb,Velizhanin:2010cm,Fioravanti:2010ge,Frolov:2010wt,Gromov:2010km}. Both gauge perturbation theory for short operators and string perturbation theory in the form of L\"uscher corrections have proceeded to a tremendous degree of sophistication. A very convincing matching has been shown\footnote{Notably, the issue concerning some mismatches \cite{Roiban:2009aa}, which were still announced to affect the strong coupling regime, has very recently been resolved \cite{Gromov:2011de,Roiban:2011fe,Vallilo:2011fj}.}. This
remarkable result has strengthened the expectation that the entire
planar sector of the theory may in fact be integrable, and
accessible {\it via} the so-called Thermodynamic Bethe Ansatz
(TBA) method. The latter consists in obtaining a set of master
equations, whose solutions encode the spectral {\it data} of the theory.
This program has the potential of providing a set of exact analytic
results for an interacting four-dimensional quantum field theory,
and, with it, a new insight in our understanding of
strongly-coupled nonperturbative phenomena in gauge theories. 
Once more, the study of two-dimensional models is showing its power in modelling our understanding of four-dimensional theories (cf. \cite{Faddeev:1996iy}, Introduction, lines 37-58). Currently, a remarkable effort is being put into the construction and test of such a TBA system of equations \cite{Gromov:2009tv,Bombardelli:2009ns,Arutyunov:2009ur}.

Despite the progress obtained, several fundamental questions are still left unanswered. First of all, a
systematic way of taking into account the above-mentioned wrapping
corrections has not yet been provided, due to their highly
complicated nature \cite{Sieg:2005kd}. Furthermore, no rigorous proof of
integrability is yet available, and the quantum Hamiltonian of the system is not known in closed form, but only to a certain order in perturbation theory. Instead, so far the approach has been (in the philosophy of the inverse scattering method) to assume integrability and S-matrix factorization, deduce the entire integrable structure, and {\it a posteriori} check the validity of the assumptions (see also \cite{Puletti:2007hq}). However, with long-range Hamiltonians (as the one emerging from gauge perturbation theory actually is) even setting up an asymptotic scattering theory is problematic, and it is still a challenge to rigorously prove the integrability of the asymptotic problem. Perhaps, with the help of the algebraic methods we are going to describe in this review, the knowledge of the complete Hamiltonian will eventually become accessible\footnote{The so-called `dressing phase' (see formula (\ref{eqn;FullPhase}) and subsequent text) is essential for the Hamiltonian. In \cite{Bargheer:2008jt}, the presence of this phase has been connected to boosts and general twist transformations for the long-range spin-chain, see also section \ref{spi} and references therein.}.
The full algebraic structure
is still, in many respects, mysterious, and higher correlation functions of the theory are just starting to be explored from the point of view of integrability. Three-point functions\footnote{Because of quantum conformal invariance, one-, two- and three-point functions contain all the information one needs.} are still quite a virgin territory, and it is still unclear if the power of integrability will provide a systematic way of computing them. When appropriately normalized, these three-point functions scale as the two-point functions in the planar limit, and one would like to compute them with spin-chain techniques. In this respect, the universal R-matrix of quantum groups has been used in the past \cite{LeClair:1991cf} to encode the braiding relations of quantum field multiplets in an integrable $1+1$-dimensional QFT, thereby extending ``off-shell" the ``on-shell" quantum-group symmetry of the S-matrix. Along the same lines, correlation functions and 
form factors\footnote{Form factors are matrix elements of field operators. They satisfy algebraic relations, called {\it form-factor axioms} \cite{Karowski:1978vz,Smirnov}, depending locally on the fields and their sectors.} could be studied with the help of the universal R-matrix. 
 
Not fully
understood is also the nature of certain fascinating dualities
that have been observed in Wilson loops and $n$-point functions.
These dualities have recently been related to algebraic structures
very similar to those responsible for the integrability of the
spectral problem, in particular to an infinite-dimensional
symmetry of the so-called Yangian type \cite{Drummond:2009fd}. It is plausible that all the Yangians we will progressively encounter in this review (sigma model, spin-chain, S-matrix, spacetime $n$-point functions) all share a common origin deeply inside the integrable structure of the theory.

Hopf algebras and quantum groups provide a suitable mathematical framework where to study these properties. Quantum groups are certain mathematical structures that emerged in Physics in the context of quantum integrable systems and  the quantum inverse scattering method developed by
the Leningrad school \cite{Faddeev:1980zy}. These structures were later axiomatized by Drinfeld and by Jimbo in terms of Hopf algebras. For standard textbook-references on Hopf algebras / quantum groups, see for instance \cite{Abe,Drin,jimbo,Chari,Kassel}. The algebraic reason for integrability can often be singled out in the existence of an infinite-dimensional non-abelian symmetry algebra (such as the Yangian) that severely constrains the dynamics. Like the angular momentum in quantum mechanics, a non-abelian algebra commuting with the Hamiltonian generates the subspaces of equal-energy states, and the spectrum re-organizes itself in terms of the corresponding irreducible representations. The S-matrix is nearly fixed purely by the symmetry algebra, and it displays very specific features \cite{Dorey:1996gd}. For a review on how Hopf algebras systematize the scattering problem in integrable systems, we refer to \cite{Delius:1995tc}.
According to an idea of Zamolodchikov's, the infinite dimensional quantum group symmetry of massive integrable field theories plays the same role in their exact solution as that of the Virasoro algebra for conformal field theories.

An accurate knowledge of the quantum algebra governing the integrability of the asymptotic problem might reveal crucial insights into the structure of the finite-size corrections as well (see, for instance, \cite{Fioravanti:1996rz}). The almost miraculous results described earlier for short operators in ${\cal{N}}=4$ SYM are a strong motivation for the search of deep algebraic structures responsible for such a matching. These structures should ideally take over the job of completing the proof of spectral equivalence to an arbitrary loop order, where the direct computation will be challenged.  

The Yangian has already turned out to be very useful to derive some results and check others, which would have otherwise taken a perhaps prohibitive amount of work. Even before the explicit derivation of all bound state S-matrices \cite{Arutyunov:2009mi}, Yangian symmetry had been used to derive the bound state Bethe equations \cite{deLeeuw:2008ye} without the need of an explicit diagonalization \cite{Arutyunov:2009iq} of the corresponding transfer matrices\footnote{One striking features of these Bethe equations is that, when expressed in terms of the appropriate bound state variables, they basically assume the same form as the Bethe equations for fundamental particles.}. Such diagonalization also makes use of the Yangian, and turns out to be essential to prove important conjectures put forward in the literature \cite{Beisert:2006qh}. These conjectures, in turn, play a very important role in deriving equations for the finite-size problem (Thermodynamic Bethe Ansatz and Y-system), and one may wonder if the Yangian could play a role in a possible group-theoretical proof of the proposals that have been so far advanced in the literature \cite{Bajnok:2010ke}, and in describing the system even at finite length \cite{Sieg:2010jt}.   

One will then be able to see if it is possible to apply this
algebraic framework to the quantization of the (dual)
two-dimensional sigma model, a formidable problem where all conventional
methods have failed so far. On the other hand, its understanding
is believed to be instrumental in order to clarify the
relationship between strings and nonperturbative phenomena in
gauge field theories. This fascinating connection has been long
sought-for through the work of many generations of theoreticians.

The point of view we would like the reader to take away from the present exposition is that there is a deep and beautiful algebraic structure, not entirely understood, which underlies the integrability of the AdS/CFT system. Fully understanding this structure will most likely provide not only a way of testing the proposals put forward so far for an exact solution, and possibly deriving them from first principles (see also \cite{Bazhanov:2010ts}), but may also represent a significant progress in Mathematics. The quantum group behind the complicated beauty of this integrable system most probably represents a new structure mathematicians have not come across so far\footnote{P. Etingof, private communication.}. 

The review is structured as follows. In section \ref{Yanu}, we briefly display two of the traditional realizations of the Yangian algebra, namely Drinfeld's first and second realization, as those that have been mostly used in the AdS/CFT context so far. In section \ref{224}, we review the Yangian symmetry of the perturbative super Yang-Mills spin-chain (section \ref{spi}), and of the classical string sigma model (section \ref{sect;sigma}), both related to the superconformal symmetry algebra $\alg{psu}(2,2|4)$. We also discuss general features of classical integrability, higher charges and Lax pairs, using as a toy model the theory of the principal chiral field (\ref{lpc}). Starting from section \ref{4}, we enter the core of the topic of this review, {\it i.e.} the quantum group structure of the AdS/CFT S-matrix, based on the centrally-extended $\alg{psl}(2|2)$ Lie superalgebra. In section \ref{sch}, we describe in detail the relevant quasi-triangular Hopf algebra and how it emerges from the spin-chain and from the string sigma model picture, together with some general notions of Lie superalgebras.  
In section \ref{sec:YS}, we describe the $\alg{psl}(2|2)$ Yangian symmetry of the S-matrix. In section \ref{classr}, we focus on the semiclassical limit of the quasi-triangular Hopf algebra. Section \ref{ssec:gen} contains standard notions related to classical $r$-matrices, Belavin-Drinfeld theorems, quantum doubles and loop-algebras, and various technology connected to the classical Yang-Baxter equation. Sections \ref{ssec:clpsu} to  \ref{diferenza} describe the corresponding AdS/CFT case, and highlight the main similarities and the important new phenomena one encounters, such as the presence of the so-called {\it secret symmetry} (section \ref{segreto}). In section \ref{stleg}, we describe bound state representations, providing details about the differential-operator formalism of \cite{Arutyunov:2008zt}, and show how to construct the corresponding S-matrices. We also briefly discuss the issue of `fusion'. This discussion is then expanded upon in section \ref{lunghe}, where long ({\it i.e.} typical) representations are treated. After recalling some notions of the representation theory of Lie superalgebras, we display the construction of long representations for the centrally-extended $\alg{psl}(2|2)$ case (section \ref{2s}), and discuss their reducibility properties. In the same section, we study the quasi-triangular structure in these representations and discuss general rectangular Young tableaux. In section \ref{ampl}, we quickly mention recent progress connected to Yangian symmetry in spacetime $n$-point amplitudes, where structures similar to those presented in this review for the spectral problem are being observed right now. In fact, very recent is the discovery of the above-mentioned `secret symmetry' also in this context \cite{Beisert:2011pn}, with the role of the secret Yangian generator played, in perfect analogy with the spectral problem we will be treating here (see section \ref{segreto}), by the helicity generator of $\alg{u}(2,2|4)$. Section \ref{concl} contains a list of conclusions that one can draw in the light of the results obtained so far, in particular for what concerns deriving general character formulas, finding the universal R-matrix and elucidating the role of the secret symmetry. All these are priorities for future investigation.   

{\it Note.} Section \ref{ampl}, which lies slightly outside the main topic, might be skipped during a first reading. The Ackowledgments (section \ref{ackn}) can also be considered as a family album.  

{\it Note.} We will not care to specify a reality condition for the algebra generators, since it will be inessential to our treatment (apart from a few instances, where it will be duly specified in order to make contact with the literature).  

{\it Note.} A few reviews concerning Yangians in AdS/CFT are already available in the literature, see for instance \cite{Beisert:2010jq,Drummond:2010ep,Torrielli:2010kq}.  

\section{Yangians}\label{Yanu}
In this section, we summarize the definitions of the Yangian
$\Ya{\alg{g}}$ of a simple Lie algebra\footnote{A Lie algebra is simple when it has no non-trivial ideals, or, equivalently, its only ideals are $\{ 0 \}$ and the algebra itself. An {\it ideal} is a subalgebra such that the commutator of the whole algebra with the ideal is contained in the ideal.} $\alg{g}$ in the so-called
Drinfeld's first and second realizations. We also give the isomorphism
between the two realizations\footnote{The reader is referred to the standard literature (see for example
\cite{Etingof,MacKay:2004tc,Molev}) for a treatment of this
subject. For the generalization to simple Lie
superalgebras, see for instance \cite{Curtright:1992kp,gz2,2003CMaPh.240...31A,stuko,Gow,Gowthesi,Creutzig:2010hr}.}. The first realization is the one originally given in
\cite{Drin}, which naturally emerges from the spin-chain
point of view \cite{Bernard:1992ya}. The second realization
\cite{Dsecond} is more suitable for constructing the universal
R-matrix \cite{Khoroshkin:1994uk}. We will not discuss here the so-called {\it RTT} realization\footnote{In the AdS/CFT case, attempts to formulate the Yangian in this fashion meet some obstacles. We thank G. Arutyunov and M. de Leeuw for discussions on this point.} and its relevance to the study of irreducible representations of Yangians and of their underlying Lie subalgebras \cite{Faddeev:1987ih,KR1,KR2}. A collection of results on the representation theory of Yangians (cf. Drinfeld polynomials) is contained in \cite{CPr}. 

In \cite{Nazarov,GowBer}, the quantum Berezinian of the Yangian of the $\alg{gl}(m|n)$ Lie superalgebra was studied, and its relation with the center elucidated. This is the analog of the relation one has between the center of the Yangian of standard Lie algebras and the quantum determinant \cite{Kulish:1981bi}, see section \ref{sect;sigma}.

\subsection{Drinfeld's first realization}\label{ssec:drinf1}

The Yangian $\Ya{g}$  is a deformation of the universal enveloping algebra
of the loop algebra $\alg{g}[u]$ associated to a Lie algebra $\alg{g}$. We remind that $\alg{g}[u]$ is the algebra of
$\alg{g}$-valued polynomials in the variable $u$. Let
$\alg{g}$ be a finite dimensional simple Lie algebra generated by
$\gen{J}^A$ with commutation relations
$\comm{\gen{J}^A}{\gen{J}^B} = f^{AB}_C\gen{J}^C$, equipped
with a non-degenerate invariant consistent supersymmetric\footnote{We remind that an {\it invariant} form (,) is such that $([X, Y ], Z) = (X, [Y, Z]) \, \forall \, X, Y, Z \in \alg{g}$, with $[,]$ the graded commutator, see for instance \cite{Kac,Frappat:1996pb}. {\it Supersymmetric} means $(X,Y)=(-)^{deg(X) deg(Y)} \, (Y,X)$, $deg$ denoting the fermionic grading, while {\it consistent} means $(even,odd)=0$.} bilinear form defined by a metric $\kappa^{AB}$. The main example of such a form is the Killing form $\kappa^{AB} = f^{AC}_D \, f^{BD}_C$, namely the trace of the product of two generators taken in the adjoint representation (see also footnote \ref{Kfor}). The
Yangian is defined by the following commutation
relations between the level-zero generators $\gen{J}^A$ (forming
$\alg{g}$) and the level-one generators $\genY{J}^A$: \<
\label{def1}\comm{\gen{J}^A}{\gen{J}^B} = f^{AB}_C\gen{J}^C,\nonumber\\
\label{rels}\comm{\gen{J}^A}{\genY{J}^B} = f^{AB}_C\genY{J}^C. \> The
generators of higher levels are defined recursively by subsequent
commutation of these basic generators, subject to the following Serre
relations (for $\alg{g} \neq \alg{sl(2)}$): \<
\label{Serr}\comm{\genY{J}^{A}}{\comm{\genY{J}^B}{\gen{J}^{C}}} +
\comm{\genY{J}^{B}}{\comm{\genY{J}^C}{\gen{J}^{A}}} +
\comm{\genY{J}^{C}}{\comm{\genY{J}^A}{\gen{J}^{B}}} \eq
\frac{1}{4} f^{AG}_{D}f^{BH}_Ef^{CK}_{F}f_{GHK}
\gen{J}^{\{D}\gen{J}^E\gen{J}^{F\}}. \> Curly brackets enclosing indices indicate complete symmetrization. Indices are raised or lowered 
with $\kappa^{AB}$ or its inverse, respectively. For the algebra $\alg{sl(2)}$, the above Serre relations are trivial, and one needs to impose a more complicated set of relations. The reader can find a detailed description of these relations in section 2.1.1 of \cite{MacKay:2004tc}. The Yangian is {\it not} a Lie algebra, as, for instance, the commutator of two level-one generators contains, in addition to a level-two generator, also a cubic combination of the level-zero generators.   

From the commutation relations (\ref{def1}) one can easily notice the existence of a {\it shift} automorphism
\begin{equation}
\label{shau}
\gen{J}^{A} \rightarrow \gen{J}^{A}, \qquad \qquad \genY{J}^A \rightarrow \genY{J}^A + c \, \gen{J}^{A},
\end{equation}
with $c$ a constant. This extends to an automorphism of the whole Yangian  $\Ya{\alg{g}}$. 

The Yangian is
equipped with a Hopf algebra structure. The coproduct is uniquely
determined for all generators by specifying it on the level-zero
and -one generators as follows: \<
\label{coptr}\copro (\gen{J}^A) \eq\gen{J}^A\otimes \mathbbmss{1}+\mathbbmss{1}\otimes\gen{J}^A, \\
\label{cop}\copro( \, \genY{J}^A )\eq\genY{J}^A\otimes \mathbbmss{1}+\mathbbmss{1}\otimes\genY{J}^A+\half
\struc^{A}_{BC}\gen{J}^B\otimes\gen{J}^C. \> Antipode and counit are easily obtained
from the Hopf algebra definitions. We remind that the antipode $\Sigma$ is an anti-involution (with a fermionic sign for superalgebras, {\it i.e.} $\Sigma(AB) = (-)^{deg(A) deg(B)} \, \Sigma(B) \, \Sigma(A)$).

\subsection{Drinfeld's second realization}\label{ssec:drinf2}

Drinfeld's second realization explicitly solves the
recursion that is implicit in the first realization. It defines
$\Ya{\alg{g}}$ in terms of generators $\kappa_{i,m},
\xi^\pm_{i,m}$, $i=1,\dots, \text{rank} \alg{g}$, $m=0,1,2,\dots$,
and relations
\begin{eqnarray}
\label{def:drinf2}
&[\kappa_{i,m},\kappa_{j,n}]=0,\quad [\kappa_{i,0},\xi^\pm_{j,m}]=\pm a_{ij} \,\xi^+_{j,m},\nonumber\\
& \comm{\xi^+_{j,m}}{\xi^-_{j,n}}=\delta_{i,j}\, \kappa_{j,n+m},\nonumber\\
&[\kappa_{i,m+1},\xi^\pm_{j,n}]-[\kappa_{i,m},\xi^\pm_{j,n+1}] = \pm \frac{1}{2} a_{ij} \{\kappa_{i,m},\xi^\pm_{j,n}\},\nonumber\\
&\comm{\xi^\pm_{i,m+1}}{\xi^\pm_{j,n}}-\comm{\xi^\pm_{i,m}}{\xi^\pm_{j,n+1}} = \pm\frac{1}{2} a_{ij} \acomm{\xi^\pm_{i,m}}{\xi^\pm_{j,n}},\nn\\
&i\neq j,\, \, \, \, n_{ij}=1+|a_{ij}|,\, \, \, \, \, Sym_{\{k\}} [\xi^\pm_{i,k_1},[\xi^\pm_{i,k_2},\dots [\xi^\pm_{i,k_{n_{ij}}}, \xi^\pm_{j,l}]\dots]]=0.
\end{eqnarray}
In these formulas, $a_{ij}$ is the Cartan matrix, which we will assume to be symmetric.

Yangians are quite different from affine Kac-Moody algebras\footnote{The affine Kac-Moody algebra associated to a finite-dimensional Lie algebra has defining relations 
\begin{eqnarray}
\label{aff}
[\gen{J}^A\otimes t^n, \gen{J}^B \otimes t^m]=[\gen{J}^A,\gen{J}^B]\otimes t^{n+m} + (\gen{J}^A,\gen{J}^B) \, n\, \delta_{n,-m} \, \gen{C}, 
\end{eqnarray}
with $\gen{C}$ a central element, and $(,)$ the Killing form. One usually adjoins a derivation to the algebra, in order to remove a root-degeneracy (see {\it e.g.} \cite{Goddard:1986bp}).}, although they share a Lie subalgebra (for $n=m=0$ in (\ref{def:drinf2}) and (\ref{aff})). Yangians can be obtained as certain quotients of the quantized version of affine Kac-Moody algebras (see \cite{Dsecond,Chari}). 

{\rm Drinfeld's first and second realization are isomorphic to each other. Let ${H}_i,
{E}_i^\pm$ be a Chevalley-Serre basis for $\alg{g}$, and
denote by ${\widehat H}_i, {\widehat E}_i^\pm$ the corresponding
level-one generators in the first realization of the Yangian.
Drinfeld \cite{Dsecond} gave the isomorphism
\begin{eqnarray}
\label{def:isom}
&\kappa_{i,0}={H}_i,\quad \xi^+_{i,0}={E}^+_i,\quad \xi^-_{i,0}={E}^-_i,\nonumber\\
&\kappa_{i,1}=\widehat{{H}}_i-v_i,\quad \xi^+_{i,1}=\widehat{{E}}^+_i-w_i,\quad \xi^-_{i,1}=\widehat{{E}}^-_i-z_i,
\end{eqnarray}
where
\<\label{def:specialel}
v_i &=& \frac{1}{4} \sum_{\beta\in\Delta^+}\left(\alpha_i,\beta\right)(\cwgen{E}_\beta^-\cwgen{E}_\beta^+ +  \cwgen{E}_\beta^+\cwgen{E}_\beta^-) - \half\cwgen{H}_i^2, \\
w_i &=& \frac{1}{4}\sum_{\beta\in\Delta^+}  \left(\cwgen{E}_\beta^-\adjb{{E}_i^+}{\cwgen{E}_\beta^+} + \adjb{{E}_i^+}{\cwgen{E}_\beta^+}\cwgen{E}_\beta^- \right) -  \frac{1}{4}\acomm{{E}_i^+}{{H}_i}, \\
z_i &=& \frac{1}{4}\sum_{\beta\in\Delta^+}
\left(\adjb{\cwgen{E}_\beta^-}{{E}_i^-}\cwgen{E}_\beta^+ +
\cwgen{E}_\beta^+ \adjb{\cwgen{E}_\beta^-}{{E}_i^-} \right) -
\frac{1}{4}\acomm{{E}_i^-}{{H}_i} . \> $\Delta^+$
denotes the set of positive root vectors, $\cwgen{E}^\pm_\beta$
are generators of the Cartan-Weyl basis constructed from ${ H}_i, \, { E}_i^\pm$, and the
adjoint action is defined as $\adjb{x}{y} = [x,y]$. For references on the connection between the two
realizations for the related case of quantum affine algebras, see
for instance \cite{kt2,DingFrenkel,Khoroshkin:1994uj,jing,HaMi}.}  

\section{The Yangian of $\alg{psu}(2,2|4)$}\label{224}
\subsection{${\cal{N}} = 4$ SYM spin chain}\label{spi}

Generically, the level-zero generators are realized on a spin-chain as local charges
\begin{eqnarray}
\gen{J}^A \, = \, \sum_k \, \gen{J}^A (k),
\end{eqnarray}
where the index $k$ runs over the spin-chain sites. In a spin chain of infinite length, the level-one Yangian generators are
typically realized in terms of bilocal combinations such as
\begin{eqnarray}
\label{lev1}
\genY{J}^A \, = \, \sum_{k<n} \, \struc^{A}_{BC} \, \gen{J}^B (k) \, \gen{J}^C (n).
\end{eqnarray}
Level-$n$ generators are $(n+1)$-local expressions. At finite length, while Casimirs of the Yangian may still be well defined, boundary effects usually prevent from
having conserved charges of the type (\ref{lev1}). For instance, if one tries to impose periodic boundary conditions, one can have that a charge like (\ref{lev1}) gives two inequivalent results when acting on two states that are related to each other by a cyclic permutation of the spins. However, 
we recommend to consult \cite{Bernard:1992ya} and references therein for notable exceptions, and for a review of this subject.

The Yangian charges (\ref{lev1}) for the ${\cal{N}} = 4$ SYM spin chain, at
infinite length and at the leading order in 't
Hooft's coupling, have been constructed in \cite{Dolan:2003uh,Dolan:2004ps}. They are based on the Lie superalgebra (superconformal algebra) $\alg{psu}(2,2|4)$. In
\cite{Dolan:2004ys}, the first two Casimirs of the Yangian have been
computed and identified with the first two local abelian
Hamiltonians of the spin-chain with periodic boundary conditions.

Perturbative corrections to the Yangian charges in definite subsectors have
been studied in
\cite{Serban:2004jf,Agarwal:2004sz,Agarwal:2005ed,Zwiebel:2006cb,Beisert:2007sk}.
The integrable structure of spin-chains with long-range
interactions, such as the one describing the perturbation
theory of ${\cal{N}} = 4$ SYM, is not entirely understood. In order to prove integrability, one has to explicitly
construct the higher Hamiltonians, or engineer a method of generating
them (see for instance
\cite{Beisert:2003tq,Beisert:2005wv,Beisert:2007jv,Zwiebel:2008gr}).
In absence of other standard tools, Yangian symmetry would
constitute a formal proof of integrability order by order in
perturbation theory. The suitable two-loop
expression of the Yangian charges (\ref{lev1}) for the
$\alg{su}(2|1)$ sector has been derived in \cite{Zwiebel:2006cb}.
In \cite{Beisert:2007sk}, a large degeneracy of states in the
$\alg{psu}(1,1|2)$ sector has been explained by finding nonlocal
charges related to the loop-algebra of the $\alg{su}(2)$
automorphism of $\alg{psu}(1,1|2)$. Further references include
\cite{Agarwal:2004cb,Mikhailov:2004ca,Agarwal:2006nv,Ihry:2008gm,Mansson:2008xv,Zwiebel:2009vb,Zoubos:2010kh}.

\subsection{Sigma model}\label{sect;sigma}
The emergence of higher non-local charges of Yangian type from a two-dimensional
classically integrable field theory\footnote{The literature devoted to this subject is extensive. We mention here, as a starting point for the interested reader, the early papers \cite{Luscher:1977rq,Brezin:1979am}, and the papers \cite{Curtright:1979am,Ridout:2011wx} for the supersymmetric case.} can be
understood {\it via} the example of the so-called Principal Chiral
Model (PCM). This is the theory of a field $g=g(x,t)$ taking
values in a connected simple compact\footnote{Compactness is assumed in order to have finite-dimensional unitary representations.} Lie group $G$, with a Lagrangian given by
\begin{eqnarray}
\label{lpc}
\mathscr{L} = {\rm tr} [\partial_\mu \, g^{-1} \, \partial^\mu \,
g].
\end{eqnarray}
This Lagrangian has left and right global symmetries $g \rightarrow h g, \, g  h$, with $h\in G$. The corresponding Noether currents are given by
\begin{eqnarray}
\label{lr}
J_{\mu}^{L,R} = -(\partial_\mu g) g^{-1},  \, g^{-1} \, (\partial_\mu g).
\end{eqnarray}
These currents (cf. Cartan 1-forms on group manifolds) belong to the Lie algebra $\alg{g}$ of $G$, which
is generated by certain $T^A$'s satisfying $[T^A, T^B]=\,
f^{A B}_C \, T^C$. This means that one can write these currents
(and the corresponding charges $\alg{J}$) as
\begin{eqnarray}
J_\mu = J_\mu^A \, T_A, \qquad \, \, \partial^\mu J_\mu^A \, = \, 0, \qquad \alg{J}^A \, = \, \int_{-\infty}^{\infty} dx \, \, J_0^A.
\end{eqnarray}
By subsequent integration by parts and disregarding boundary terms, the action associated to the Lagrangian (\ref{lpc}) can be brought to a form quadratic in the Noether currents.
It is easy to check that, upon using the equations of motion, such currents satisfy the condition of ``flatness" (cf. Maurer-Cartan equation):
\begin{eqnarray}
\label{piat}
\partial_0 J_1 - \partial_1 J_0 + [J_0,J_1] \, = \, 0.
\end{eqnarray}
 $(J_0,J_1)$ form a so-called {\it Lax pair}\footnote{A typical example of a Lax pair is the following. Consider the equation $\frac{d}{dt} A = [A,B]$, with $A,B$ two matrices. It is straightforward to show that $\tr A^n$ is a conserved charge for arbitrary $n$. The generating function for all these charges is $\tr \exp (A)$. $(A,B)$ form a Lax pair, and one can directly generate conserved charges from the Lax pair by a suitable trace operation. The condition (\ref{piat}) is also the consistency (or, {\it integrability}) condition for the system of equations 
\begin{eqnarray} 
\label{lineare}
&&\partial_1 \, F \, = \, J_0 \, F, \nonumber\\
&&\partial_0 \, F \, = \, J_1 \, F,
\end{eqnarray}
where $F$ is an arbitrary vector. The system (\ref{lineare}) defines the so-called {\it auxiliary linear problem}, and it constitutes the starting point of the classical inverse scattering method. The very existence of a Lax pair representation for the dynamical equations can often be taken as a synonym of integrability.}. 

Together with the conservation of $J$, the flatness condition
automatically implies that the following non-local currents are
conserved\footnote{Indeed, the currents themselves satisfy $\partial^\mu \, {\widehat{J}_\mu}^{{}\, \, A} = 0$.}:
\begin{eqnarray}
\label{nnl}
{\widehat{J}_\mu}^{{}\, \, A} = \epsilon_{\mu \nu} \, J^{\nu, \, A} + \frac{1}{2} \, f^A_{BC} \, J_\mu^B \int_{-\infty}^x dx' \, J_0^C (x'),
\end{eqnarray}
\begin{eqnarray}
\label{nlc}
\frac{d}{dt} \, \widehat{\alg{J}}^A =  \frac{d}{dt} \int_{-\infty}^{\infty} dx \, {\widehat{J}_0}^{{}\, \, A} (x) =0.
\end{eqnarray}
Recursive
application of the same argument leads to the conservation of an
infinite tower of non-local charges. Existence of such higher
non-local charges implies the classical integrability of the
model. These charges have non-trivial Poisson brackets among
themselves and with the Noether charges. In the absence of
anomalies, the quantum version of these charges
\cite{Luscher:1977uq} forms the non-abelian structure of the
Yangian. Typically, one can find a family of flat connections depending on a continuous parameter $\lambda$, often called {\it spectral parameter}. 

A remark is in order. Looking at the charges $\widehat{\alg{J}}^A$ in (\ref{nlc}), one may wonder whether the non-local part is actually just one half of the square of the local charges ${\alg{J}}^A$ (which would mean that  one has not really found new independent conserved charges). In fact, one could be tempted to rewrite the nested integral in (\ref{nlc}) as half of the same expression, plus half of the expression where a change of variables has been performed to swap the integration variables $x$ and $x'$. This would reconstruct the square of ${\alg{J}}^A$, were it not for a minus sign coming from the structure constants.

We also notice that, since we have left and right currents (\ref{lr}), two copies of the Yangian, constructed according to the above procedure, will actually be present in the PCM.

\smallskip

The path-ordered exponential of the spatial part of the Lax connection is called  the {\it monodromy matrix}. Its trace, called the {\it transfer matrix}, is a generating function for the tower of (non-local) conserved charges. One recovers these charges as a Taylor expansion around a specific value of $\lambda$, for instance $\lambda =\infty$. Expansion around a different point, say, $\lambda=0$,  and usually after taking the logarithm and a suitable combination of derivatives w.r.t. $\lambda$, may instead generate the tower of local commuting charges giving rise to the integrable Hamiltonians \cite{Tarasov:1983cj,Bernard:1992ya}. The latter expansion point is typically a special point for the {\it $R$-matrix} of the problem (see formula (\ref{vari}) and subsequent discussion), and it is usually located where the $R$-matrix degenerates into a projector\footnote{\label{simmetr} Consider the following $R$-matrix (proportional to the so-called {\it Yang's $R$-matrix}):
$$
R=\frac{u}{u\pm 1}\( \mathbbmss{1}\otimes \mathbbmss{1} \, + \, \frac{P}{u}\),
$$
with $P$ the permutation operator $P \, a \otimes b = (-)^{deg(a) deg(b)} \, b \otimes a$. One can see that the residue at the pole $u=\mp 1$ is proportional to $\mathbbmss{1}\otimes \mathbbmss{1} \, \mp \, P$, which projects onto the antisymmetric (resp., symmetric) tensor-product representation. In this review, we will always assume that the S-matrix, whether it will be denoted by $R$ or S, will act as a map from $V_1 \otimes V_2$ to $V_1 \otimes V_2$, with $V_1$ and $V_2$ two algebra modules. For all practical purposes, we will think of $R$ and S as one and the same mathematical object, and indifferently use either letters in order to mantain the text populated with symbols familiar to both physicists (S) and mathematicians ($R$).}. Often, however, extracting the commuting charges is not a straightforward operation. For the case of the PCM, for instance, we refer to the specific treatment of \cite{Evans:1999mj}.

\smallskip

Let us restrict to the case of models with $\alg{gl}(n)$ Yangian symmetry for a moment. The local commuting charges form a commutative (Cartan) subalgebra of the Yangian, and they have determinantal expressions (see for instance \cite{KR}). The center of the Yangian belongs to this commutative subalgebra and it is generated by one of this determinantal expressions, called the {\it quantum determinant}. For supersymmetric theories, the trace and the supertrace of the monodromy matrix can generate two different families of commuting Hamiltonians.

\smallskip

The classical integrability of the Green-Schwarz superstring sigma
model in the \AdSxS\  background has been established in
\cite{Bena:2003wd}. There, the corresponding infinite set of
non-local classically conserved charges has been found, according
to a logic very close to the one described above (similar
observations for the bosonic part of the action were made in
\cite{Mandal:2002fs}). The fact that the string sigma model is actually based on a coset group makes the treatment slightly more involved, but conceptually quite similar. Further work in this context can be found
in
\cite{Alday:2003zb,Arutyunov:2003rg,Hou:2004ru,Hatsuda:2004it,Das:2004hy,Alday:2005gi,Frolov:2005dj,Das:2005hp,Arutyunov:2008if,Stefanski:2008ik}.

\smallskip

{\rm We conclude this section with a remark on the Hopf
algebra structure of the non-local charges (\ref{nlc}). How expressions like
(\ref{nnl}), (\ref{nlc}) can give rise to the coproduct (\ref{cop})
is the outcome of a contour integral analysis contained {\it e.g.} in
\cite{Bernard:1992mu}. There exists also a
semiclassical argument \cite{Luscher:1977rq,MacKay:1992rc}, which we will now present. One
can imagine two well-separated solitonic excitations (see Figure
1)
as the classical version of a scattering state.  The principal chiral model has such solutions (see for instance \cite{Zakharov:1978wc}). Soliton $1$ is
located inside the region $(-\infty,0)$, while soliton $2$ is
inside $(0,\infty)$. If one defines the {\it semiclassical action}
of a charge on such a solution as the charge itself evaluated on the
profile, one can conveniently split the integral of the current in
the individual domains which are most relevant for each of the two
solitons, respectively. In other words,

\centerline{\setlength{\unitlength}{.9cm}
\begin{picture}(6,4)(-3,-2)
\put(-4.5,0){\vector(1,0){9}}
\put(0,-1.5){\vector(0,1){3}}
\qbezier(1.5,0)(2.5,2.5)
(3.5,0)
\qbezier(-3.5,0)(-2.5,2.5)
(-1.5,0)
\end{picture}}

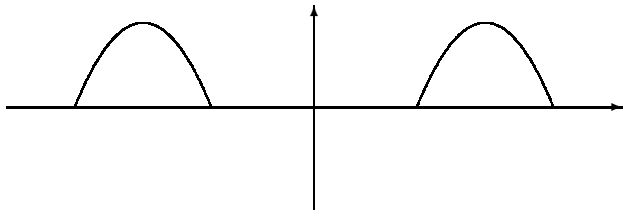
{Fig. 1: A semiclassical scattering state with two well-separated solitons.}

\begin{eqnarray}
\alg{J}^A_{|profile} &=&  \int_{-\infty}^{\infty} dx \, {J_0^A}_{|profile}
=\int_{-\infty}^{0} dx \, J_0^A \, + \, \int_{0}^{\infty} dx \, J_0^A \, \nonumber \\
&\sim & \, \, \alg{J}^A_1 \, + \alg{J}^A_2 \,\, \, \, \,  \longrightarrow \, \,
\Delta (\alg{J}^A) = \alg{J}^A \otimes \mathbbmss{1} + \mathbbmss{1} \otimes \alg{J}^A
\end{eqnarray}
and, from (\ref{nnl}) and (\ref{nlc}),

\begin{eqnarray}
\widehat{\alg{J}}^A_{|profile} &=&  \int_{-\infty}^{0} dx \, J_1^A + \frac{1}{2} \, f^A_{BC} \, \int_{-\infty}^{0} dx \, J_0^B (x) \, \int_{-\infty}^{x}  dy \, J_0^C (y) \, \nonumber\\
&&+ \, \int_{0}^{\infty} dx \, J_1^A  \, + \, \frac{1}{2} \, f^A_{BC} \, \int_{0}^{\infty} dx \, J_0^B (x) \, \int_{0}^{x} dy \,  J_0^C (y) \nonumber\\
&&+ \, \frac{1}{2} \, f^A_{BC} \, \int_{0}^{\infty} dx \, J_0^B (x) \, \int_{-\infty}^{0} dy \, J_0^C (y),
\end{eqnarray}
which schematically reproduces (\ref{cop}). Upon quantization in absence of anomalies, one can promote this action to the action of charge-operators on the Hilbert space of the asymptotic states. One can therefore directly link the non-locality of the classical charge to the ``non-triviality" of the corresponding coproduct\footnote{\rm One calls `` trivial" a coproduct of the (local) type (\ref{coptr}).}.
}

\section{The centrally-extended $\alg{psl}(2|2)$ Yangian}\label{4}

\subsection{The Hopf algebra of the S-matrix}\label{sch}

As we will shortly motivate, the algebra we will focus our attention on is given by (two copies of\footnote{In what follows, it will be sufficient to consider one copy of this algebra, as the two copies can be treated independently.}) the centrally-extended $\alg{psl}(2|2)$ Lie superalgebra  (which we will call $\alg{psl}(2|2)_c$ for short). This algebra emerges upon choosing a vacuum for the spin-chain  \cite{Beisert:2005tm} (see the discussion following formula (\ref{mild})), and the same algebra arises in the decompactification limit of the string sigma-model \cite{Arutyunov:2006ak}.

We begin by reporting
the commutation relations of (a single copy of) the algebra. For convenience of the reader, we first explicitly spell out the commutators of the two sets of $\alg{sl}(2)$ bosonic generators, in order to display our conventions for the Cartan matrix entry in these two sectors:
\begin{eqnarray}
\begin{array}{ll}
\ [\mathbb{L}_{1}^{\ 1},\mathbb{L}_{1}^{\ 2}] = 2 \mathbb{L}_{1}^{\ 2}, \ \, \,
\, \, \, \, \, [\mathbb{L}_{1}^{\ 1},\mathbb{L}_{2}^{\ 1}] = - 2
\mathbb{L}_{2}^{\ 1}, &  \ [\mathbb{L}_{1}^{\ 2},\mathbb{L}_{2}^{\ 1}] =
\mathbb{L}_{1}^{\ 1},\\
\ [\mathbb{R}_{3}^{\ 3},\mathbb{R}_{3}^{\ 4}] = 2 \mathbb{R}_{3}^{\ 4}, \ \, \,
\, \, \, \, \, [\mathbb{R}_{3}^{\ 3},\mathbb{R}_{4}^{\ 3}] = - 2
\mathbb{R}_{4}^{\ 3}, &  \ [\mathbb{R}_{3}^{\ 4},\mathbb{R}_{4}^{\ 3}] =
\mathbb{R}_{3}^{\ 3}.\\
\end{array}
\end{eqnarray}
The remaining commutators are as follows (Latin indices refer in our conventions to the $\mathbb{L}$-type of $\alg{sl}(2)$ generators, while Greek indices to the $\mathbb{R}$-type): 
\begin{eqnarray}
\label{tnsr}
\begin{array}{ll}\ [\mathbb{L}_{a}^{\ b},\mathbb{G}^{\alpha}_{c}] =
\delta_{c}^{b}\mathbb{G}^{\alpha}_{a}-\frac{1}{2}\delta_{a}^{b}\mathbb{G}^{\alpha}_{c}, &   \
[\mathbb{R}_{\alpha}^{\ \beta},\mathbb{Q}_{\gamma}^{a}] =
\delta_{\gamma}^{\beta}\mathbb{Q}_{\alpha}^{a}-\frac{1}{2}\delta_{\alpha}^{\beta}\mathbb{Q}_{\gamma}^{A},\\
\ [\mathbb{L}_{a}^{\ b},\mathbb{Q}_{\alpha}^{c}] =
-\delta_{a}^{c}\mathbb{Q}_{\alpha}^{b}+\frac{1}{2}\delta_{a}^{b}\mathbb{Q}_{\alpha}^{c}, & \
[\mathbb{R}_{\alpha}^{\ \beta},\mathbb{G}^{\gamma}_{a}] =
-\delta^{\gamma}_{\alpha}\mathbb{G}^{\beta}_{a}+\frac{1}{2}\delta_{\alpha}^{\beta}\mathbb{G}^{\gamma}_{a},\\
\ \{\mathbb{Q}_{\alpha}^{\ a},\mathbb{Q}_{\beta}^{\
b}\}=\epsilon_{\alpha\beta}\epsilon^{ab}\mathbb{C},&\ \{\mathbb{G}^{\
\alpha}_{a},\mathbb{G}^{\
\beta}_{b}\}=\epsilon^{\alpha\beta}\epsilon_{ab}\mathbb{C}^{\dag},\\
\ \{\mathbb{Q}_{\alpha}^{a},\mathbb{G}^{\beta}_{b}\} =
\delta_{b}^{a}\mathbb{R}_{\alpha}^{\ \beta} +
\delta_{\alpha}^{\beta}\mathbb{L}_{b}^{\ a}
+\frac{1}{2}\delta_{b}^{a}\delta_{\alpha}^{\beta}\mathbb{H}.&
\end{array}
\end{eqnarray}
The elements $\mathbb{H}$, $\mathbb{C}$ and
$\mathbb{C}^{\dag}$ commute with all the generators. The `dagger' symbol on the third central element is to remind that, in unitary representations, $\mathbb{C}$ and $\mathbb{C}^{\dag}$ are one the complex conjugate of the other.  

As usual for Lie superalgebras, $[even,odd]\subset odd$, therefore the odd part forms a representation of the even subalgebra. In this case, the even part is given by the $\alg{sl}(2) \oplus \alg{sl}(2)$ subalgebra, with generators $\mathbb{L}$ and $\mathbb{R}$ satisfying $\sum_a \, \mathbb{L}_{a}^{\ a} =0$ and $\sum_\alpha \, \mathbb{R}_{\alpha}^{\ \alpha} =0$, together with the center $\{ \mathbb{H}, \mathbb{C},
\mathbb{C}^{\dag} \}$. The odd part forms the representation $(2,\bar{2})\oplus (\bar{2},2)$ ($\mathbb{Q}$ and $\mathbb{G}$, respectively) \cite{Frappat:1996pb}.

The fact that a simple Lie superalgebra can admit such a large central extension is peculiar to $\alg{psl}(2|2)$. In fact, $A(1,1)\equiv \alg{psl}(2|2)$ is the only basic classical \footnote{\label{Kfor}Let us focus on simple Lie superalgebras. We remind that a {\it classical} Lie superalgebra is such that its even subalgebra is a {\it reductive} Lie algebra, namely a direct sum of semisimple and abelian Lie algebras.  A classical Lie superalgebra is called {\it basic} if it admits a non-degenerate invariant supersymmetric bilinear form, otherwise it is called {\it strange}. One usually takes as such a form the Killing form, {\it i.e.} the supertrace of the product of two generators in the adjoint representation (although any representation besides the adjoint would provide a form with the necessary properties, see \cite{Kac:1977qb} for details). The Killing form is proportional to the dual Coxeter number. The dual Coxeter number $\alg{c}_2$ is defined as $f^{A B}_C \, f^{A B}_D =\alg{c}_2 \, \delta_{C D}$, and it is related to trace of the quadratic Casimir in the adjoint representation. When the dual Coxeter number is zero, one can (in the light of the remark in brackets we just made above) take the supertrace in any other representation for which the form does not give an identically vanishing result (one could try, for instance, the fundamental representation). The Lie superalgebras $A(n,n)\equiv \alg{psl}(n+1|n+1)$, $n\geq 1$, and $D(2,1;\epsilon)$, in particular, have zero Killing form, but they are basic. For a very direct way of exhibiting a non-degenerate bilinear form for $A(n,n)$, $n\geq 1$, one can consult for instance \cite{Evans:1990qq}, Appendix B. One can take a {\it distinguished} simple root system ({\it i.e.}, with the least number of fermionic simple roots). Notice that the Cartan matrix of $A(n,n)$ in this system is degenerate \cite{Frappat:1996pb}. This has to do with the number of Cartan elements needed to achieve a Chevalley-Serre realization, which forces one row in the Cartan matrix to be dependent on the other ones. Notice also that, after centrally-extending $A(n,n)$ to $\alg{sl}(n+1|n+1)$ (the algebra of supertraceless $n+1|n+1 \times n+1|n+1$ matrices) by adding one central element (see the discussion in the text immediately following this footnote), the supertrace in the defining $n+1|n+1 \times n+1|n+1$ representation immediately becomes degenerate, since the product of any generator with the central element is still supertraceless. For more details in the case of coset supergroups, especially those relevant to AdS/CFT, see for instance \cite{SchaferNameki:2009xr}, where the suitable decomposition of $\alg{psu}(2,2|4)$ (related to $A(3,3)$) and related coset reductions of the bilinear form are studied, and \cite{Babichenko:2009dk}. We refer to \cite{FabianThesis,Frappat:1996pb} for further details and explanations.} simple Lie superalgebra for which this happens \cite{IoharaKoga}. Leaving aside affine extensions, in fact, one either has no central extensions at all, or, for the series $A(n,n)$ with $n\neq 1$, one has a one-dimensional central extension to $\alg{sl}(n+1|n+1)$, the algebra of supertraceless matrices of dimension $n+1|n+1$  (in a bosons$|$fermions notation). This is because the $n+1|n+1 \times n+1|n+1$-identity matrix is also supertraceless. But only for $A(1,1)$ one can simultaneously use two epsilon-tensors, and allow, besides the $\alg{sl}(2|2)$ generator $\mathbb{H}$, two further independent central charges $\mathbb{C}$ and $\mathbb{C}^\dag$ to appear on the r.h.s. of the two `same-type' anticommutators of supercharges, $\{ \alg{\mathbb{Q} },\alg{\mathbb{Q} } \}$ and $\{ \alg{\mathbb{G} },\alg{\mathbb{G} } \}$ respectively, as shown in (\ref{tnsr}). 

The representation relevant to super Yang-Mills, and which we will call ``fundamental", is that of a {\it dynamical} spin-chain, {\it i.e.} sites can be created or
destroyed as a byproduct of the action of the Lie superalgebra generators (``length-changing" action). In the basis of \cite{Beisert:2005tm}, the length-changing action of, for instance,
the central charges goes as follows:
\begin{eqnarray}
\label{mild}
\mathbb{H} \, |p\rangle \, &=& \, \epsilon (p) \, |p\rangle, \nonumber\\
\mathbb{C} \, |p\rangle \, = \, c (p) \, |p \, Z^-\rangle, \qquad &&\qquad \mathbb{C}^{\dag} \, |p\rangle \, = \, \bar{c} (p) \, |p \, Z^+\rangle,
\end{eqnarray}
where $Z^{+ (-)}$ adds (removes) one site to (from) the chain. The length-changing action of the symmetry generators is easily justified when realizing that they, as well as the Hamiltonian / mixing matrix of anomalous dimensions, can mix operators with different numbers of bosonic and fermionic fields. In the case of the Hamiltonian, this mixing is restricted to operators which have the same bare scaling dimension. 

A {\it magnon} is a spin-wave excitation on the spin-chain. We denote as $|p\rangle$ the one-magnon state $|p\rangle = \sum_n \, e^{i p n} \, |\cdots Z \, Z \, \phi(n) \, \, Z \cdots \rangle$. $Z$ is a chosen complex combination of two of the six real scalar fields in ${\cal{N}}=4$ SYM. $\phi$ is one of the $4$ possible orientations of the ``spin" (or ``polarizations") in the fundamental representation of $\alg{psl}(2|2)_c$, here taken at position $n$ along the chain. The two bosonic polarizations are denoted as $w_1$, $w_2$ and the two fermionic ones as $\theta_3$, $\theta_4$. In the absence of magnonic excitations, one simply obtains the vacuum state $|\cdots Z \, Z \cdots \rangle$. Indeed, operators of the form $\tr \, Z^J$ in the SYM theory are {\it half BPS}, in that they are annihilated by half of the supersymmetries. Their scaling dimension is therefore protected from receiving quantum corrections. For fixed $J$,  $\tr \, Z^J$ corresponds to a ferromagnetic vacuum\footnote{At one loop, the ferromagnetic nature is essentially due to the presence of the squared coupling constant $g_{YM}^2$ in front of the (Heisenberg-like) Hamiltonian. One would eventually like to have this squared coupling real and positive.}.
The algebra $\alg{psu}(2|2)$ (and its central extension) is the algebra that preserves such a vacuum, and the excitations on the vacuum form irreducible representations of this residual algebra. One of the $\alg{su}(2)$'s corresponds to the residual R-symmetry\footnote{R-symmetry is the symmetry that rotates the generators of the extended (${\cal{N}}=4$) supersymmetry. Choosing a complexified scalar breaks the original $\alg{so}(6)$ R-symmetry to two copies of $\alg{su}(2)$.}, the other $\alg{su}(2)$ to the residual Lorentz algebra\footnote{The vacuum preserves the Lorentz algebra, which provides other two copies of $\alg{su}(2)$. In total, one sees how two copies of $\alg{psu}(2|2)$ are bound to arise. These two copies are also related to the two wings of the $\alg{psu}(2,2|4)$ Dynkin diagram, for an appropriate choice of simple root-system.}. 

A state like $|\cdots Z \, Z \cdots \rangle$ is obtained from $\tr \, Z^J$ in the limit $J \to \infty$ (``asymptotic problem"). On the string theory side of the correspondence, this amounts to relaxing the level-matching condition and effectively dealing with open-string excitations (the `giant magnons' of \cite{Hofman:2006xt}). The analysis of the finize-size effects, which concerns the true gauge-invariant SYM operators at finite $J$ (dual to closed strings), is postponed to the solution of the asymptotic problem. The asymptotic problem is in fact easier to attack, as it can be treated in terms of scattering {\it data}.

\smallskip

The length-changing property can be interpreted, at the Hopf
algebra level, as a non-local modification of the (otherwise
trivial) coproduct \cite{Gomez:2006va,Plefka:2006ze}. One can see
how this works, for instance, in the case of the central charges\footnote{It is worth noticing that, in sectors larger than the one corresponding to the $\alg{psl(2|2)}_c$ excitations, a similar Hopf algebra interpretation is far less direct, if possible at all, given that the length-changing pattern may be wilder than (\ref{mild}).}.
When acting on a two-particle state, one needs to compute
\begin{eqnarray}
\label{resc}
&&\mathbb{C} \otimes \mathbbmss{1} \, |p_1\rangle \otimes |p_2\rangle = \nonumber\\
&&\mathbb{C} \otimes \mathbbmss{1} \, \sum_{n_1 << n_2} \, e^{i \, p_1 \, n_1 \, + \, i \, p_2 \, n_2} \, |\cdots Z \, Z \, \phi_1 \, \underbrace{Z \cdots Z}_{n_2 - n_1 - 1} \, \phi_2 \, Z \cdots \rangle  \, =\nonumber\\
&& \qquad \qquad \, \, \, (n_2 \rightarrow n_2 + 1) \, \, = \, \, c (p_1) \, e^{i p_2}  \, |p_1\rangle \otimes |p_2\rangle.
\end{eqnarray}

The rescaling $n_2 \rightarrow n_2 + 1$ is needed to bring back the state to its original form with $n_2 - n_1$ vacuum sites between the two excitations, because that is what is defined as $|p_1\rangle \otimes |p_2\rangle$ from the very beginning.
We have considered the state as infinitely extended on both sides, therefore the rescaling only involves the action of $\mathbb{C} \otimes \mathbbmss{1}$, and not of $\mathbbmss{1} \otimes \mathbb{C}$. In other words, only the space in-between the two excitations matters.
Such an action is clearly non-local, as
acting on the first magnon (with momentum $p_1$) produces a result
which depends also on the momentum $p_2$ of the second magnon. 

\smallskip

The next step is to compute the S-matrix governing the scattering of the two excitations against each other. Thanks to integrability, when two particles cross paths they keep their momenta $p_1$ and $p_2$ unchanged, but their spins are transformed by means of a non-trivial matrix, the S-matrix itself. The latter therefore acts trivially on the space of momenta, but reshuffles the internal quantum numbers (see also the Introduction). The requirement of invariance under the symmetry of the problem amounts to the commutation of the S-matrix with the coproduct. The coproduct is in fact nothing else than the action of the symmetry on two-particle states. Once again, because one assumes the integrability of the problem, the two-particle scattering contains the whole information required to decipher the entire dynamics of the system. 

\smallskip

Imposing the above-mentioned 
invariance condition is equivalent to requiring $\Delta (\mathbb{C}) {\rm S} =
{\rm S} \Delta (\mathbb{C})$ for the S-matrix. In our case, this implies computing

\begin{eqnarray}
\label{smatcop}
{\rm S} \, \Delta(\mathbb{C}) \, = {\rm S} \, [\mathbb{C} \otimes \mathbbmss{1} + \mathbbmss{1} \otimes \mathbb{C}] \, = {\rm S} \, [e^{i p_2} \mathbb{C}_{local} \otimes \mathbbmss{1} + \mathbbmss{1} \otimes \mathbb{C}_{local}],
\end{eqnarray}
where $\mathbb{C}_{local}$ is the {\it local} part of
$\mathbb{C}$, acting as $\mathbb{C}_{local} |p\rangle = c(p)
|p\rangle$. An analogous argument works for $\Delta (\mathbb{C}) {\rm S}$. As $p_2$ naturally pertains to the second space in the tensor product, one is to read off (\ref{smatcop}) the following coproduct
\begin{eqnarray}
\label{coprodo}
\Delta (\mathbb{C}_{local}) = \mathbb{C}_{local} \otimes e^{i p} + \mathbbmss{1} \otimes \mathbb{C}_{local}.
\end{eqnarray}
Formula (\ref{coprodo}) is the Hopf-algebra manifestation of the
non-triviality of the coproduct. Particle labels $1,2$ being taken care of, one drops the subscript {\it local}, entirely encoding the non-locality of the action in the deformed coalgebra structure (\ref{coprodo}). 

A similar coproduct arises for all the other (super)charges of $\alg{psl}(2|2)_c$. It is
controlled by an additive quantum number $[[A]]$ such that
\begin{eqnarray}
\label{coprodot}
\Delta (\alg{J}^A) = \alg{J}^A \otimes e^{i [[A]] p} + \mathbbmss{1} \otimes \alg{J}^A
\end{eqnarray}
and $\Delta(e^{ip}) = e^{i p} \otimes e^{ip}$. In a convenient frame\footnote{\label{frame} The notion of {\it frame} will be expanded upon in the discussion preceding formula (\ref{vari}). However, let us briefly introduce the concept at this point for the convenience of the reader. As the detailed analysis of \cite{Arutyunov:2006yd} made precise, changes in the choice of basis (``gauge") for the scattering states modify the explicit form
of the S-matrix, and necessarily of the coproduct. The physical content is however unchanged. That is, these transformations {\it do not change the eigenvalues of the transfer matrix constructed with the S-matrix, and therefore the energies of the spectrum one obtains {\it via} the Algebraic Bethe Ansatz procedure}. In \cite{Arutyunov:2006yd}, these ``gauge" transformations are seen as acting on the relevant Zamolodchikov-Faddeev operators. Equivalently, these transformations can be interpreted as acting on the coproduct as certain similarity transformations or as twists. They can be non-local from the point of view of the one-particle basis, {\it i.e.} they can depend on both momenta of the two scattering particles. This feature sets them quite outside the set of innocuous changes of reference basis one normally allows for when dealing with algebra modules. Moreover, as we will display in formula (\ref{tw}) and remark in the related discussion, these twists can lack a matrix representation, and should rather be thought of as acting {\it via} differential operators. Nonetheless, the essential features of the Hopf algebra that is generated do not change (in particular, one cannot `twist away' the deformation). An appropriate choice of ``frame", or ``gauge" (basis), is essential to obtain an S-matrix that solves the traditional Yang-Baxter Equation (YBE, see equation (\ref{ybe}) and related discussion), and not a twisted version of it ({\it i.e.} with extra momentum-dependent phase factors explicitly appearing in the equation). Two important frames where the YBE is solved in its traditional untwisted form are the $\alg{psl(1|2)}$ frame \cite{Beisert:2005tm,Janik:2006dc,Plefka:2006ze}, as shown in \cite{Torrielli:2007mc}, and the (more symmetric) $\alg{sl}(2)$ frame of \cite{Arutyunov:2006yd}. In the manifest $\alg{sl(1|2)}$ frame, an entire $\alg{sl(1|2)}$ subalgebra has a trivial coproduct (while for the raising/lowering generators of one of the two $\alg{sl}(2)$'s the coproduct is non-trivial). The rest of the generators have a coproduct of the type (\ref{coprodot}) but with {\it integer} quantum numbers $[[A]]$ {\it only}. This frame is the closest one to the spin-chain picture, where the rescaling (\ref{resc}) can only produce phases with exponents which are integer portions of the momenta. The quantum numbers $[[A]]$ are then found directly from the length-changing picture originally given in \cite{Beisert:2005tm}, as shown in \cite{Plefka:2006ze}. Instead, in the $\alg{sl}(2)$ (or, ``string") frame, adopted as standard soon after its introduction, all the bosonic $\alg{sl}(2)$ generators have a trivial coproduct, and the structure of the S-matrix is mostly symmetric.} one has that the only non-zero quantum numbers are $[[\mathbb{Q}]]=\frac{1}{2}$,  $[[\mathbb{G}]]=- \frac{1}{2}$, $[[\mathbb{C}]]=1$,  $[[\mathbb{C}^\dag]]=- 1$, from which (\ref{coprodot})  can easily be shown to be a (Lie) algebra
homomorphism. The corresponding counit $\alg{e}$ and antipode $\Sigma$ are straightforwardly
derived from the Hopf algebra axioms, and the whole structure can be
proven to define a consistent Hopf algebra. In particular,
\begin{equation}
\label{u1}
\Sigma (\alg{J}^A ) = - e^{- i [[A]] p} \alg{J}^A.
\end{equation}
This antipode is idempotent, {\it i.e.} it squares to the identity (in fact,  
 $\Sigma (e^{i p}) = e^{- i p}$). The antipode is an anti-involution\footnote{This means, $\Sigma (AB) \, = (-)^{deg(A) deg(B)} \Sigma (B) \Sigma (A)$.} related to crossing symmetry\footnote{Crossing symmetry is usually required in relativistic scattering. In the AdS/CFT case, where the spin-chain / gauge fixed sigma model is non-relativistic, the existence of a charge conjugation map acting on the fundamental representation, and of the associated crossing symmetry of the scattering matrix with scalar factor (relevant for deriving the asymptotic Bethe equations), was a crucial discovery of \cite{Janik:2006dc}. We also remark that the $R$-matrix one associates to the inverse scattering problem and, possibly, to the exact (finite-size) Bethe equations, need not be crossing symmetric. We thank D. Fioravanti for discussions on this point.}. Since $[[\mathbb{H}]]=0$, the energy simply changes sign under crossing, but the other central charges have non-zero ``$[[A]]$" quantum number, and  (\ref{u1}) implies that they undergo an additional $U(1)$ rotation \cite{Arutyunov:2007tc}.

As anticipated in footnote \ref{frame},
(non-local) changes of basis (`frame') for the scattering states can make
the factors $e^{i [[A]] p}$ appear in different places in the coproduct
(possibly with a different power), without significantly changing
the fundamental Hopf algebra structure. Some of these non-local changes of basis can be implemented by formally defining an operator $\mathbb{J}$ such that, for example, 
\begin{equation}
\label{nomat1}
[\mathbb{J},\mathbb{C}]= \mathbb{C}.
\end{equation}
In this way, one can show that\footnote{The author thanks Peter Schupp, Jan Plefka and Fabian Spill for an early collaboration on this problem.}
\begin{eqnarray}
\label{tw}
\mathbb{C} \otimes e^{i p} + \mathbbmss{1} \otimes \mathbb{C} =
e^{i(\mathbb{J} \otimes p - p \otimes \mathbb{J})}(\mathbb{C} \otimes \mathbbmss{1} + e^{i p} \otimes \mathbb{C}) e^{-i(\mathbb{J} \otimes p - p \otimes \mathbb{J})}.
\end{eqnarray}
In other words, a formal twist can move the length-changing operators $Z^\pm$ in (\ref{mild}) from the left to the right of the local action of the algebra generators on the spin-chain. Of course, the operator $\mathbb{J}$ will have to do a similar job for all the other generators besides  $\mathbb{C} $. This means that $\mathbb{J}$ will have to satisfy additional commutation relations besides and of the type (\ref{nomat1}). One complication is given by the fact that $p$ and $\mathbb{J}$ have to be taken to commute with one another in (\ref{tw}), which apparently clashes with (\ref{nomat1}). A way around this obstacle is found in \cite{FabianThesis} in one particular frame. In the frame chosen there, in fact, one can express the generator $\mathbb{J}$ in terms of derivatives with respect to other free parameters that label the representation in that particular frame, without the explicit appearance in $\mathbb{J}$ of the derivative with respect to the momentum $p$. At any rate, one can already see that no four-dimensional matrix can realize (\ref{nomat1}) for the fundamental representation of the centrally-extended algebra $\alg{psl(2|2)}_c$, since $\mathbb{C}$ is proportional to the identity matrix. One should rather use a differential operator to realize $\mathbb{J}$ \cite{FabianThesis}. 

In fact, $\mathbb{J}$ is the Cartan element of the $\alg{sl}(2)$ algebra of outer automorphisms of $\alg{psl(2|2)}_c$, inherited from $\alg{psl(2|2)}$ \cite{Serganova}. An explicit description of the action of these automorphisms on the supercharges and on the central charges can be found in \cite{Beisert:2006qh}.  `Outer' means that these automorphisms cannot be written as (anti)commutators of the algebra with particular elements of the algebra itself\footnote{The matrices corresponding to  plus or minus the identity in the associated $SL(2)$ automorphism group of $\alg{psl(2|2)}$ turn out to be actually inner \cite{Serganova}.}. Much in the same way as for the triple central extension, also the presence of a continuous outer automorphism group is peculiar to $A(1,1)$ amongst all simple basic classical Lie superalgebras.

After reinterpreting the dynamical action of the symmetry algebra in terms of a deformed coproduct, the local (cf. discussion below (\ref{coprodo})) representation of the algebra turns out to be a particular atypical representation (see section \ref{bsf} for bound state number $\ell=1$), parameterized by the values taken by the central charges. This representation is four-dimensional, and its explicit matrix description also easily follows from the one we will present in section \ref{bsf} for bound states, when restricting to bound state number equal to $1$. Strictly speaking, this representation is not highest weight, since there is no state annihilated by all positive roots.

Let us also stress again that the coproducts corresponding to different frames for the spin-chain states give rise to slightly different S-matrices, the main difference among them obviously being various phase factors $e^{i p_{1,2}}$ with various powers appearing in or disappearing from their entries. This ambiguity is no surprise, since, in this context, the S-matrix is ultimately a gauge-dependent quantity (where `gauge' now refers to some original gauge symmetry of the model), unlike the spectrum that one derives from it. For instance, in the worldsheet gauge used in \cite{Klose:2006zd}, the diagonal entries of the tree-level S-matrix depend explicitly on the gauge parameter. This connection with gauge transformations is also pointed out in \cite{Beisert:2005tm}, this time w.r.t. the SYM theory. The central charges themselves, while vanishing on physical states (cyclic spin-chains), can be seen having an action quite reminiscent of gauge symmetries (here, the familiar gauge transformations one has in any Yang-Mills theory). This may give a clue on how they are ultimately embedded in the symmetry group of AdS/CFT, yet being outside $\alg{psu}(2,2|4)$ \cite{Arutyunov:2011uz}. In fact, the relation between the centrally-extended algebra (and its Yangian) emerging from the worldsheet after fixing the light-cone gauge \cite{Arutyunov:2006ak}, and the original superconformal (Yangian) algebra, is an outstanding problem\footnote{We thank Tristan McLoughlin for exchanges on this point.}. If it is true that one can derive the Bethe Ansatz equations in subsectors from first principles using the S-matrix of the  $\alg{psl(2|2)}_c$ algebra ({see {\it e.g.} the reviews \cite{Staudacher:2010jz,Ahn:2010ka}), the celebrated Beisert-Staudacher equations \cite{Beisert:2005fw} ({\it alias}, the Bethe equations for the bigger  $\alg{psu}(2,2|4)$ algebra) instead, although tested beyond doubt, still remain a conjecture, and it would be desirable to have an {\it a priori} derivation\footnote{We thank A. Doikou and D. Fioravanti for discussions on this point, see also the recent \cite{Balog:2011nm}.}.

The condition of invariance
of the S-matrix under the symmetry algebra should be casted in the form (see footnote \ref{simmetr})
\begin{eqnarray}
\label{vari}
\Delta^{op} R = R \, \Delta.
\end{eqnarray}
The {\it opposite} coproduct $\Delta^{op}$ is defined as $\Delta^{op} \, = \, P \Delta$, with $P$ the graded pernutation operator $P \, a \otimes b \, = \, (-)^{deg(a) deg(b)} \, b \otimes a$. 
In a physical picture, if the coproduct acts, say, on {\it in} scattering states, its opposite acts on {\it out} states, and {\it vice versa}. Formula (\ref{vari}) represents the very definition of the $R$-matrix (S-matrix), as the transformation matrix between {\it in} and {\it out} states. In the theory of quantum groups, the existence of such an object makes the Hopf algebra {\it quasi-cocommutative}\footnote{The prefix {\it quasi} indicates that the coproduct would {\it almost} be cocommutative, were it not for a similarity transformation represented by the conjugation {\it via} the (invertible) $R$-matrix itself, $\Delta^{op} = R \Delta R^{-1}$.}. As usual, quasi-cocommutativity represents the similarity between the two representations obtained tensoring two modules  using the coproduct or its opposite. The two ways give representations of the same dimension, but these ought not be the same. The relation (\ref{vari}) establishes when the two are similar to each other. The element $R$ is often called the {\it intertwiner} between the two tensor product representations. The $R$-matrix for {\it quasi-triangular} Hopf algebras satisfies the famous Yang-Baxter equation (YBE), also called `star-triangle' equation:

\begin{equation}
\label{ybe}
R_{12} \, R_{13} \, R_{23} \, = \, R_{23} \, R_{13} \, R_{12},
\end{equation}
where $R_{ij}$ indicates the two spaces on which the $R$-matrix acts in the triple tensor product of representations.
 
The $\alg{sl}(2)\oplus \alg{sl}(2)$ generators have zero $[[A]]$ quantum number, therefore their coproduct is trivial. This implies that the $R$-matrix intertwining the coproduct (\ref{coprodot}) is $\alg{sl}(2)\oplus \alg{sl}(2)$-invariant in the traditional sense\footnote{{\it I.e.},$ [R,\Delta]=0$.}, and it can be decomposed as a sum of projectors onto irreducible representations of $\alg{sl}(2)\oplus \alg{sl}(2)$. It also means that the eigenvalues of the Cartan generators of the $\alg{sl}(2)\oplus \alg{sl}(2)$ subalgebra are conserved in the scattering. From the specific form of such matrices in the fundamental representation (and, in general, in all the bound states representations, see section \ref{stleg}) one deduces the conservation of the total numbers of fermionic excitations of type $\theta_3$ and, separately, of type $\theta_4$ in the scattering, in addition to the total number of excitations (bosonic plus fermionic). In this counting, one has to pay attention to the fact that a boson of type $w_2$ counts as a pair of fermions $\theta_3 \, \theta_4$.

In the presence of central elements, there is a special consistency requirement one has to consider. Since $\Delta (\mathbb{C})$ is also central,
and $R$ is invertible,
\begin{eqnarray}
\label{coco}
\Delta^{op} (\mathbb{C}) \, R \, = \, R \, \Delta (\mathbb{C}) \, = \, \Delta (\mathbb{C}) \, R \qquad \implies \qquad
\Delta^{op} (\mathbb{C}) \, = \, \Delta (\mathbb{C}).
\end{eqnarray}
This can be equivalently stated by recalling the discussion on tensor product representations just above (\ref{ybe}). Specifically, since they are Lie algebra homomorphisms, both maps $\Delta$ and $\Delta^{op}$ define Lie algebra representations of the same dimension. The defining equation for the (invertible) $R$-matrix, namely  $\Delta^{op}  \, = \, R \, \Delta \, R^{-1}$ just tells us that these two representations are related to each other by a similarity transformation. If so, then they have to share the center.

In our case, (\ref{coco}) is guaranteed by the physical requirement
\begin{equation}
\label{interpr}
U^2 \equiv e^{i p} \, \mathbbmss{1}= \kappa \, \mathbb{C} \, \, + \, \mathbbmss{1}
\end{equation}
for a certain constant $\kappa$ related to the coupling $g_{YM}$
\cite{Beisert:2005tm}. Combining (\ref{interpr}) with (\ref{coprodot}), one has in fact (see also \cite{Torrielli:2011zz})
\begin{eqnarray}
\Delta(\mathbb{C}) = \mathbb{C} \otimes \mathbbmss{1} + \mathbbmss{1} \otimes \mathbb{C} + \kappa \, \, \mathbb{C} \otimes \mathbb{C} = \Delta^{op} (\mathbb{C}).
\end{eqnarray}
An analogous relation works for $\mathbb{C}^{\dag}$. These requirements are equivalent to imposing that the total value of the central charges $\mathbb{C}$ and $\mathbb{C}^{\dag}$ vanishes when the total momentum is set to zero. Vanishing total momentum, in turn, corresponds to periodic boundary conditions, which have to be asked for when dealing with the true single-trace operators of SYM. For two-particle states, vanishing of the total central charges means $\Delta (\mathbb{C})=\Delta (\mathbb{C}^{\dag})=0$ when $p_1 + p_2 =0$, which is realized by (\ref{interpr}), (\ref{coprodot}). 
 
By interpreting (\ref{interpr}) as an algebraic condition linking the central charges to the coproduct-deformation, one ensures (\ref{coco}) holds at the Hopf algebra level. All the axioms of a quasi-co\-commutative
Hopf algebra are therefore satisfied. We also notice that, even after the change of basis (\ref{tw}), the condition of cocommutativity of the central charges would boil down to the same relation (\ref{interpr}) .

The S-matrix in the fundamental representation turns out to be completely fixed (apart from an overall scalar phase) by the condition (\ref{vari}). The reason for this fact is that the coproduct (\ref{coprodot}) for the supercharges (that is, already at the Lie superalgebra level) is non-trivial. Another reason relates to the irreducibility / indecomposability of the tensor product of two fundamental representations (see section \ref{lunghe}).

The coproduct (\ref{coprodot}) was shown to emerge\footnote{After
a non-local change of basis, see the previous discussion.} also from the
dual string-theory sigma model. In \cite{Klose:2006zd}, the result was
reproduced by applying the standard Bernard-LeClair procedure
\cite{Bernard:1992mu} to the light-cone worldsheet Noether
charges obtained in \cite{Arutyunov:2006ak}.

{\rm Let us give here an alternative semi-classical
argument for the emergence of such a deformed coproduct from the
worldsheet theory, based on the same type of reasoning presented
at the end of section \ref{sect;sigma}.  The light-cone worldsheet
Noether supercharges have a non-local contribution in the
worldsheet fields:
\begin{eqnarray}
\alg{J}^A = \int_{-\infty}^{\infty} d \sigma \, J_0^A (\sigma ) \,
e^{i \, [[A]] \, \int_{-\infty}^{\sigma} \, d \sigma' \, \partial \chi^-
(\sigma')}.
\end{eqnarray}
This is due to the fact that, although the Noether charges are originally integrals of local densities, the light-cone field $\chi^-$ is not physical in the gauge chosen, and one should rather use its derivative. 
If we consider the two well-separated solitonic excitations of Figure
1, the {\it semiclassical action} of these charges
on such a scattering state is again obtained by
splitting the integrals:
\begin{eqnarray}
{\alg{J}^A}_{|profile} &=&  \int_{-\infty}^{\infty} d \sigma \, J_0^A (\sigma )_{|profile}\,  e^{i \, [[A]] \, \int_{-\infty}^{\sigma} \, d \sigma' \, \partial \chi^- (\sigma')_{|profile}} \,  \nonumber\\
&=&\int_{-\infty}^{0} d \sigma \, J_0^A (\sigma ) e^{i \, [[A]] \, \int_{-\infty}^{\sigma} d \sigma' \, \partial \chi^- (\sigma')} \,  + \nonumber \\
&& \qquad \qquad \qquad \, \, \, \, \, \, \, \, \, \, \int_{0}^{\infty} d \sigma \, J_0^A (\sigma ) \, e^{i \, [[A]] \, \int_{-\infty}^{0} d \sigma' \, \partial \chi^- (\sigma')} \, e^{i \, [[A]] \, \int_{0}^{\sigma} d \sigma' \, \partial \chi^- (\sigma')}\,  \nonumber\\
&\sim &\alg{J}^A_1 \, + e^{i [[A]] \, p_1} \alg{J}^A_2 \,\, \, \, \,  \longrightarrow \, \,
\Delta (\alg{J}^A) = \alg{J}^A \otimes \mathbbmss{1} + e^{i [[A]] p} \, \otimes \alg{J}^A, \nonumber
\end{eqnarray}
where one has used the definition of the worldsheet momentum in terms of the  field $\chi^-$ 
applied to the first excitation\footnote{Notice that the alternative expression  $$
\alg{J}^A = \int_{-\infty}^{\infty} d \sigma \, J_0^A (\sigma ) \,
e^{i \, [[A]] \, \int_{\sigma}^{\infty} \, d \sigma' \, \partial x^-
(\sigma')}$$
would produce, with analogous reasonings, the twisted coproduct on the l.h.s. of (\ref{tw}). This alternative expression should correspond to a non-local field redefinition on the worldsheet.}.

Let us conclude with some further comments on crossing symmetry. From the Hopf-algebra antipode $\Sigma$ it is easy to
derive the so-called `antiparticle'
representation $\tilde{\alg{J}}^A$, and the corresponding
charge-conjugation matrix $C$:
\begin{eqnarray}
\label{pode} \Sigma (\alg{J}^A) \, = \, C^{-1} \, [ \,
\tilde{\alg{J}}^{A} ]^{st} \, C.
\end{eqnarray}
One denotes with $M^{st}$ the supertranspose\footnote{The supertranspose is defined as $[M^{st}]_{ij} = (-)^{deg(i) deg(j) + deg(j)} M_{ji}$. The reason for such definition is that, in this way, one has $[AB]^{st} = (-)^{deg(A) deg(B)} \, B^{st} \, A^{st}$.} of the matrix $M$. In the appropriate representation variables (see the definitions for general bound states in (\ref{parametro})) the ``tilde" is given by the map 
\begin{equation}
\label{til}
x^\pm \rightarrow \frac{1}{x^\pm}.
\end{equation}
Since $\frac{x^+}{x^-} = e^{i p}$, the map (\ref{til}) changes sign to the momentum.
Indeed, such a map {\it also} changes sign to the energy of the particle.

Since the antipode map is a Lie algebra homomorphisms, both the antipode and the supertranspose operation (possibly composed with a transformation of the parameters, such as the tilde operation on the r.h.s. of (\ref{pode})) define Lie algebra representations of the same dimension. The relation (\ref{pode}) just tells us that these two representations are related to each other by a similarity transformation. One can choose a frame where the charge-conjugation matric $C$ has integer entries, and its square is the diagonal matrix\footnote{Wherever applicable and not otherwise specified, we will assume the ordering $(w_1, w_2, \theta_3,\theta_4)$.} $diag(1,1,-1,-1)$ \cite{Arutyunov:2007tc}. 

Those just described are the ingredients entering the crossing-symmetry relations
originally written down in \cite{Janik:2006dc}, where the
existence of an underlying Hopf-algebra symmetry of the S-matrix was first
conjectured. Such relations naturally follow from (\ref{pode}) combined with the general formula
\begin{eqnarray}
(\Sigma \otimes \mathbbmss{1}) \, R \, = \, (\mathbbmss{1} \otimes \Sigma^{-1}) \, R \,= \, R^{- 1},
\end{eqnarray}
where the (invertible) antipode is derived from the coproduct (\ref{coprodot}).

As we already mentioned in footnote \ref{frame}, a 
reformulation in terms of a
Zamolodchi\-kov-Faddeev (ZF) algebra has been given in
\cite{Arutyunov:2006yd}. In the ZF presentation, the basic objects
are creation and annihilation operators, whose
commutation relations are determined in terms of the S-matrix of
the problem. Connections with $q$-deformations (at root of unity) have been pointed out in \cite{Gomez:2007zr,Young:2007wd,Beisert:2008tw,Beisert:2011wq} (\cite{Gomez:2006va}).

Notice that the $R$-matrix we are discussing becomes equal to the identity for equal values of the two momenta. Also, one can show \cite{Martins:2007hb,Beisert:2006qh} that this $R$-matrix is equivalent to Shastry's $R$-matrix $R_S$ for the Hubbard model \cite{Shastry:1986zz} {\it via} a spectral-parameter dependent transformation which preserves the Yang-Baxter equation:
\begin{eqnarray}
R_S (\lambda_1, \lambda_2) = G_1 (\lambda_1) G_2 (\lambda_2) \, R(\lambda_1, \lambda_2)\, G_1 (\lambda_1)^{-1} \, G_2 (\lambda_2)^{-1}.
\end{eqnarray}
For more on the relationship with the Hubbard model, see for instance \cite{Rej:2005qt,Feverati:2006hh}

\subsection{Yangian symmetry of the S-matrix}\label{sec:YS}
The S-matrix in the fundamental representation has been shown to
possess $\alg{psl}(2|2)_c$ Yangian-type symmetry \cite{Beisert:2007ds}:
\begin{equation} \label{coprodottoN} \Delta^{op} ( \, \widehat{\mathbb{J}}
\, ) \, \, R = R \, \Delta ( \, \widehat{\mathbb{J}} \, ). \end{equation}
This can be proved by explicit computation\footnote{One can  check the invariance of the S-matrix on a restricted set of generators, as many as they are enough to generate the remaining ones {\it via} commutators. Invariance under the remaining generators will then automatically follow. Such minimal set of generators is given, for instance, by a simple root system, as it is used in Drinfeld's second realization (see section \ref{ssec:drinf2}).}, given the list of coproducts for all the $\alg{psl}(2|2)_c$ Yangian generators provided in \cite{Beisert:2007ds}.
In order to be a Lie algebra homomorphism, the coproduct should
respect (\ref{rels}). Therefore, the structure of the Yangian
coproduct has to take into account the deformation in
(\ref{coprodot}). If one requires a minimal modification of (\ref{cop}) in order to accommodate this deformation, one is led to the following formula:
\begin{eqnarray}
\label{Niklas}
\Delta (\, \widehat{\alg{J}}^A) = \widehat{\alg{J}}^A \otimes \mathbbmss{1} + U^{[[A]]} \, \otimes \widehat{\alg{J}}^A + \frac{1}{2} \, f^A_{BC} \, \alg{J}^B \, U^{[[C]]} \, \otimes \alg{J}^C,
\end{eqnarray}
where we denote 
$$U=e^{ip}.$$
In \cite{Beisert:2007ds}, the list of coproducts for each individual generator, satisfying the above-mentioned compatibility requirement, and following the pattern (\ref{Niklas}), is explicitly given. 
The relevant representation of $\widehat{\alg{J}}^A$ is the so-called {\it evaluation} representation, which is obtained by multiplying the level-zero generators by an {\it evaluation} (sometimes also called `spectral') parameter\footnote{The tensor product of Yangian evaluation representations is typically irreducible (as a Yangian representation), except for special values of the spectral parameters. These values usually correspond to singularities of the Yangian rational R-matrix. At these poles, the intertwiner becoming singular means that the coproduct and its opposite are no longer related by similarity, and the tensor product representation becomes reducible (but generically indecomposable) as a Yangian representation. Let us also remark that an {\it evaluation} representation is often a representation which has a tail additional to just being the level zero generators multiplied by a spectral parameter, as it has to satisfy the Serre relations. The precise definition of evaluation representations involves a pull-back ({\it evaluation}) map, and can be found for instance in \cite{Chari}. Evaluation representations are very important. For instance, in the case of ${\cal{Y}}(\alg{sl}(2))$, every finite-dimensional irreducible representation is isomorphic to a tensor product of evaluation representations, see \cite{Chari}. The same is not true for bigger Lie algebras, and it is related to the (im)possibility of splitting Drinfeld polynomials into products of minimal ones (we thank C. Young for explanations on this point).}. In this case one has
\begin{eqnarray}
\label{u}
\widehat{\alg{J}}^A \, = \, u \, \alg{J}^A \, = \, \frac{g}{4i}
\left(x^++\frac{1}{x^+}+x^-+\frac{1}{x^-}\right) \, \alg{J}^A,
\end{eqnarray}
for a suitably normalized coupling constant $g$. The variables $x^\pm$, parameterizing the fundamental
representation, are to be defined in (\ref{parametro}). Notice that, in general, not all representations of a Lie algebra $\alg{g}$ can be extended to evaluation representations of the Yangian, since the Serre relations need to be satisfied (see the general treatment of the Yangian at the beginning of this review, section \ref{ssec:drinf1}).

{\it The reason for (\ref{u}) is again related to the fact that all the central charges at level one also have a central coproduct and, therefore, ought to be cocommutative, {\it i.e.} $\Delta^{op} (\widehat{\mathbb{C}}) = \Delta (\widehat{\mathbb{C}})$, {\it etc.}. This fixes the dependence of the evaluation parameter on the representation labels (up to an additive numerical constant which we have omitted).}

It is immediate to notice how the shift automorphism (\ref{shau}) becomes, in the evaluation representation (\ref{u}), a simple shift of the evaluation parameter by a constant:
\begin{equation}
\label{shcon}
u \longrightarrow u + c.
\end{equation}
In two-dimensional relativistic integrable models, the evaluation parameter $u$ is often interpreted as the particle-rapidity, which is defined in terms of the energy $E$, momentum $p$ and mass $m$ of the particle as
\begin{equation}
E = m \cosh u, \qquad \qquad p = m \sinh u.
\end{equation}
This way, the shift transformation (\ref{shcon}) corresponds to a Lorentz boost of the rapidity by an amount $c$ \cite{Bernard:1992mu}. 

\smallskip

The antipode reads 
\begin{equation}
\label{Yant}
\Sigma (\widehat{\alg{J}}^A ) = - U^{-[[A]]} \widehat{\alg{J}}^A.
\end{equation}
The traditional `tail' which arises when deriving the antipode from the coproduct (\ref{cop}), namely the tail in $\Sigma (\widehat{\alg{J}}^A ) = - \widehat{\alg{J}}^A + \frac{1}{4} \, f^A_{B C} \, f^{B C}_D \, \alg{J}^D$, is absent when deriving  (\ref{Yant}) from (\ref{Niklas}) (related to the vanishing of the $\alg{psl}(2|2)_c$ dual Coxeter number {\it via} footnote \ref{Serg}). 

A special remark concerns the `dual' structure constants
$f^A_{BC}$ appearing in (\ref{Niklas}). They should reproduce the
general form (\ref{cop}), and analogous structure constants with all indices
lowered should be used to prove the Serre relations (\ref{Serr}).
However, since the Killing form of $\alg{psl}(2|2)_c$ is zero, one
encounters a problem in defining these structure constants. In
\cite{Beisert:2007ds}, the quantities $f^A_{BC}$ are explicitly
given as a list of numbers, without necessarily referring to an
index-lowering procedure\footnote{\label{Serg}An argument was provided in
\cite{Beisert:2007ds}, according to which one can make sense of
these quantities as dual structure constants in an enlarged
non-degenerate algebra, endowed with an invertible bilinear form (see also \cite{FabianThesis,Spill:2008zz}). This algebra is obtained by
adjoining the $\alg{sl}(2)$ automorphism of $\alg{psl}(2|2)_c$
\cite{Serganova,Beisert:2006qh} to the algebra of generators.
Apart from allowing the inversion of the bilinear form and
the determination of $f^A_{BC}$, these extra
generators would drop out of the final form of the Yangian
coproduct when the latter is applied to the Lie superalgebra generators as in (\ref{Niklas}).}. The table of coproducts is in this
way fully determined. We will return to this point in 
section \ref{ssec:clpsu}. 

\smallskip

Another remark concerns the dependence of the spectral parameter
$u$ on the representation variables $x^\pm$, or, equivalently, on
the eigenvalues of the central charges of $\alg{psl}(2|2)_c$. For
simple Lie algebras, the spectral parameter is typically an
additional variable one attaches to the evaluation representation.
Together with the existence of the shift-automorphism $u
\rightarrow u + c$ of the Yangian in evaluation
representations, this implies that a Yangian-invariant S-matrix
depends only on the difference of the spectral
parameters\footnote{An alternative proof of this fact can be found
in \cite{Beisert:2007ds}, based on the form (\ref{cop}) of the coproduct. Schematically, the Yangian coproduct is of the form $\Delta (\widehat{x}\, ) = u_1 x \otimes \mathbbmss{1} + \mathbbmss{1} \otimes u_2 x + indep.\,  on \, \, u_{1,2}$. Rewriting it as $\Delta (\widehat{x}\, ) = (u_1 - u_2)x \otimes \mathbbmss{1} + u_2 \, \Delta(x) +indep. \, on \, \, u_{1,2}$, and using the fact that $\Delta(x)$ is a symmetry of the S-matrix, one deduces that the S-matrix depends on the spectral parameters only through the combination $u_1 - u_2$. This argument can be easily extended to the case of the deformed coproduct (\ref{Niklas}), (\ref{coprodot}) (but of course only as long as one is allowed to consider the spectral parameters as independent variables).}:
$$
R = R(u_1 - u_2).
$$
On the other hand, the dependence of $u$ on the variables
parameterizing the central extension alters this property, and one
does not observe a difference form in the fundamental S-matrix. We
will come back to this issue in section \ref{diferenza} (see also \cite{BazhanovTalk}).

We finally remark that there usually exists a way of reconstructing the (infinite-dimensional) symmetry algebra in a specific representation, from the knowledge of the S-matrix satisfying the Yang-Baxter equation in that representation (see for instance section 8.3 in \cite{Arutyunov:2006yd}).  

\section{The classical $r$-matrix}\label{classr}
\subsection{From quantum to classical, and return}{\label{ssec:gen}}
The form of the Yangian discussed in the previous section closely resembles the standard one, but it also displays several unconventional features. In order to gain a deeper understanding, and according to a well-established mathematical procedure, it is useful to study the problem in
certain limits. One important instance, whenever available, is the {\it classical} limit, {\it i.e.} one studies
perturbations of the $R$-matrix around the identity:
\begin{eqnarray}
\label{classexp}
R = \mathbbmss{1}\otimes \mathbbmss{1} \, + \,
\hbar \, r \, + \, {\cal{O}} (\hbar^2 ),
\end{eqnarray}
$\hbar$ being a small parameter. The first-order term $r$ is called the {\it classical} $r$-matrix\footnote{Formula (\ref{classexp}) can be thought of as a sort of exponential map, see also \cite{Freidel:1992cd}. In fact, usually $r$ lives in $\alg{g}\otimes \alg{g}$, for $\alg{g}$ a Lie algebra, while $R$ in $U(\alg{g})\otimes U(\alg{g})$, $U(\alg{g})$ being the universal enveloping algebra of $\alg{g}$. We will be dealing with $r$-matrices depending on spectral parameters, which we simply call {\it $r$-matrices}. Those which do not have such a dependence are called {\it constant} $r$-matrices. One can usually obtain a constant $r$-matrix by suitably holding the arguments of an $r$-matrix fixed to certain values.}. One can easily prove that, if $R$ satisfies the Yang-Baxter equation (YBE), then $r$ satisfies the so-called {\it classical} YBE (CYBE):
\begin{eqnarray}
\label{cYBE}
[r_{12},r_{13}] + [r_{12},r_{23}] + [r_{13},r_{23}] =0.
\end{eqnarray}
The notation $r_{ij}$ is the same as in formula (\ref{ybe}). 
In standard situations, it turns out that the study of (\ref{cYBE}) can bring to a classification of solutions of the YBE itself, and of the possible quantum group structures underlying such solutions. Let us see how this works starting with a famous theorem  \cite{BD1,BD2}.

{\small \begin{itemize} \item {\it Theorem} (Belavin-Drinfeld I): Consider a finite-dimensional simple Lie algebra $\alg{g}$, and a solution $r(u_1, u_2)$ of
the CYBE, taking values in $\alg{g}\otimes \alg{g}$. Let such a solution be
of difference form, $r = r(u_1 - u_2)$. Furthermore, let one of the
following three equivalent conditions be satisfied: $(i)$ r has at least one pole in the complex variable $\delta u = u_1 - u_2$, and there is no Lie subalgebra $\alg{g}'\subset \alg{g}$ such that $r$ is an element of $\alg{g}' \otimes \alg{g}'$ for any $\delta u$, or $(ii)$ $r$ has
a simple pole in $\delta u = 0$, with residue proportional to
$\sum_\mu I_\mu \otimes I_\mu$, $I_\mu$ being a basis in $\alg{g}$ orthonormal with respect to a chosen nondegenerate invariant bilinear form\footnote{Such a residue can be identified with the quadratic Casimir $C_2$ in $\alg{g} \otimes \alg{g}$.}, or $(iii)$
the determinant of the matrix $r_{\mu \nu}(\delta u)$ formed by the coordinates of the
tensor $r(\delta u) = \sum_{\mu \nu} r_{\mu \nu}(\delta u)\, I_\mu \times I_\nu$ is not identically zero. Under these requirements, such a solution
satisfies the unitarity condition $r_{12}(\delta u) = - r_{21}(-\delta u)$, and extends meromorphically to the entire
complex $\delta u$-plane. All the poles of $r(\delta u)$ are simple, and form a lattice $\Gamma$ in the $\delta u$-plane.
Furthermore, modulo automorphisms, one has three possible types of solutions: elliptic (if $\Gamma$ is a
two-dimensional lattice), trigonometric (if $\Gamma$ is one-dimensional), or rational (if $\Gamma = \{0\}$).
\end{itemize}
}

From the knowledge of the $r$-matrix, there is a standard procedure how to construct an
associated Lie bialgebra, and obtain a quantization of it. This procedure involves the so-called `Manin triples' (see for example \cite{Etingof} and references
therein). The term `quantization' has here the mathematical meaning of completing the classical structure to a quantum group, or, equivalently, to complete a classical $r$-matrix to a solution of the YBE.
In the case of integrable systems based on such quantum groups, this coincides with what physicists understand as quantization, namely, going from the semiclassical regime to the quantum one\footnote{This is advertised by the following correspondence:
\begin{eqnarray}
\label{qcomm}
\{ A,B\} = \lim_{\hbar \to 0} \, \frac{[A,B]}{i \, \hbar}.
\end{eqnarray}
In a nutshell, we could say that the theory of integrable systems provides us, in certain standard examples, with the analytical knowledge of what the r.h.s. of (\ref{qcomm}) {\it exactly is, as a function of $\hbar$} (cf. Sklyanin algebras \cite{Sklyanin:1980ij}).}.
The associated quantum group structures emerging from the quantization are, in the three cases described by the above theorem, elliptic quantum groups ($dim(\Gamma)=2$), (trigonometric) quantum groups ($dim(\Gamma)=1$), and Yangians ($\Gamma=\{0\}$, respectively). Investigations of analogous theorems in the case of superalgebras (and an exposition of some additional subtleties that emerge in that case) can be found in \cite{Leites:1984pt,Zhang:1990du,Karaali1,Karaali2}.

A convenient way of understanding how this quantization procedure works in the case of Yangians is by studying the so-called Yang's $r$-matrix \cite{Yang:1967bm}:
\begin{eqnarray}
\label{Yang}
r=\frac{C_2}{u_2 - u_1}~.
\end{eqnarray}
This is the prototypical rational solution of the CYBE\footnote{Since, by definition of the Casimir $C_2$, one has $[C_2,\alg{J}^A \otimes \mathbbmss{1} + \mathbbmss{1} \otimes \alg{J}^A]=0 \, \, \forall A$, one can easily prove that (\ref{Yang}) solves the CYBE.}. By making use of the geometric series expansion, we can rewrite this $r$-matrix as follows:
\begin{eqnarray}
\label{Tayl}
&&r=\frac{C_2}{u_2 - u_1} = \frac{\alg{J}^A \otimes \alg{J}_A}{u_2 - u_1}=\sum_{n\geq 0} \alg{J}^A u_1^n \otimes \alg{J}_A u_2^{-n-1}=\sum_{n\geq 0} \alg{J}^A_n \otimes \alg{J}_{A,-n-1},
\end{eqnarray}
where we have assumed $|u_1/u_2|<1$ (the reverse would just switch the two copies of the Yangian in the Yangian double, see the discussion following formula (\ref{loo})), and we have used the bilinear form $\kappa_{A B}$ to express the quadratic Casimir in terms of the Lie algebra generators $\alg{J}^A \in \alg{g}$. The above rewriting is necessary in order to be able to attribute the dependence on the spectral parameter $u_1$ (respectively, $u_2$) to operators in the first (respectively, second) space. We will call this procedure ``factorization". This gives the $r$-matrix a meaning in terms of tensor products of algebra representations and, at the same time, suggests a universal interpretation. The assignment $\alg{J}^A_n \, = \, u^n \, \alg{J}^A$ in (\ref{Tayl}), in fact, entails the following loop-algebra commutation relations:
\begin{eqnarray}
\label{loo}
[\alg{J}^A_m , \alg{J}^B_n] = \, f^{AB}_C \, \alg{J}^C_{m+n}.
\end{eqnarray}
One can check that, with these commutation relations, the classical Yang-Baxter equation is satisfied by $r =   \sum_{n\geq 0} \alg{J}^A_n \otimes \alg{J}_{A,-n-1}$ (cf. (\ref{Tayl})) in a purely abstract way ({\it i.e.}, independently on specific representations of (\ref{loo})).

It is easy to show\footnote{\label{ft}One just needs to use the properties of fractions (or, alternatively, expand the rational $r$-matrix near the simple pole at the origin) and impose the CYBE. Namely, if $r = r_{\mu \nu} (\delta u) \, I^\mu \otimes J^\nu$, one has that near the pole $u_1 = u_2$ the CYBE reduces to $$\frac{c_{\mu \nu}(u_1)}{u_1 - u_2} \, r_{\rho \lambda} (u_1 - u_3) \, ([I^\mu , I^\rho]\otimes J^\nu \otimes J^\lambda + I^\mu \otimes [I^\nu , J^\rho]\otimes J^\lambda)=0,$$ for some residual function $c_{\mu \nu}(u_1)$. This implies in particular $$[I^\mu , I^\rho] = f^{\mu \rho}_\lambda I^\lambda$$ for some constants $f^{\mu \rho}_\lambda$. In \cite{BD1}, the Jacobi identity is shown, which proves that the two spans discussed above form Lie subalgebras of $\alg{g}$.} that the spans of the generators appearing separately on each factor of $r$ must form two Lie subalgebras of $\alg{g}$. The two span subalgebras, together with the original algebra $\alg{g}$, form a so-called {\it Manin triple}. Characterization of these subalgebras is an essential pre-requisite which the subsequent construction and characterization of the quantum group is based upon. 

In order to proceed to the quantization, one then needs to explore the spectral-parameter dependence of the two span subalgebras. The specific decomposition (\ref{Tayl}) corresponds to $\alg{g}[[u_1]] \otimes u_2^{-1} \alg{g}[[u_2^{-1}]]$, where $\alg{g}[[x]]$ is the algebra of $\alg{g}$-valued polynomials in the variable $x$.
In turn, the loop algebra is nothing else than the `classical' limit of the Yangian $\Ya{\alg{g}}$, the latter being a (quantum) deformation of the former (see section \ref{ssec:drinf1}). {\it Via} this example, one can realize how {\it rational} solutions of the CYBE, such as (\ref{Yang}), give rise to Yangian algebras upon quantization. Namely, the quantized versions of such $r$-matrices take values in the tensor product of the Yangian (or, rather, of its double, as we will shortly discuss). It is clearly of the utmost importance to be able to identify and characterize as precisely as possible the Manin triple corresponding to a given $r$-matrix, since it provides the {\it germ} of the quantization.  

One can also notice quite clearly a feature of the Yangian to be.
The Yangian on its own does not admit a universal R-matrix. What one has in mind when searching for a universal R-matrix is actually the {\it double} of the Yangian ${\cal{DY}}(\alg{g})$. Following Drinfeld, the canonical element $R=\sum_I \, e_I \otimes e^I$ in the tensor product of the direct and dual copy of the relevant quantum algebra inside the double, is just the universal R-matrix\footnote{The double construction is very general, and it is in fact the standard way to derive universal R-matrices for quasi-cocommutative Hopf algebras (possibly followed by a suitable identification procedure performed on the two copies of the double, like in the case of quantized Lie algebras).}. The two copies inside the double are conjugated {\it via} a suitable pairing compatible with the Hopf algebra structure, and it is with respect to this pairing that the dual basis $e^I$ is defined. From the above geometric series expansion we already see that the double of the Yangian conjugates elements with a  positive integer level $n$ to elements of the opposite copy of the Yangian, labeled by a `negative integer level' $-n-1$ \cite{KT}.  

It is hard to overestimate the importance of the classical $r$-matrix in the theory of integrable systems. Most notably, the classical $r$-matrix controls the Poisson brackets of the $\cal{L}$-operators in the inverse scattering method ({\it Sklyanin bracket}), and it appears in the theory of Poisson-Lie groups. A large literature is devoted to its properties, see for instance \cite{Faddeev:1987ph,Sklyanin:1980ij,Chari} and references therein. 

{\rm A final remark concerns another theorem \cite{BD3}:
\begin{itemize} \item {\it Theorem} (Belavin-Drinfeld II): With the hypothesis of Belavin-Drinfeld I theorem, let $r$ not be of difference form, but the dual Coxeter number of $\alg{g}$ be non-zero. Then, there exists a change of variables that reduces $r$ to a difference form. \end{itemize}
}

\subsection{The classical $r$-matrix of $\alg{psl}(2|2)_c$}{\label{ssec:clpsu}
In the case of the S-matrix found in \cite{Beisert:2005tm}, the parameter controlling the classical expansion is naturally associated with the inverse of the suitably normalized coupling constant $g$:
\begin{eqnarray}
R = \mathbbmss{1} \otimes \mathbbmss{1} \, + \, \frac{1}{g} \, r
\, + {\cal{O}}(\frac{1}{g^2}).
\end{eqnarray}
The unitary classical $r$-matrix $r$ is identified with the tree-level string scattering matrix computed in \cite{Klose:2006zd}. The following  parameterization \cite{Arutyunov:2006iu} of the variables $x^\pm$ (satisfying the non-linear constraint (\ref{parametro})) makes it easier to take the classical limit:
\begin{equation}
\label{clAF}
x^\pm (x) = x \sqrt{1 - \frac{1}{g^2 (x - \frac{1}{x})^2}} \pm \frac{i x}{g (x - \frac{1}{x})} \, \, \to \, \, x.
\end{equation}
The limit is taken by sending $g$ to $\infty$, while keeping $x$
fixed. The quantity $x$ can therefore be interpreted as an unconstrained
`classical' variable. This classical limit was studied in
\cite{Torrielli:2007mc}. It is clear from section \ref{ssec:gen} that the main target is to give a precise characterization of the algebra the $r$-matrix takes values in, as the quantization of this algebra can reveal the full quantum symmetry of the
S-matrix. 

The fundamental representation of $\alg{psl}(2|2)_c$
 tends, in the classical limit, to a certain representation of $\alg{psl}(2|2)_c$,
with generators parameterized by $x$. The classical $r$-matrix
$r=r(x_1,x_2)$ is not of difference form. This, together with (and
related to) the fact that we are dealing with a non-simple Lie
superalgebra (with vanishing dual Coxeter number), immediately makes the
application of Belavin-Drinfeld type of theorems not possible.

However, one can get an inspiration from those standard results. The
classical $r$-matrix has a simple pole at the origin  $x_1 - x_2 =
0$, which consistently fits into the picture of an underlying
Yangian symmetry. An easy exercise shows that the residue of a
solution of the CYBE at such a simple pole {\it must be} an invariant of
the tensor product algebra\footnote{To be more precise, the residue must be an invariant of the two span subalgebras singled out by the two factors of $r$, which have been discussed in footnote \ref{ft}. One just collects $c_{\mu \nu}(u_1) \,  I^\mu \otimes I^\nu$ as the residue, and invariance follows directly from the first equation in footnote \ref{ft}.} $\alg{g}\otimes \alg{g}$, if $r \in
\alg{g}\otimes \alg{g}$. This means that, if the two span subalgebras coincide with $\alg{g}$ itself, one has to have $[residue,\alg{J}^A \otimes \mathbbmss{1} + \mathbbmss{1} \otimes \alg{J}^A]=0 \, \, \forall A$. Close investigation reveals that the
residue of the classical $r$-matrix at the pole $x_1 - x_2 = 0$ is
actually the Casimir $C_2$ of the Lie superalgebra
$\alg{gl}(2|2)$:
\begin{eqnarray}
C_2 = \sum_{i,j=1}^4 \, (-)^{deg(j)} \, E_{ij}\otimes E_{ji},
\end{eqnarray}
with $E_{ij}$ being the unit-matrices with all zeros but $1$ in
position $(i,j)$, and $deg(j)$ being once again the fermionic grading of the index
$j$. One observes that, in absence of a quadratic
Casimir for $\alg{psl}(2|2)_c$, the classical $r$-matrix displays
on the pole (with a somewhat rough terminology, we will say it
`borrows') the quadratic Casimir of a bigger
algebra. Indeed, $\alg{gl}(2|2)$, the algebra of $2|2 \times 2|2$ matrices, is obtained by adjoining to
$\alg{sl}(2|2)$ the non-supertraceless Cartan element
\begin{equation}\label{bee}\mathbb{B}=diag(1,1,-1,-1).\end{equation} For this bigger algebra, a non-degenerate form exists
and the quadratic Casimir can be
constructed\footnote{Let us remark that, consistently with (a
supersymmetric version of) the Belavin-Drinfeld II theorem,
on the pole of the classical $r$-matrix (and only there) one can find a change of
variables to a difference form \cite{Torrielli:2007mc}.}. However, one cannot conclude from here that
the quantum symmetry algebra 
includes $\alg{gl}(2|2)$ with a trivial coproduct for $\mathbb{B}$. In fact,
$\Delta(\mathbb{B}) = \mathbb{B}\otimes \mathbbmss{1}
+ \mathbbmss{1} \otimes \mathbb{B}$ is not a symmetry of the S-matrix. The classical $r$-matrix has a ``tail" (I. Cherednik, private communication), corresponding to amplitudes in the quantum R-matrix which violate this symmetry (see the discussion concerning the {\it secret symmetry} in section \ref{segreto}).

Nevertheless, this property of `borrowing' is reminiscent of a
prescription well known in the theory of quantum groups, due to Khoroshkin and
Tolstoy \cite{KT,Khoroshkin:1994uk}. The universal R-matrix for the Yangian double based
on a simple Lie (super)algebra $\alg{g}$ can very schematically be written
as
\begin{eqnarray}
\label{KTformula}
R \, = \, \prod_{roots} e^{\xi^+ \otimes \xi^-} \, \, \, \, e^{a_{ij}^{-1} \kappa^i \otimes \kappa^j}  \, \, \, \, \prod_{roots} e^{\xi^- \otimes \xi^+},
\end{eqnarray}
with $\xi^\pm$ positive (resp., negative) roots of $\alg{g}$,
$\kappa_i$ Cartan generators and $a_{ij}$ the corresponding
(non-degenerate) Cartan matrix (cf. section \ref{ssec:drinf2}).
Whenever $a_{ij}$ is degenerate, as for $\alg{psl}(n|n)$, the
prescription is to adjoin to the Cartan subalgebra as many extra
Cartan generators as they are needed to reach a non-degenerate Cartan
matrix. At that point, one can take the inverse of $a_{ij}$. All the extra Cartan
elements will therefore appear in the exponent of
(\ref{KTformula}). One could then expect that, if a universal
$R$-matrix exists for the AdS-CFT problem at hand, and if it has
to be of the Khoroshkin-Tolstoy type, an extra
Cartan element such as $\mathbb{B}$ has to come into play. The
question is how to consistently embed this new generator in the (classical and quantum) Yangian symmetry algebra of the S-matrix.

{\rm We notice that the Lie superalgebra $\alg{gl}(2|2)$
already appeared at one-loop in gauge theory. When the
coupling $g$ goes to zero, in fact, the $R$-matrix becomes a
twisted version of
\begin{eqnarray}
R_{\, 1 \, loop} \, \sim \,  \mathbbmss{1} \otimes \mathbbmss{1} \ + \, \frac{C_2}{u_1 - u_2},
\end{eqnarray}
(namely, a quantum $R$-matrix of the so-called {\it Yang's type}, see \cite{Cai:q-alg9709038,Beisert:2005wm,Arutyunov:2009ce,Rej:2010mu} for the $\alg{gl}(1|1)$ case), with
$C_2$ the quadratic Casimir of $\alg{gl}(2|2)\otimes
\alg{gl}(2|2)$ (see, for instance, \cite{Arutyunov:2007tc}). Because of the twist, the difference form is lost even in the one loop limit.
}

{\it Note.} The function defined by the eigenvalue of the universal R-matrix $R$ acting on the highest weight of a highest weight tensor product irreducible representation $\rho$ is called the {\it character of $R$ in $\rho$} \cite{KT,Khoroshkin:1994uk}, and it is related to the overall scalar factor that the universal R-matrix produces when evaluated in that representation (see also \cite{Rej:2010mu}).

\subsection{Universal formulations}\label{sec:univ}
In order to gain understanding of the role of the new generator $\mathbb{B}$,
one may try a factorization procedure, analogous to the example of
Yang's classical $r$-matrix in section \ref{ssec:gen}. In the present case, the expression of $r$ is more complicated than in Yang's example, and one has to work harder to find a
suitable `geometric-like' series expansion which factorizes it. A
first proposal \cite{Moriyama:2007jt} was later seen to work only
for the fundamental representation, while it fails to reproduce the
bound state classical $r$-matrix \cite{deLeeuw:2010nd}. Nevertheless, this proposal
had the merit of showing how the new generator $\mathbb{B}$ could be allocated
in an expression not much dissimilar from
Yang's form.

{\footnotesize
We report the expression found in \cite{Moriyama:2007jt} with the sole purpose of
displaying the new generator (and its hypothesized higher
loop-algebra/Yangian partners). With a proper regularization and
resummation, one has
\begin{eqnarray}
\label{MT}
&&r=\sum_{n\geq 0} \, \mathbb{G}^\alpha_{a,n} \otimes \hat{\mathbb{Q}}^a_{\alpha,-n-1} \, - \, \mathbb{Q}^a_{\alpha,n} \otimes \hat{\mathbb{G}}^\alpha_{a,-n-1} \, + \, \mathbb{H}_n \otimes \mathbb{B}_{-n-1} \, +  \, \mathbb{B}_n \otimes \hat{\mathbb{H}}_{-n-1} \nonumber\\
&&+ \, (\mathbb{L}^a_{b,n} \otimes \hat{\mathbb{L}}^b_{a,-n-1} \, - \,  \mathbb{L}^a_{b,-n-1} \otimes \hat{\mathbb{L}}^b_{a,n}) \, - \, (\mathbb{R}^\alpha_{\beta,n} \otimes \hat{\mathbb{R}}^\beta_{\alpha,-n-1} \, - \,  \mathbb{R}^\alpha_{\beta,-n-1} \otimes \hat{\mathbb{R}}^\beta_{\alpha,n}).\nonumber
\end{eqnarray}
We will not report here the explicit expressions of the generators appearing in this rewriting as functions of the classical variable $x$ (\ref{clAF}).

One useful thing to notice is that the Cartan part of the above expression corresponds to a $\alg{gl}(2|2)$ Cartan matrix such that
(cf. (\ref{KTformula}))
\begin{eqnarray}
a^{-1}_{ij} \kappa^i \kappa^j = 4 \mathbb{H} \mathbb{B} + \mathbb{L}^2 - \mathbb{R}^2.
\end{eqnarray}
The new generator $\mathbb{B}$ is needed to perform the
factorization, and, precisely as in $\alg{gl}(2|2)$, it couples to
the central charge $\mathbb{H}$ (here seen as the magnon energy). Another feature of this
proposal is the formula $\mathbb{B}_n =
\frac{1}{2} (x^n - x^{-n}) \, diag (1,1,-1,-1)$ \cite{Moriyama:2007jt}. One notices
that $\mathbb{B}_0$ vanishes, which in this representation may be related to the absence
of a Lie algebra symmetry of the S-matrix of type $\mathbb{B}$
(with trivial coproduct, cf. section \ref{ssec:clpsu}). However,
one can see from this proposal how (higher Yangian)
generators $\mathbb{B}$ of $\alg{gl}(2|2)$-type are needed in order
to reach a universal formula. This will be a consistent feature of all subsequent attempts at factorization (including in the so-called {\it near flat space} limit, see the Conclusions). The natural question is
whether such symmetries can be found for the quantum S-matrix of
\cite{Beisert:2005tm}.}

Before answering this question, we present another proposal of factorization of the classical $r$-matrix \cite{Beisert:2007ty}, which has been shown to reproduce the classical limit of the bound state S-matrix as well \cite{deLeeuw:2008dp,Arutyunov:2009mi}. One can show that the same $r$-matrix can in fact be rewritten as
\begin{eqnarray}
\label{eqn;Rmat}
&&r = \frac{\mathcal{T}-\tilde{\mathbb{B}}\otimes
\mathbb{H}-\mathbb{H}\otimes
\tilde{\mathbb{B}}}{i(u_{1}-u_{2})}-\frac{\tilde{\mathbb{B}} \otimes \mathbb{H} }{iu_{2}}
+\frac{\mathbb{H}\otimes \tilde{\mathbb{B}}}{iu_{1}} - \frac{\mathbb{H}\otimes \mathbb{H}}{\frac{2 i u_1 u_2}{u_1 - u_2}},\nonumber\\
&&\mathcal{T}=2\left(\mathbb{R}^{\
\alpha}_{\beta}\otimes\mathbb{R}^{\ \beta}_{\alpha}- \mathbb{L}^{\
a}_{b}\otimes\mathbb{L}^{\ b}_{a}+ \mathbb{G}^{\
\alpha}_{a}\otimes\mathbb{Q}^{\ a}_{\alpha}- \mathbb{Q}^{\
a}_{\alpha}\otimes\mathbb{G}^{\ \alpha}_{a}\right),\nonumber\\
&&\tilde{\mathbb{B}} =\frac{1}{4} \frac{1}{\epsilon_{lim}(x)} \, \, diag (1,1,-1,-1).
\end{eqnarray}
The variable $u$ appearing in the above formulas is the classical limit of the `quantum' evaluation parameter $u$ in (\ref{u}), appropriately rescaled by the coupling constant to make it finite. Also, all generators are taken in their classical limit (cf. section \ref{ssec:clpsu}), and $\epsilon_{lim}(x)$ is the classical limit of the energy eigenvalue $\epsilon(p)$ in (\ref{mild}). 

As one can see, one of the main advantages of (\ref{eqn;Rmat}) resides in its being quite close to Yang's form. All classical Yangian generators are simply obtained as $\alg{J}_n = u^n \alg{J}$ after factorizing {\it via} the geometric series expansion. In these way, $r$ can be casted in terms of infinite sums of abstract generators directly as in (\ref{Tayl}). These generators, together with the abstract factorized form of $r$ one obtains, can be shown to originate a consistent Lie bialgebra structure \cite{Beisert:2007ty}. This structure certainly deserves further study. In particular, its quantization is a fascinating open problem. A very important feature is that ${\tilde{\mathbb{B}}}_0$ lives in the opposite copy of the classical double with respect to the copy the level-zero $\alg{psl}(2|2)_c$ generators live in.  

We end this section by referring to interesting studies of the classical $r$-matrix and of the $r,s$ non-ultralocal structure of the $\alg{psu}(2,2|4)$ sigma-model \cite{Dorey:2006mx,Aoyama:2007tz,Mikhailov:2007eg,Vicedo:2008jy,Vicedo:2008jk,Magro:2008dv,Vicedo:2010qd,Magro:2010jx}. It is still an open question how to relate these studies to the results for $\alg{psl}(2|2)_c$ which we have described here. Interesting connections to quantum deformations and the Hubbard model can be found in \cite{Beisert:2010kk}.

\subsection{The `secret symmetry'}\label{segreto}

The answer to the question posed in the previous section, namely,
whether there exist quantum symmetries of type $\mathbb{B}$,
turns out to be in the affirmative. One can in fact prove that the full
quantum S-matrix is invariant under the following exact symmetry, found in
\cite{Matsumoto:2007rh} and shortly afterwards confirmed in \cite{Beisert:2007ty}:

\begin{eqnarray}
\label{secretsymmetry}
&&\Delta (\hat{\mathbb{B}})= \hat{\mathbb{B}} \otimes \mathbbmss{1} \, + \, \mathbbmss{1} \otimes \hat{\mathbb{B}}
+\frac{i}{2g} (\mathbb{G}^\alpha_{a} \otimes \mathbb{Q}^a_{\alpha} \, + \, \mathbb{Q}^a_{\alpha} \otimes \mathbb{G}^\alpha_{a} ),\nonumber\\
&&\Sigma (\hat{\mathbb{B}}) \, = \, - \hat{\mathbb{B}} \, + \, \frac{2i}{g} \mathbb{H},\nonumber\\
&&\hat{\mathbb{B}} \, = \, \frac{1}{4} (x^+ + x^- - 1/x^+ - 1/x^-) \, diag(1,1,-1,-1).
\end{eqnarray}
There is a similar symmetry for all symmetric bound state representations \cite{deLeeuw:2008dp}.
The coproduct above is somehow reminiscent of a level-one
Yangian symmetry (cf. (\ref{cop})), in particular if one thinks of
$\mathbb{B}$ as the generator `dual' to $\mathbb{H}$ (in the sense
of section \ref{sec:YS}). One can also notice the similarity with the coproduct of the analogous generator in the $\alg{gl}(1|1)$ Yangian, see \cite{Cai:q-alg9709038}. The eigenvalues of $\mathbb{B}$ are consistent with
(both) the classical limits $\mathbb{B}_1$, $\tilde{\mathbb{B}}_1$
described in the previous section, in their respective normalizations. By commuting the secret symmetry with the (level-zero) supersymmetries, one generates new types of Yangian supercharges \cite{Matsumoto:2007rh}, which are automatically exact symmetries of the S-matrix. These new supersymmetries act on bosons and fermions with two {\it different} spectral parameters, respectively,
much like the
charges of the classical proposal of  \cite{Moriyama:2007jt}. The {\it commutant} of all these symmetries turns out to be quite a wild-looking algebra, and it has been so far rather hard to characterize this commutant in any more specific way. 

We stress that there is no ``level zero" analog of the Yangian charge we have discussed in this section. With a trivial coproduct (as expected from a Cartan generator), a matrix like $diag (1,1,-1,-1)$ is simply not a symmetry of the S-matrix. The reason is that there exists a non-zero amplitude for two fermions going into two bosons, and {\it viceversa} \cite{Beisert:2005tm}. 

The appearance of a symmetry generator starting from the first Yangian level on, even if reinterpreted as an ``indentation" in the juxtaposition of the two copies of the Yangian inside the Yangian double (as it seems to emerge from \cite{Beisert:2007ty} where $\mathbb{B}_0$ is attributed to the `negative' copy), remains to date quite a bizarre and new phenomenon.

\subsection{Remarks on the difference form}\label{diferenza}
Let us conclude with a few remarks concerning the (absence of)
difference form, as mentioned in section \ref{sec:YS} (for illuminating insights on this point, we urge the reader to consult \cite{BazhanovTalk}). 

\smallskip

First, the most
promising proposal available for the classical $r$-matrix
\cite{Beisert:2007ty}, described in section \ref{sec:univ},
displays an explicit dependence on the
spectral parameters which is almost purely of difference form. The non-difference form is mostly encoded in the classical representation labels
$x^\pm_{lim}(u)$ appearing in the symmetry generators\footnote{When dealing with Lie superalgebras and their representations, let alone with central extensions thereof, this dependence is ultimately not surprising. We thank P. Sorba for discussions about this point.}.

\smallskip

Moreover, Drinfeld's second realization (cf. section
\ref{ssec:drinf2}) for the $\alg{psl}(2|2)_c$ Yangian discussed in
section \ref{sec:YS} has been obtained in \cite{Spill:2008tp},
together with the suitable evaluation representation. This has been done for an all-fermionic Dynkin diagram\footnote{It would be interesting to do the same for distinguished Dinkyn diagrams, which have in this case only one fermionic simple root. We thank Fabian Spill for exchanges on this point, see also \cite{Dobrev:2009bb}.}. As
common in Drinfeld's second realization, different generators come
equipped with different spectral parameters:
$\kappa_{j,n} = (u + c_j)^n \, \kappa_{j,0}$, $\xi^\pm_{j,n} = (u + c_j)^n \, \xi^\pm_{j,0}$.
In this case, the coefficients $c_j$ depend on
the representation labels $x^\pm$. The map between the first and second realization of the $\alg{psl}(2|2)_c$ Yangian has a form very similar to the standard expression (or, rather, to its natural graded analog), although there are a few differences. In some sense these differences could be related to the following fact. In the strict mathematical sense, odd roots of the Lie superalgebra $\alg{psl}(2|2)$ are simultaneously positive and negative \cite{Cornwell}. 

\smallskip

The second realization given in \cite{Spill:2008tp} indeed
possesses a shift-automorphism $u \rightarrow u + const$,
which normally guarantees the difference form of the S-matrix.
This corroborates the idea that one may be able to achieve a rewriting of
the quantum S-matrix where the dependence on $u_1$ and $u_2$
is purely of difference form, the rest being taken care of
by suitable combinations of the algebra generators\footnote{In the
fundamental representation, such a rewriting has been shown to be
possible in \cite{Torrielli:2008wi}. The resulting expression is vaguely
reminiscent of what a Khoroshkin-Tolstoy type of formula
(\ref{KTformula}) (or some natural quantization of the classical
$r$-matrix (\ref{eqn;Rmat})) would look like in this
representation.}. {\it Via} this rewriting, one would expect it to become manifest that the S-matrix is the result of
evaluating a hypothetical Yangian universal R-matrix in this
particular representation (see also \cite{Spill:2008yr,Rej:2010mu}). This expectation seems to be consistent
with independent studies concerning the relationship with the
exceptional\footnote{$\alg{psl}(2|2)_c$
can be obtained by In\"on\"u-Wigner contraction of ${\alg{D}}(2,1;\epsilon)$, when
one sends $\epsilon$ to $-1$ and suitably scales the
algebra generators with $\epsilon$  in the limit. For this reason, sometimes  in the mathematical literature $\alg{psl}(2|2)_c$ is indicated with the symbol ${\alg{D}}(2,1;-1)$. For work related to Drinfeld's second realization of the quantum affine supergroup associated to 
${\alg{D}}(2,1;\epsilon)$ see also \cite{Heckenberger:2007ry}.} Lie superalgebra ${\alg{D}}(2,1;\epsilon)$
\cite{Beisert:2005tm,Matsumoto:2008ww,Matsumoto:2009rf}. Furthermore, this idea appears to be
corroborated by the explicit form of the bound state S-matrix,
which we discuss in the next section. However, we will found out later (when dealing with long representations, see section \ref{lunghe}) that the situation is actually more complicated than what is suggested by these expectations.

\section{The bound state S-matrix}\label{stleg}
The discussion of the previous section highlights the importance of investigating the structure of the S-matrix for generic representations of $\alg{psl}(2|2)_c$. One motivation is related to the issue of the existence of a universal R-matrix. Another motivation is understanding the role the Yangian and the secret symmetry have to play in the algebraic solution to the spectral problem. There is also a more stringent need of constructing S-matrices in more complicated representations, which has to do with computing finite-size corrections to the energies. A guiding criterion to attack the finite-size problem is suggested by the TBA approach.
The TBA maps the model on a finite circle to a {\it mirror} model defined on an infinite line at finite temperature. In the mirror theory, one can meaningfully speak of asymptotic states and S-matrices, only one needs to know all the possible bound state S-matrices. This idea goes back to \cite{Zamolodchikov:1989cf}, and, in the context of AdS/CFT, it has been
discussed in \cite{Ambjorn:2005wa} and developed in \cite{Arutyunov:2007tc} (for a review, see \cite{Bajnok:2010ke}).

\smallskip

According to this philosophy, it becomes crucial to have a concrete realization of the bound state S-matrices. Usually, these can be {\it bootstrapped} once the S-matrix of fundamental constituents is known \cite{Zamolodchikov:1978xm,Dorey:1996gd}. However, the present case is more complicated. Bound state representations appear in the tensor product of two fundamental representations. Such tensor product is generically irreducible, but at some special values of the momenta it becomes reducible but still indecomposable. For such values of the momenta a bound state is exchanged in the direct channel, and its representation fills one of the two blocks in which the indecomposable tensor product re-organizes itself. As a vector space, its polarizations correspond to the symmetrization of the fundamental states, and the representation is dubbed {\it symmetric} in the same sense as in footnote \ref{simmetr}. The other block/representation, as we will discuss at length in section \ref{lunghe}, is the antisymmetrized version, and it actually corresponds to a physical bound state particle not of the original string theory, but of the {\it mirror} model \cite{Arutyunov:2007tc}.

The fundamental S-matrix does not reduce to a projector on the bound state pole. In fact, the S-matrix is lower rank on the pole, and the structure would be the correct one for projecting onto the bound state representation, but the residue does not square to itself. This fact prevents a straightforward application of the so-called {\it fusion} procedure  to build the bound state S-matrix starting from the fundamental one\footnote{Roughly speaking, since the bound state is in the symmetric representation, one may think of tensoring two fundamental S-matrices and symmetrizing (see also \cite{Kulish:1981gi,Kulish:1981bi}). Curiously, a quite `practical' fusion procedure is at work in the Bethe equations and transfer matrix eigenvalues for bound states. The basic mechanism seems to be as follows. One observes the concatenation of objects of the form (say, for two excitations)
$$
e^{-i p_1}\, \frac{x_1^+ - y}{x_1^- - y} \, e^{-i p_2}\, \frac{x_2^+ - y}{x_2^- - y}.
$$
On the bound state pole $x_1^+ = x_2^-$, therefore one obtains from the above
\begin{eqnarray}
\label{fsu1}
e^{-i (p_1 + p_2)}\, \frac{x_2^+ - y}{x_1^- - y} .
\end{eqnarray}
Notice now that $p_1 + p_2$ is the total (bound state) momentum. Also, summing together the fundamental constraints (see the last formula in (\ref{parametro}))
\begin{eqnarray}
&&x_1^+ + \frac{1}{x_1^+} - x_1^- - \frac{1}{x_1^-} = \frac{2 i}{g},\nonumber\\
 &&x_2^+ + \frac{1}{x_2^+} - x_2^- - \frac{1}{x_2^-} = \frac{2 i}{g},
\end{eqnarray}
and using the bound state condition, one obtains
\begin{eqnarray}
\label{fsu}
x_2^+ + \frac{1}{x_2^+} - x_1^- - \frac{1}{x_1^-} = \frac{4 i}{g}.
\end{eqnarray}
This means that the `fused' block (\ref{fsu1}) has the same form as the fundamental ones, but with variables satisfying the bound state constraint. The author thanks M. de Leeuw for explanations. 
For the case of the dressing phase (see formula (\ref{eqn;FullPhase}) and subsequent text), something similar is going on, although a careful treatment is needed to properly take into account crossing and the `direct versus mirror' region of the momenta, see \cite{Arutyunov:2009kf}. For S-matrices and Bethe wave functions, as we were pointing out, the concatenation is not so straightforward. See section \ref{lunghe} for more details.}. The most practical way to construct the S-matrix for bound states seems to be a direct derivation from the invariance under the symmetry algebra in each bound state representation. This becomes rapidly quite cumbersome \cite{Arutyunov:2008zt} (see also \cite{Ahn:2010xa,MacKay:2010zb,Palla:2011eu} in similar contexts). Moreover, the algebra does not uniquely fix the S-matrix (up to a scalar factor) when the bound state number increases, and one needs to resort to the YBE, or, as shown in \cite{deLeeuw:2008dp}, to Yangian invariance. The Yangian ultimately provides the solution to this complicated problem, since it allows to uniquely determine the S-matrix for arbitrary bound state numbers, as done in \cite{Arutyunov:2009mi} and as we will now discuss.

Notice that the bound state S-matrices now satisfy also mixed Yang-Baxter type equations as the bound state number increases, corresponding to three-particle scattering involving bound states with different bound state numbers, see {\it e.g.} \cite{Arutyunov:2008zt}. 

\subsection{Bound state formalism and S-matrix structure}\label{bsf}
The bound state representations are atypical (short) completely symmetric representations of dimension $4 \ell$, $\ell=1,2,...$. They are all BPS, and indeed their dispersion relation is the shortening condition for representations of $\alg{psl}(2|2)_c$. This guarantees their stability as particles in the asymptotic spectrum, and allows us to develop their scattering theory. 

All these representations extend to evaluation representations of the Yangian, with an appropriate evaluation parameter $u$ (\ref{fattoreG}) \cite{deLeeuw:2008dp}. A convenient realization is given in terms of differential operators acting on the space of degree $\ell$ polynomials (superfields) in two bosonic ($w_a, \, a=1,2$) and two fermionic
($\theta_\alpha, \, \alpha=1,2$) variables. This realization is only possible for symmetric representations, where, for instance, one symmetrizes two  fundamentals in the bosonic polarization as
\begin{equation}
\label{simme}
w_1 \otimes w_1, \qquad \frac{1}{2} (w_1 \otimes w_2 + w_2 \otimes w_1), \qquad w_2 \otimes w_2
\end{equation}
(analogously, one antisymmetrizes the fermionic polarizations). (\ref{simme}) can indeed be interpreted in terms of the polynomials $w_1^2, w_1 w_2, w_2^2$. In the antisymmetric representation, one would for instance have the combination $w_1 \otimes w_2 - w_2 \otimes w_1$, and the corresponding polynomial would simply be zero.
All the details about this formalism, and on the derivation of the bound state S-matrix that will be sketched in the rest of this section, can be found in \cite{Arutyunov:2009mi}.

{\rm
We report here the algebra action on superfields of degree $\ell$ (number of bound state constituents) \cite{Arutyunov:2008zt}:
\begin{eqnarray}
\Phi_{\ell} &=& \phi^{a_{1}\ldots a_{\ell}}w_{a_{1}}\ldots
w_{a_{\ell}} +\phi^{a_{1}\ldots a_{\ell-1}\alpha}w_{a_{1}}\ldots
w_{a_{\ell-1}}\theta_{\alpha}+\phi^{a_{1}\ldots a_{\ell-2}\alpha\beta}w_{a_{1}}\ldots
w_{a_{\ell-2}}\theta_{\alpha}\theta_{\beta},\nonumber
\end{eqnarray}
\begin{eqnarray}
\begin{array}{lll}
  \mathbb{L}_{a}^{\ b} = w_{a}\frac{\partial}{\partial w_{b}}-\frac{1}{2}\delta_{a}^{b}w_{c}\frac{\partial}{\partial w_{c}}, &\qquad& \mathbb{R}_{\alpha}^{\ \beta} = \theta_{\alpha}\frac{\partial}{\partial \theta_{\beta}}-\frac{1}{2}\delta_{\alpha}^{\beta}\theta_{\gamma}\frac{\partial}{\partial \theta_{\gamma}}, \\
  \mathbb{Q}_{\alpha}^{\ a} = \mbox{a} \, \theta_{\alpha}\frac{\partial}{\partial w_{a}}+\mbox{b} \, \epsilon^{ab}\epsilon_{\alpha\beta} w_{b}\frac{\partial}{\partial \theta_{\beta}}, &\qquad& \mathbb{G}_{a}^{\ \alpha} = \mbox{d} \, w_{a}\frac{\partial}{\partial \theta_{\alpha}}+\mbox{c} \, \epsilon_{ab}\epsilon^{\alpha\beta} \theta_{\beta}\frac{\partial}{\partial w_{b}},\nonumber
\end{array}
\end{eqnarray}
\begin{eqnarray}
\begin{array}{ll}
 \mathbb{C} = \mbox{ab} \left(w_{a}\frac{\partial}{\partial w_{a}}+\theta_{\alpha}\frac{\partial}{\partial
 \theta_{\alpha}}\right), \, \, \, \, \, \, \, \, \, \qquad \mathbb{C}^{\dag} = \mbox{cd} \left(w_{a}\frac{\partial}{\partial w_{a}}+\theta_{\alpha}\frac{\partial}{\partial
 \theta_{\alpha}}\right), \nonumber
\end{array}
\end{eqnarray}
\begin{eqnarray}
\label{parametro}
 \mathbb{H}= (\mbox{ad +bc})\left(w_{a}\frac{\partial}{\partial w_{a}}+\theta_{\alpha}\frac{\partial}{\partial
 \theta_{\alpha}}\right),
\end{eqnarray}
\begin{eqnarray}
&&\mbox{a} = \sqrt{\frac{g}{2\ell}}\eta, \quad \mbox{b} = \sqrt{\frac{g}{2\ell}}
\frac{i}{\eta}\left(\frac{x^{+}}{x^{-}}-1\right), \qquad \qquad \eta = e^{i \frac{p}{4}} \, \sqrt{i x^- \, - i x^+},\nonumber\\
&&\mbox{c} = -\sqrt{\frac{g}{2\ell}}\frac{\eta}{x^{+}},  \quad
\mbox{d}=\sqrt{\frac{g}{2\ell}}\frac{x^{+}}{i\eta}\left(1-\frac{x^{-}}{x^{+}}\right), \qquad
x^{+} +
\frac{1}{x^{+}}-x^{-}-\frac{1}{x^{-}}=\frac{2i\ell}{g}.\nonumber
\end{eqnarray}
Intuitively, a bound state composed of a definite number of bosons of type $1$ and $2$, and a definite number of fermions of type $3$ and $4$, corresponds to an ordered monomial made out of those same numbers of bosonic $w_1$, $w_2$ and fermionic $\theta_3$, $\theta_4$ variables, respectively. 

We also report the $\alg{su}(2) \oplus \alg{su}(2)$ block-diagonal structure of the S-matrix, ensuing from the fact that the coproduct is trivial on $\alg{su}(2) \oplus \alg{su}(2)$:

Case I $(i),(ii)$:\ \ \ $2\times \ell_1 \ell_2$ \ vectors $\in V^{\rm{I}}$ \ ($(i),(ii)$ for $\alpha=3,4$ resp.)
\begin{equation}
\label{eqn;BasisCase1}
\stateA{k,l}\equiv\underbrace{\theta_{\alpha}w_1^{\ell_1-k-1}w_2^{k}}_{\rm{Space
1}}~\underbrace{\vartheta_{\alpha}v_1^{\ell_2-l-1}v_2^{l}}_{\rm{Space
2}},
\end{equation}
Case II $(i),(ii)$: \ \ $2\times 4 \ell_1 \ell_2$ \ vectors $\in V^{\rm{II}}$ \ ($(i),(ii)$ for $\alpha=3,4$ resp.)
\begin{eqnarray}\label{eqn;BasisCase2}
\stateB{k,l}_1&\equiv& \underbrace{\theta_{\alpha}w_1^{\ell_1-k-1}w_2^{k}}~\underbrace{v_1^{\ell_2-l}v_2^{l}},\nonumber\\
\stateB{k,l}_2&\equiv&\underbrace{w_1^{\ell_1-k}w_2^{k}}~\underbrace{\vartheta_{\alpha}v_1^{\ell_2-l-1}v_2^{l}},\nonumber\\
\stateB{k,l}_3&\equiv&\underbrace{\theta_{\alpha}w_1^{\ell_1-k-1}w_2^{k}}~\underbrace{\vartheta_{3}\vartheta_{4}v_1^{\ell_2-l-1}v_2^{l-1}},\nonumber\\
\stateB{k,l}_4&\equiv&\underbrace{\theta_{3}\theta_{4}w_1^{\ell_1-k-1}w_2^{k-1}}~\underbrace{\vartheta_{\alpha}v_1^{\ell_2-l-1}v_2^{l}},
\end{eqnarray}
Case III: \ $6 \ell_1 \ell_2$ \ vectors $\in V^{\rm{III}}$
\begin{eqnarray}\label{eqn;BasisCase3}
\stateC{k,l}_1&\equiv&\underbrace{w_1^{\ell_1-k}w_2^{k}}~\underbrace{v_1^{\ell_2-l}v_2^{l}},\nonumber\\
\stateC{k,l}_2&\equiv&\underbrace{w_1^{\ell_1-k}w_2^{k}}~\underbrace{\vartheta_{3}\vartheta_{4}v_1^{\ell_2-l-1}v_2^{l-1}},\nonumber\\
\stateC{k,l}_3&\equiv&\underbrace{\theta_{3}\theta_{4}w_1^{\ell_1-k-1}w_2^{k-1}}~\underbrace{v_1^{\ell_2-l}v_2^{l}},\nonumber\\
\stateC{k,l}_4&\equiv&\underbrace{\theta_{3}\theta_{4}w_1^{\ell_1-k-1}w_2^{k-1}}~\underbrace{\vartheta_{3}\vartheta_{4}v_1^{\ell_2-l-1}v_2^{l-1}},\nonumber\\
\stateC{k,l}_5&\equiv&\underbrace{\theta_{3}w_1^{\ell_1-k-1}w_2^{k}}~\underbrace{\vartheta_{4}v_1^{\ell_2-l}v_2^{l-1}},\nonumber\\
\stateC{k,l}_6&\equiv&\underbrace{\theta_{4}w_1^{\ell_1-k}w_2^{k-1}}~\underbrace{\vartheta_{3}v_1^{\ell_2-l-1}v_2^{l}},
\end{eqnarray}
{\rm
\begin{eqnarray}
R=\begin{pmatrix}
  \fbox{\small{$\mathscr{X}$}} & ~ & ~ & ~ & ~ \\
  ~ & \fbox{\LARGE{$\mathscr{Y}$}} & ~ & \mbox{\Huge{$0$}} & ~ \\
  ~ & ~ & \fbox{\Huge{$\mathscr{Z}$}} & ~ & ~ \\
  ~ & \mbox{\Huge{$0$}} & ~ & \fbox{\LARGE{$\mathscr{Y}$}} & ~ \\
  ~ & ~ & ~ & ~ & \fbox{\small{$\mathscr{X}$}}~
\end{pmatrix},
\end{eqnarray}}
\begin{eqnarray}
&&\mathscr{X}:V^{\rm{I}}\longrightarrow V^{\rm{I}},\qquad \qquad \qquad \qquad \qquad \qquad \, \mathscr{Y}:V^{\rm{II}}\longrightarrow V^{\rm{II}}\nonumber\\
&&\stateA{k,l}\mapsto \sum_{m=0}^{k+l}
\mathscr{X}^{k,l}_m\stateA{m,k+l-m},\qquad \qquad \stateB{k,l}_j\mapsto \sum_{m=0}^{k+l}\sum_{j=1}^{4}
\mathscr{Y}^{k,l;j}_{m;i}\stateB{m,k+l-m}_j,\nonumber
\end{eqnarray}
\begin{eqnarray}
&&\mathscr{Z}:V^{\rm{III}}\longrightarrow V^{\rm{III}}\nonumber\\
&&\stateC{k,l}_j\mapsto \sum_{m=0}^{k+l}\sum_{j=1}^{6}
\mathscr{Z}^{k,l;j}_{m;i}\stateC{m,k+l-m}_j.
\end{eqnarray}
We recall that the full S-matrix consists of two copies of the above matrix times {\it the square of} the following factor\footnote{As usual,
the overall scalar factor is essential to determine the physical poles of the S-matrix.} \cite{Chen:2006gq,Roiban:2006gs,Arutyunov:2008zt}
\begin{eqnarray}\label{eqn;FullPhase}
S_{0}(p_{1},p_{2})&=&\left(\frac{x_{1}^{-}}{x_{1}^{+}}\right)^{\frac{\ell_2}{2}}\left(\frac{x_{2}^{+}}{x_{2}^{-}}\right)^{\frac{\ell_1}{2}}\sigma(x_{1},x_{2})\,\nonumber\\ &&\qquad \qquad  \times \, \sqrt{G(\ell_2-\ell_1)G(\ell_2+\ell_1)}\prod_{q=1}^{\ell_1-1}G(\ell_2-\ell_1+2q),
\end{eqnarray}
\begin{eqnarray}
\label{fattoreG}
G(\ell ) = \frac{u_1 - u_2 + \frac{\ell }{2}}{u_1 - u_2 - \frac{\ell }{2}},\qquad \, \, u=\frac{g}{4i}
\left(x^++\frac{1}{x^+}+x^-+\frac{1}{x^-}\right).
\end{eqnarray}
$\sigma(x_{1},x_{2})$ is the so-called `dressing phase' \cite{Beisert:2006ib,Beisert:2006ez}. The square of the last factor in (\ref{eqn;FullPhase}) is related to the so-called {\it anomalous thresholds} \cite{Coleman:1978kk}. These are peculiar double-poles occurring in two-dimensional scattering, corresponding to two intermediate bound states $\ell_2 + q$ and $\ell_2 - q$, $q=1,...,\ell_1-1$, that go on-shell. Instead, the square-roots clearly produce (after squaring) regular (or, with abuse of terminology, {\it physical}) $s$- (at bound state number $\ell_2+\ell_1$) and $t$- (at bound state number $\ell_2-\ell_1$) channel poles.
}

\subsubsection{Case I}
We present the derivation of the S-matrix only for Case I states, in order to exemplify how the Yangian is used to uniquely fix the expression for the entries. The other two cases are obtained from Case I, by using the fact that Case II and III vectors are related to Case I vectors by application of $\Delta (\alg{J}^A)$, $\Delta ( \, \widehat{\alg{J}}^A )$, for suitable combinations of supersymmetry generators $\alg{J}^A$ and $\widehat{\alg{J}}^A$.

The exact solution for Case I is given as follows. First, one defines a `vacuum'
\begin{eqnarray}
|0\rangle \equiv w_1^{\ell_1}\ v_1^{\ell_2} \, \, \in \, V^{\rm III}\nonumber
\end{eqnarray}
\bigskip
such that $R |0\rangle =|0\rangle$, and then one uses the fundamental relation $\Delta^{op} \, R \, = \, R \, \Delta$
to determine the action of $R$ on the vector $|0,0\rangle^{\rm{I}}$ (\ref{eqn;BasisCase1}):

{\small 
\begin{eqnarray}\label{eqn;Sfor00}
&&R|0,0\rangle^{\rm{I}}= \frac{R \Delta(\mathbb{Q}^{1}_{3})\Delta(\mathbb{G}^{4}_{2})|0\rangle}{(\mbox{a}_2\mbox{c}_1-\mbox{a}_1\mbox{c}_2)\ell_1\ell_2}
=\frac{\Delta^{op}(\mathbb{Q}^{1}_{3})\Delta^{op}(\mathbb{G}^{4}_{2})R|0\rangle}{(\mbox{a}_2\mbox{c}_1-\mbox{a}_1\mbox{c}_2)\ell_1\ell_2}=\frac{x_1^-
-x_2^+}{x_1^+ -x_2^-}\frac{e^{i\frac{p_1}{2}}}{e^{i\frac{p_2}{2}}}\, |0,0\rangle^{\rm{I}} \equiv {\cal{D}} |0,0\rangle^{\rm{I}}.\nonumber\\
&&
\end{eqnarray}
}
One can generate the entire Case I from $|0,0\rangle^{\rm{I}}$, using Yangian charges:
{\small
\begin{eqnarray}
&&|k,l\rangle^{\rm{I}}=\nonumber \\
&&
\frac{\prod_{i=1}^k\left[\Delta(\hat{\mathbb{L}}^1_2)+\frac{\ell_1-2u_2-2i+1}{2}
\Delta(\mathbb{L}^1_2)\right]\prod_{j=1}^l\left[-\Delta(\hat{\mathbb{L}}^1_2)-\frac{1+2j-2u_1-\ell_2}{2}
\Delta(\mathbb{L}^1_2)\right]}
{\prod_{r=1}^{k}(\ell_1-r)\prod_{p=1}^{l}(\ell_2-p)\prod_{q=1}^{k+l}\left(\delta u+\frac{\ell_1+\ell_2}{2}-q\right)}|0,0\rangle^{\rm{I}},\nonumber
\end{eqnarray}
}
from which, using the same argument as in (\ref{eqn;Sfor00}), one gets \ (for $\delta u = u_1 - u_2$)
{\small
\begin{eqnarray}
&&R|k,l\rangle^{\rm{I}} = \, \nonumber {\cal{D}} \,  \times \\
&&\frac{\prod_{i=1}^k\left[\Delta^{op} (\hat{\mathbb{L}}^1_2)+\frac{\ell_1-2u_2-2i+1}{2}
\Delta^{op} (\mathbb{L}^1_2)\right]\prod_{j=1}^l\left[-\Delta^{op} (\hat{\mathbb{L}}^1_2)-\frac{1+2j-2u_1-\ell_2}{2}
\Delta^{op} (\mathbb{L}^1_2)\right]}
{\prod_{r=1}^{k}(\ell_1-r)\prod_{p=1}^{l}(\ell_2-p)\prod_{q=1}^{k+l}\left(\delta u+\frac{\ell_1+\ell_2}{2}-q\right)} \, |0,0\rangle^{\rm{I}}, \nonumber
\end{eqnarray}
}

\begin{eqnarray}
R|k,l\rangle^{\rm{I}} =
\sum_{n=0}^{k+l}\mathscr{X}^{k,l}_n|n,k+l-n\rangle^{\rm{I}},\nonumber
\end{eqnarray}
\begin{eqnarray}
&&\mathscr{X}^{k,l}_{n} =
{\cal{D}}\frac{\prod_{i=1}^{n}(\ell_1-i)\prod_{i=1}^{k+l-n}(\ell_2-i)}{\prod_{r=1}^{k}(\ell_1-r)\prod_{p=1}^{l}(\ell_2-p)\prod_{q=1}^{k+l}(\delta u +\frac{\ell_1+\ell_2}{2}-q)}\times \nonumber \\
&&\times \sum_{m=0}^{k}\left\{ {k\choose k-m }{l\choose n-m
}\prod_{p=1}^{m}\mathfrak{c}^+_p
\prod_{p=1-m}^{l-n}\mathfrak{c}^-_p
\prod_{p=1}^{k-m}\mathfrak{d}_{\frac{k-p+2}{2}}
\prod_{p=1}^{n-m}\tilde{\mathfrak{d}}_{\frac{k+l-m-p+2}{2}}\right\},\nonumber
\end{eqnarray}
\begin{eqnarray}
&&\mathfrak{c}^{\pm}_m = \delta u \pm \frac{\ell_1-\ell_2}{2} -m+1, \qquad
\tilde{\mathfrak{c}}^{\pm}_m = \delta u \pm
\frac{\ell_1+\ell_2}{2} -m+1, \nonumber \\
&&\mathfrak{d}_i = \ell_1+1-2i, \qquad \qquad \qquad \, \,
\tilde{\mathfrak{d}}_i = \ell_2+1-2i.\nonumber
\end{eqnarray}
{\rm This amplitude\footnote{Sometimes the S-matrix entries, which we occasionally refer to as ``amplitudes", are also called {\it Boltzmann weights}, as a remainder of their role in vertex models of statistical mechanics.} is the restriction to suitable integer parameters of a hypergeometric function:
\begin{eqnarray}
\label{hypergeom} &&\mathscr{X}^{k,l}_n = (-1)^{k+n} \, \pi D
\frac{\sin [(k-\ell_1) \pi ] \, \Gamma (l+1)}{\sin [\ell_1 \pi]
\sin [(k +l -\ell_2-n) \pi ] \, \Gamma (l-\ell_2+1) \Gamma
   (n+1)} \nonumber\\
&& \times \frac{\Gamma
   (n+1-\ell_1) \Gamma
   \left(l+\frac{\ell_1-\ell_2}{2}-n-\delta u
   \right) \Gamma \left(1-\frac{\ell_1+\ell_2}{2}-\delta u \right)}{\Gamma
   \left(k+l-\frac{\ell_1+\ell_2}{2}-\delta u +1\right) \Gamma \left(\frac{\ell_1-\ell_2}{2}- \delta u \right)} \times \\
&& _4\tilde{F}_3
   \left(-k,-n,\delta u
   +1-\frac{\ell_1-\ell_2}{2} ,\frac{\ell_2-\ell_1}{2}-\delta u; 1-\ell_1,\ell_2-k-l,l-n+1;1 \right),\nonumber
\end{eqnarray}
where $
_4\tilde{F}_3 (x,y,z,t;r,v,w;\tau) = {_4F_3}
(x,y,z,t;r,v,w;\tau)/[\Gamma (r) \Gamma (v) \Gamma (w)]
$.
Due to a special relation between the parameters, the above hypergeometric function is actually a $6j$-symbol. In fact, this has a simple explanation. The states in Case I carry a representation of the `bosonic' $\alg{sl}(2)_{\mathbb{L}}$ subalgebra (meaning, generators of type $\mathbb{L}$), and of the associated restriction of the Yangian. As demonstrated in \cite{Arutyunov:2009ce}, the amplitude (\ref{hypergeom}) is precisely obtained by evaluating the universal R-matrix \cite{Khoroshkin:1994uk} of such $\alg{sl}(2)_{\mathbb{L}}$ Yangian in the relevant bound state evaluation representation (up to an overall scalar factor). $6j$-symbols are obviously related to the intertwining of $\alg{sl}(2)$ (highest weight) representations.
}

As one can see from this example and from the study of other subsectors \cite{Beisert:2005wm,Arutyunov:2009ce}, upon restriction to suitable subspaces of states, the difference form of the $R$-matrix in each of those specific blocks is restored, when using the appropriate variables. On the complete space, achieving a difference form is not possible. However, the $R$-matrix displays an interesting property that we will shortly point out.

We remark that the pole structure of this amplitude, which is studied in detail in  \cite{Arutyunov:2009mi}, is consistent with the fact that, in this $\alg{su}(1|1)$ sector, one does not expect any physical $s$-channel bound-state poles. The factor $\cal{D}$, in fact, cancels the $s$-channel pole at bound-state rank $\ell_1 + \ell_2$ coming from
the overall scalar factor (\ref{eqn;FullPhase}), and no physical $s$-channel poles are left in this amplitude. 

\subsubsection{Other Cases}
As we said, the $R$-matrix for Case II and III states is uniquely obtained by using the fact that symmetry generators allow one to reach these states starting from Case I states. Schematically, one has, on one hand
\begin{eqnarray}
R \, \Delta(\mathbb{Q}) \, |Case \, {\rm II}\rangle_i \, = \, R \, Q_i \, |Case \, {\rm I}\rangle \, = \, Q_i \, R \, |Case \, {\rm I}\rangle  Q_i \mathscr{X} |Case \, {\rm I}\rangle.\nonumber
\end{eqnarray}
On the other hand,
\begin{eqnarray}
R \, \Delta(\mathbb{Q}) \, |Case \, {\rm II}\rangle_i \, = \, \Delta^{op} (\mathbb{Q}) \, R \, |Case \, {\rm II}\rangle_i = \, R_i^j  \, \Delta^{op} (\mathbb{Q}) \, |Case \, {\rm II}\rangle_j \, = \, R_i^j \, Q^{op}_j \, |Case \, {\rm I}\rangle.\nonumber
\end{eqnarray}
Combining the two one obtains
\begin{eqnarray}
\label{fact}
R_i^j = Q_i \, \mathscr{X} \, \big( [Q^{op}]^{-1} \big)^j.
\end{eqnarray}
For the explicit derivation of Case II and III (and a few subtleties thereof, related to the continuation of the formulas to small bound state numbers), we refer the reader to \cite{Arutyunov:2009mi}. The final result is completely explicit, although immediately not very communicative. Few features, however, are straightforwardly noticed.

First,
this construction automatically provides a {\it factorizing twist} \cite{twi} (see also, for instance, \cite{pfeiffer,gz2}) 
for the $R$-matrix in the bound state representations (and, therefore, {\it also for the fundamental representation})\footnote{$\mathscr{X}$ in (\ref{fact}) is naturally factorisable in a (\ref{factor}) fashion, being the universal R-matrix of ${\cal{Y}}(\alg{sl}(2))$.}:
\begin{eqnarray}
\label{factor}
R = F_{21} \times {F_{12}}^{-1}.
\end{eqnarray}
However, we remark that the coproduct twisted with $F_{12}$ is by construction cocommutative, but, as expected, not at all trivial.

Second, apart perhaps from the overall factor, the final result
depends only on $\delta u$, on combinatorial factors involving
integer bound-state components, and on specific combination of
the algebra labels $\mbox{a}_i,\mbox{b}_i,\mbox{c}_i,\mbox{d}_i$, $i=1,2$ labeling the two scattering bound states. These combinations are the same noticed in
\cite{Torrielli:2008wi}. It remains quite hard to figure out a
universal formula reproducing this S-matrix in the various bound
state representations. Nevertheless, it looks like such a universal
object would treat the evaluation parameters of the Yangian as
truly independent variables, appearing only in difference form due
to the Yangian shift-automorphism. The rest of the labels would
appear because of the presence in the universal R-matrix of the (super)charges in the typical `positive
$\otimes$ negative'-root combinations (\ref{KTformula}), breaking
the difference form due to the constraint that links the
evaluation parameter to the central charges. In the next chapter, we will see that even this expectation has to face some challenges.

The bound state S-matrices we have described have been utilized in \cite{Arutyunov:2009iq}. There, by means of the Algebraic Bethe Ansatz technique, the corresponding transfer matrices (taken as ordered products of S-matrices\footnote{Notice that transfer matrices built in this way, because of the fact that the S-matrix satisfies the YBE, automatically obey the {\it RTT} relations, and are therefore good transfer matrices for the inverse scattering problem.})  have been diagonalized for arbitrary bound state numbers, and certain conjectures on the generating functions for the transfer-matrix eigenvalues \cite{Beisert:2006qh} have been verified. 

\section{Long Representations}\label{lunghe}

As far as the AdS/CFT spectral problem is concerned, long representations do not correspond to particles in the spectrum. However, long (typical) representations naturally
enter in the construction of the large-$L$ asymptotic solution of the TBA equations, via the
so-called Y-functions\footnote{To give a proper account of the standard literature would be an overwhelming task. We refer to \cite{Bajnok:2010ke,Gromov:2010kf,Kuniba:2010ir} for reviews.} \cite{Gromov:2009tv}. The string hypothesis \cite{Korepin,Arutyunov:2009zu} is related to finite dimensional representations of the quantum group symmetry of the transfer matrix \cite{BazhanovTalk}. The Y-functions entering the string hypothesis are related to the nodes of the Dynkin diagram associated to the relevant symmetry algebra,  {\it via} the auxiliary roots in the Bethe equations. Furthermore, {\it rectangular} representation (in the sense specified in the discussion following formula (\ref{split})) are those for which the bilinear relations (Hirota, fusion) traditionally have the simplest closed form \cite{Krichever:1996qd,Kazakov:2007fy}. Rectangular representations form the smallest sector which includes the physical short representations and for which it is possible to solve the Y-system. After finding such a solution, one restricts to the physical short representations.

Let us consider highest weight representations of simple Lie superalgebras. A representation is called {\it atypical} if there exists another weight vector, different from the highest weight one, annihilated by all positive roots. Equivalently, the eigenvalue of a certain Casimir element identically vanishes in that representation. When this happens, in the process of constructing the multiplet by subsequently applying  positive roots, one encounters a zero and the multiplet truncates. The dimension of the multiplet remains smaller than what it would be if the special condition on the Casimir eigenvalue would not be met. These representations are called {\it BPS} or {short} multiplets in Physics, and the other representations are called {\it long} multiplets. In supersymmetric field theories, if a multiplet is BPS, then  its anomalous dimension is protected from receiving quantum corrections\footnote{Notice that, in ${\cal{N}}=4$ SYM, operators that are not protected nevertheless have their anomalous dimensions encoded in short representations of the centrally-extended algebra $\alg{psl}(2|2)_c$ (we are being cavalier on issues related to the infinite length of the operators). The magnon dispersion relation is in fact a shortening condition for the fundamental representation of $\alg{psl}(2|2)_c$.}. 

For finite-dimensional simple Lie algebras,
irreducible representations have only two closed invariant subspaces, $\{0\}$ and the whole module. 
{\it Indecomposable} means that a representation is not expressible as a direct sum of non-trivial representations. {\it Not fully reducible} means that it is not a direct sum of irreducible representations. If a representation is indecomposable, it is also not fully reducible. The converse is not true, since a not fully reducible representation could be a direct sum of indecomposables.  Irreducible representations are necessarily indecomposable. Finite-dimensional indecomposable representations of ordinary simple Lie algebras are irreducible.  However, in the case of superalgebras, a representation can be reducible but indecomposable. Any matrix of such a reducible representation can only be cast in upper-triangular form. Given a reducible but indecomposable representation, one calls {\it subrepresentation} the one singled out by the block corresponding to the subset of states that indeed transform among themselves under the action of the algebra. Let us call this subset $J$. Then, the set of equivalence classes defined as the elements of the complement of $J$ modulo elements of $J$ gives another representation, called {\it factor representation}.

The standard situation we will encounter shall be that an irreducible module $W$ will admit a maximal invariant subspace\footnote{Namely, a subspace $I$ such that the only invariant subspace that strictly contains $I$ is $W$ itself.} for certain values of the parameters on which the module depends. This subspace $I$ can be irreducible (as it will be for us) or indecomposable. The factor module is an atypical representation.  

The tensor product of two fundamental representations of $\alg{psl}(2|2)_c$  is generically irreducible (and a long representation), apart from special values of the central charges, when it becomes reducible but indecomposable \cite{Beisert:2006qh}. At these values, the S-matrix has a simple pole, corresponding to a bound state in the spectrum. However, the residue of the S-matrix at the bound state pole is not a projector. As we already anticipated, it is of lower rank (equal to $8$) and it has non-zero components only in the subspace corresponding to the bound state representation, but it does not square to itself. This can be easily seen in the manifest $\alg{sl}(1|2)$-invariant frame (see footnote \ref{frame}). In this frame, the coproduct is trivial for an  $\alg{sl}(1|2)$ subalgebra of $\alg{psl}(2|2)_c$, and the S-matrix takes the form
$$
R=\sum_{i=1}^3 \, c_i \, P_i.
$$
$P_i$ are orthogonal projectors onto irreducible components in the tensor product of the two relevant $\alg{sl}(1|2)$ representations, and $c_i$ are coefficients fixed by requiring invariance under the non-trivial coproduct characterizing, in this frame, the $\alg{psl}(2|2)_c$ generators outside $\alg{sl}(1|2)$ (cf. {\it Jimbo equations}). At the bound state pole, only $P_1$ and $P_3$ have a simple pole, but the coefficients are such that the residue takes the form
$$
Res \, = \, c (P_1 + e^{i \varphi} \, P_3)
$$ 
for some overall factor $c$, and phase $\varphi$ related to the central extension {\it via} the two momenta of the scattering magnons. Such residue is lower rank, but $(Res/c)^2 \neq (Res/c)$. This prevents the application of the standard fusion procedure. There exists no (coproduct-)invariant projector on either of the 8-dimensional spaces forming the indecomposable tensor product of two fundamentals at the bound state pole \cite{Arutyunov:2008zt}. 

\subsection{Synopsis}
In our discussion we will follow \cite{Arutyunov:2009pw}.

The long representations we will be interested in can be constructed by applying an outer $\alg{sl}(2)$ automorphism (see section \ref{sch}) to
the representations of the unextended $\alg{sl}(2|2)$ superalgebra. The latter representations can in turn be
obtained {\rm from} those constructed for $\alg{gl}(2|2)$ by Gould
and Zhang \cite{zhang-2005-46}, {\rm see also
\cite{Kamupingene:1989wj,Palev:1990wm}}. They are parameterized by
a continuous parameter $q\in \mathbb{C}$, which is the value of
the unique central charge (the Hamiltonian) in a given
representation. An outer $\alg{sl}(2)$ automorphism acting on
$\alg{sl}(2|2)$ can be used to generate two extra central charges,
depending on additional parameters $P$ and $g$. Here $P$ is
identified with the (generically complex) `particle momentum', while
$g$ is the coupling constant. We will focus on the
lowest (16-dimensional) long representation. The explicit realization in terms of $16\times 16$ matrices
depending on $q,P$ and $g$ can be found in \cite{Arutyunov:2009pw}. Special values of $q$
correspond to the shortening conditions. In particular, $q=1$
corresponds to an indecomposable formed out of two short
8-dimensional representations.

Given an explicit realization of the long 16-dimensional
representation, one can construct the corresponding evaluation
representation for the Yangian of section \ref{sec:YS}. We will refer to this Yangian, exclusively built upon $\alg{psl}(2|2)_c$, as the {\it minimal} Yangian. Whenever the term `Yangian' will be used from now on, it will always be understood as minimal. This is because we will need to contemplate extensions of this Yangian structure at the very end.
In fact, one finds out that,
when one of the representations involved in the scattering is long {\rm evaluation}, the corresponding $R$-matrix does not exist. 

The origin of this problem can {\rm clearly be seen in Drinfeld's second realization \cite{Spill:2008tp}\footnote{\rm Given the existence of an invertible map between the generators of Drinfeld's second \cite{Spill:2008tp} and first \cite{Beisert:2007ds} realization of the $\alg{psl}(2|2)_c$ Yangian, we will use either realizations according to the needs, considering them as completely equivalent. A general proof of this fact is however missing, since the map has always been determined so far in specific representations (although there exists a seemingly universal form that works for all the representations investigated, see the discussion following formula (\ref{uu})).}, where it can be} traced back to
non-cocommutativity of the coproduct acting on higher Yangian central charges
$\mathbb{C}_n$, with $n\geq 2$, in this representation. Since the coproducts of the Yangian central charges {\rm only} involve central elements, cocommutativity of the central charges in a specific representation is a
necessary condition for the existence of an S-matrix in that representation (see section \ref{sch}). If {\rm only} some representations admit an $R$-matrix, and not others, this means that there is no universal R-matrix.

Although the Yangian evaluation representation does not admit an S-matrix}, one can look for $psl(2|2)_c$-invariant solutions of the Yang-Baxter equation.
As we said, the tensor product of two short representations
gives an {\it irreducible} long representation, {\it i.e.}
$$
V_{4d}(p_1)\otimes V_{4d}(p_2) \approx V_{16d}(P,q)\, .
$$
Here, $V_{4d}(p)$ is a fundamental 4-dimensional representation
which depends on the particle momentum and the coupling constant.
Analogously, $V_{16d}(P,q)$ is a long 16-dimensional
representation described by the momentum $P$, the coupling
constant $g$  and the parameter $q$. There is an explicit  relation
between the pairs $(p_1,p_2)$ and $(P,q)$ at fixed $g$ (in
particular, as one may intuitively expect, $P=p_1+p_2$). For a given $p_1$ and $p_2$
there is a unique corresponding long representation. However, a given long
representation can be written as a tensor product of two short
representations in two different ways (`double cover').

The observed relationship between long and short representations
suggests that the S-matrix\footnote{After constantly jumping across the dichotomy between the mathematical and the physical literature by mercilessly switching between $R$ and S (see footnote \ref{simmetr}), we now further increase the entropy and use $\mathbb{S}$ for the $R$-matrix (S-matrix) involving long representations. R is used in `universal R-matrix'.} $\mathbb{S}_{LS}$, which scatters a
long representation with a short one, can simply be composed as a
product of two S-matrices $R_{13}$ and $R_{23}$ describing the
scattering of the corresponding short representations, {\it i.e.} (see section \ref{2s})
$$
{\mathbb S}_{LS}(P,q;p_3)=R_{13}(p_1,p_3) \, R_{23}(p_2,p_3)\, .
$$
In this formula, the tensor product of two short representations
in the spaces $1$ and $2$ with momenta $p_1$ and $p_2$ gives a
long representation $(P,q)$, which scatters with a short
representation in the third space with momentum $p_3$. The two S-matrices, which one indeed finds by directly solving the Yang-Baxter
equation, turn out to precisely coincide with the product of two ``short"
S-matrices, according to the double cover.

This
also shows that the minimal Yangian symmetry can be induced on
long representation from the one defined on the short ones, {\rm and this tensor product representation}
automatically {\rm admits an S-matrix (for both branches of the double cover). This doubly branched tensor product representation of the Yangian is therefore {\it not} isomorphic to the long evaluation representation, even though the two short representations composing it are short evaluation representations of the type discussed in section \ref{sec:YS}.

Both S-matrices come with the canonical
normalization, therefore they cannot be related to each other
by a multiplicative factor. They are
not related by a similarity transformation either. However, at the special value $q=1$ where the long multiplet
becomes reducible, the two matrices ${\mathbb S}_{LS}$ become of
the form 
\begin{eqnarray}
\begin{pmatrix}
 \mu A & \, \, B + \mu C \\
 0 & D
\end{pmatrix},
\end{eqnarray}
where the block structure refers to the splitting into the
8-dimensional sub- and factor representations at $q=1$, and the scalar coefficient $\mu$ distinguishes between the two solutions. Here, $D$ corresponds to the factor representation (symmetric), and coincides with (the inverse of) the known symmetric bound-state S-matrix $\S_{AB}$ \cite{Arutyunov:2008zt}. This is in
agreement with the fact that there is a unique bound-state
S-matrix.

\subsection{Explicit construction of long representations}\label{2s}
{\it Note.} We will derive the representation theory we will be needing directly from scratch. In particular, our notation is not immediately related to the one used in \cite{Beisert:2006qh} to label representations. 

The paper \cite{zhang-2005-46} constructs all
finite-dimensional irreducible representations of $\alg{gl} (2|2)$ in
an oscillator basis. Generators of $\alg{gl} (2|2)$ are denoted by
$E_{ij}$, with commutation relations \begin{equation}
[E_{ij},E_{kl}]=\delta_{jk} E_{il} - (-)^{(deg(i)+deg(j))(deg(k)+deg(l))}
\delta_{il} E_{kj}. \end{equation} Indices $i,j,k,l$ run from $1$ to $4$,
and the fermionic grading is assigned as $deg(1)=deg(2)=0$,
$deg(3)=deg(4)=1$. The quadratic Casimir of this algebra is $C_2 =
\sum_{i,j=1}^4 (-)^{deg(j)} E_{ij}E_{ji}$. Finite dimensional irreducible representations
are labeled by two half-integers $j_1,j_2 = 0,\frac{1}{2},...$,
and two complex numbers $q$ and $y$. These numbers correspond to
the values taken by the Cartan generators on the highest weight
state $|\omega\rangle$ of the representation, defined by  \begin{eqnarray} \label{zghw} &&H_1 |\omega\rangle = (E_{11}-E_{22})
|\omega\rangle = 2 j_1 |\omega\rangle, \qquad
H_2 |\omega\rangle = (E_{33}-E_{44}) |\omega\rangle = 2 j_2 |\omega\rangle,\nonumber\\
&&I |\omega\rangle = \sum_{i=1}^4 E_{ii} |\omega\rangle = 2 q
|\omega\rangle, \, \, \, \qquad \, \, N |\omega\rangle = \sum_{i=1}^4 (-)^{deg(i)}
E_{ii} |\omega\rangle = 2 y |\omega\rangle,\nonumber \\
&& \, \, \, \,
\qquad \qquad \qquad E_{i<j}|\omega\rangle =0. \end{eqnarray} The generator $N$ never appears on
the right hand side of the commutation relations, therefore it is
defined up to the addition of a central element $\beta I$, with
$\beta$ a constant (we will drop the term $\beta I$
as inessential). This also means that we
can consistently mod out the generator $N$, and obtain $\sls
(2|2)$ as a subalgebra of the original $\alg{gl} (2|2)$
algebra\footnote{Further modding out of the center $I$ produces
the simple Lie superalgebra $\alg{psl}(2|2)$. Its finite-dimensional representations can be understood as that of $\alg{sl}(2|2)$ for which $q=0$.
Correspondingly, $\alg{sl}(2|2)$ has long irreducible representations of dimension
$16(2j_1+1)(2j_2+1)$ with $j_1\neq j_2$ and short irreducible representations with
$j_1=j=j_2$ of dimension $16j(j+1)+2$. For a discussion of the
tensor product decomposition of $\alg{psl}(2|2)$, see
\cite{Gotz:2005ka} (see also \cite{Gotz:2005jz} for the relevant notations).}. In order to construct representations of $\alg{psl}(2|2)_c$, we then first mod out $N$,
and subsequently perform a rotation by means of the
$\sls (2)$ outer automorphism \cite{Beisert:2006qh}.

Typical
(long) representations have generic values of the labels $j_1,j_2,q$, and have dimension $16(2 j_1 +1)(2 j_2 +1)$.
For atypical (short) representations, some special relations are satisfied by
these labels. Short representations occur here for $\pm q = j_1 -
j_2$ and $\pm q = j_1 + j_2 +1$. 

The fundamental $4$-dimensional short representation corresponds to $j_1=\frac{1}{2},j_2=0$ (or,
equivalently, $j_1=0,j_2=\frac{1}{2}$) and $q=\frac{1}{2}$
($q=-\frac{1}{2}$). The bound state (symmetric
short) representations 
are given by $j_2=0,q=j_1$, with $j_1 =\frac{1}{2},1,...$ and
bound state number $M \equiv s = 2 j_1$. In addition, there are
the antisymmetric short representations given by $j_1=0,q=1+j_2$,
with $j_2 =0,\frac{1}{2},...$ and bound state number $M\equiv
a =2(j_2 + 1)$. Both symmetric and antisymmetric representations have
dimension $4M$, but they are are associated with the different shortening
conditions $\pm q=j_1-j_2$ and $\pm q=1+j_1+j_2$.

\smallskip

Let us consider
the $16$-dimensional long representation characterized by
$j_1=j_2=0$, and arbitrary $q$. We denote as
$[l_1,l_2]$ the subset of states providing a representation of the Lie subalgebra
$\alg{sl} (2) \oplus \alg{sl} (2)$ with angular momentum $l_1$ w.r.t the
first $\alg{sl} (2)$, and $l_2$ w.r.t the second $\alg{sl} (2)$,
respectively. The branching rule for the $16$-dimensional long representation is \begin{equation} \label{split} (2,2) \,
\rightarrow \, 2 \times [0,0] \oplus 2 \times
[\frac{1}{2},\frac{1}{2}] \oplus [1,0] \oplus [0,1]. \end{equation} One can
 verify that the total dimension adds up to $16$,
since $[l_1,l_2]$ has dimension $(2 l_1 + 1)\times (2 l_2 + 1)$.

Notice that setting $q=0$ in this representations gives an atypical representation of $\alg{psl}(2|2)$ \cite{Gotz:2005ka}.

\smallskip

Consider now rectangular Young tableaux, with one side made of $2$
boxes, and the other side made of arbitrarily many boxes. We associate such tableaux with
certain long representations, and denote them by $(2,s)$ and $(a,2)$ according
to the length (in boxes) of the sides of the tableaux. We then associate to short
irreducible representations, denoted accordingly as $(1,s)$ (symmetric) and $(a,1)$
(antisymmetric), correspondingly shaped tableaux. These Young tableaux fit inside the so-called ``fat hook"
\cite{Kazakov:2007na}, which has branches of width equal to two
boxes. All representations $(2,s)$ (respectively, $(a,2)$) with
$s\geq 2$ (respectively, $a\geq 2$) have
dimension\footnote{The formulas which reproduce the dimension of
the representations we associate to these Young tableaux turn out to coincide with the formulas given in \cite{BahaBalantekin:1980qy}.} equal to 16.

\smallskip

The outer automorphism maps the
$\alg{gl} (2|2)$ non-diagonal generators into new generators as
follows: 

\begin{eqnarray} &&\mathbb{L}^b_a = E_{a b} \, \, \, \forall \, \, a \neq b,
\qquad \mathbb{R}^\beta_\alpha = E_{\alpha \beta} \, \, \,
\forall \, \, \alpha \neq \beta,\nonumber\\
&&\mathbb{Q}^a_\alpha = \mbox{a} \, E_{\alpha a} + \mbox{b} \, \epsilon_{\alpha \beta} \epsilon^{a b} E_{b \beta},\nonumber\\
&&\fGG^\alpha_a = \mbox{c} \, \epsilon_{a b} \epsilon^{\alpha \beta}
E_{\beta b} + \mbox{d} \, E_{a \alpha}, \end{eqnarray} subject to the constraint
\begin{equation} \mbox{ad - bc} = 1. \end{equation} Diagonal generators are
obtained. as usual, by commuting positive and negative roots. In particular,
from the explicit matrix realization, one obtains the following
values of the central charges: \begin{equation} \fHH = 2 q \, (\mbox{ad + bc}) \,
\mathbbmss{1}, \qquad \mathbb{C} = 2 q \, \mbox{ab} \, \mathbbmss{1}, \qquad
\mathbb{C}^\dagger = 2 q \, \mbox{cd} \, \mathbbmss{1}, \end{equation} ($\mathbbmss{1}$
is the $16$-dimensional identity matrix), satisfying the condition\footnote{We notice that the combination of central charges on the l.h.s. of (\ref{conditio}) is precisely left invariant by the $\alg{sl}(2)$ outer automorphisms previously discussed. Such automorphisms  therefore preserve the (a)typicality of the representations.}
\begin{equation} \label{conditio} \frac{\fHH^2}{4} - \mathbb{C}\mathbb{C}^\dagger = q^2 \,
\mathbbmss{1}. \end{equation} When $q^2 =1$, this becomes a shortening
condition. As we anticipated, for $q =1$, the $16$-dimensional representation
becomes reducible but indecomposable. Its subrepresentation
\cite{Gotz:2005ka} is a short $8$-dimensional antisymmetric
representation, its factor representation is a short $8$-dimensional symmetric one. Formula (\ref{conditio}), however, tells us
that we can conveniently think of $q$ as a {\it generalized} bound
state number, since for short representations $2 q$ would be
replaced by the bound state number $M$ in the analogous formula
for the central charges (\ref{parametro}). This is particularly useful, since it
allows us to parameterize the labels $\mbox{a,b,c,d}$ in terms of the
familiar bound state variables\footnote{We use the conventions of
\cite{Arutyunov:2009mi}.} $x^\pm$, just replacing the bound state
number $M$ by $2 q$. The explicit parameterization is given by (cf. (\ref{parametro}))
\begin{align}
\label{param}
\mbox{a} &= \sqrt{\frac{g}{4q}}\eta,  &\mbox{b} &=-\sqrt{\frac{g}{4q}}\frac{i}{\eta}\left(1-\frac{x^+}{x^-}\right) ,\nonumber\\
\mbox{c} &=-\sqrt{\frac{g}{4q}} \frac{\eta}{x^+}  , &\mbox{d}&= \sqrt{\frac{g}{4q}}\frac{x^+}{i\eta}\left(1-\frac{x^-}{x^+}\right),
\end{align}
where
\begin{eqnarray}
\label{eta} \eta = e^{\frac{ip}{4}}\sqrt{i(x^- - x^+)}
\end{eqnarray}
and
\begin{eqnarray}
x^+ + \frac{1}{x^+} - x^- - \frac{1}{x^-} \, = \, \frac{4 i q}{g}.
\end{eqnarray}
As in the case of short representations \cite{Janik:2006dc}, there exist a
uniformizing torus with variable $z$ and periods depending on $q$. The choice (\ref{eta}) for $\eta$ we carry over from the bound states is
historically preferred in the string theory analysis
\cite{Arutyunov:2006yd,Arutyunov:2007tc,Arutyunov:2008zt,Arutyunov:2009mi},
and ensures the S-matrix to be a symmetric matrix. Positive and negative values of $q$ morally correspond to
positive and negative `energy' representations, respectively.

We equip the symmetry algebra with
the deformed Hopf-algebra coproduct of section \ref{sch}:
\begin{eqnarray}
\label{copo}
&&\Delta (\mathbb{J}) = \mathbb{J} \otimes \mathbb{U}^{[[\mathbb{J}]]} +
\mathbbmss{1} \otimes \mathbb{J}, \nonumber\\
&&\Delta(\mathbb{U})=\mathbb{U}\otimes \mathbb{U}. \end{eqnarray} 
We have realized the deformation of the coproduct in terms of an abstract central generator $\mathbb{U}$ adjoined to the algebra.
$\mathbb{J}$ is any generator of $\alg{psl}(2|2)_c$. We have
$[[\mathbb{J}]]=0$ for the bosonic $\alg{sl}(2) \oplus \alg{sl}(2)$
generators and for the `energy' generator $\mathbb{H}$,
$[[\mathbb{J}]]=1$ (resp., $-1$) for the $\mathbb{Q}$ (resp.,
$\mathbb{G}$) supercharges, and $[[\mathbb{J}]]=2$ (resp., $-2$)
for the central charge $\mathbb{C}$ (resp., $\mathbb{C}^\dagger$).

According to the an argument we have repeatedly seen in the previous chapters (cf. (\ref{interpr}) and above), the value of $\mathbb{U}$ is determined by the consistency
requirement that the coproduct is cocommutative on the center. This produces the
algebraic condition \begin{equation} \mathbb{U}^2 \, = \, \kappa \, \mathbb{C}
\, + \, \mathbbmss{1} \end{equation} for some representation-independent
constant $\kappa$. With our choice of parametrization
(\ref{param}), $\kappa$ gets expressed in terms of the coupling
constant $g$ as $\kappa = \frac{2}{i g}$, and we obtain the
familiar relation \begin{equation} \mathbb{U} \, = \, \sqrt{\frac{x^+}{x^-}}
\, \mathbbmss{1} \, =\, e^{i \frac{p}{2}} \, \mathbbmss{1}. \end{equation}

The antiparticle representation $\overline{\mathbb{J}}$ is still
defined by 
\begin{eqnarray}
x^{\pm}\rightarrow \frac{1}{x^{\pm}},
\end{eqnarray}
and the explicit charge conjugation matrix can be found in  \cite{Arutyunov:2009pw}.

The next step is to study the Yangian in this representation. One
can prove that the defining commutation relations of Drinfeld's first
realization of the minimal Yangian are
satisfied (by the generators and {\it their coproducts}, see \cite{Arutyunov:2009pw}) if we assume the evaluation representation\footnote{As we pointed out after (\ref{param}), we
use the conventions of \cite{Arutyunov:2009mi} lifted to long variables as in \cite{Arutyunov:2009pw}.} \begin{equation} \label{eval}
\widehat{\mathbb{J}} \, = \, u \, \mathbb{J}, \end{equation} where the
spectral parameter $u$ assumes the familiar form \begin{equation} \label{uu}
u \, = \,
\frac{g}{4 i} (x^+ + x^-)\bigg( 1 + \frac{1}{x^+ x^-} \bigg). \end{equation}
The above value of $u$ is once again determined
by requiring cocommutativity of the Yangian central charges
$\widehat{\mathbb{C}}$, $\widehat{\mathbb{C}}^\dagger$.

Drinfeld's second realization is also obtained by applying a similar (Drinfeld's) map as in \cite{Spill:2008tp}\footnote{The map used in \cite{Arutyunov:2009pw} works equally well for the fundamental representation, and might be related to the one used in \cite{Spill:2008tp} by a redefinition of the generators.}. This ensures the
fulfillment of the Serre relations (see also \cite{Matsumoto:2009rf}). All defining relations are satisfied. The representation one
obtains after Drinfeld's map is not any longer of a simple
evaluation-type, but it is more complicated. Nevertheless, this representation one gets for Drinfeld's second
realization of the minimal Yangian is consistent, and the coproducts obtained after
Drinfeld's map respect all commutation and Serre
relations\footnote{Antipode and charge conjugation are also
perfectly consistent with Drinfeld's second realization.}.

However, as we already mentioned, the Yangian in this
representation, both for coproducts
projected into {\it long} $\otimes$ {\it short} and for {\it long} $\otimes$ {\it long}
representations, {\rm does {\it not} admit an S-matrix}. This is easily seen by considering the
Yangian central charges $\mathbb{C}_n$, $\mathbb{C}^\dagger_n$. After making sure that for
$n=0,1$, their coproducts are cocommutative, in all tested cases for $n\geq 2$ their coproducts are central, but not cocommutative. 

Only for the special case $q^2 =1$ the Yangian
central charges appear to be cocommutative also at and for the tested cases beyond $n=2$. Nevertheless, even
for the special case $q^2 =1$, the Yangian still does not seem to
admit an S-matrix in this representation. One way to see it is by noticing that the
equation \begin{equation} \label{coprodotto} \Delta^{op} ( \, \widehat{\mathbb{J}}
\, ) \, \, \S = \S \, \Delta ( \, \widehat{\mathbb{J}} \, ), \end{equation}
when applied to certain combinations of generators and on
particular states (for instance, of highest weight w.r.t. to the $\alg{sl}(2) \oplus
\alg{sl}(2)$ splitting (\ref{split})), leads to a contradiction when
the explicit matrix realization is used. This means that such an
S-matrix does not exist for this representation of the Yangian, which also implies
that a universal R-matrix for the
minimal Yangian does not exist\footnote{\label{NiklasSerre}Strictly speaking, we consider the minimal Yangian {\it together with} the two  abstract constraints that ensure cocommutativity of the level-zero and -one central charges (see also \cite{Torrielli:2011zz}). In principle, it should be possible to deduce non-cocommutativity of the higher central charges directly from the corresponding formulas for the coproducts written in terms of the algebra generators, without referring to a specific representation. These formulas should also imply that the non-cocommutative part must disappear for representations which satisfy the shortening conditions. We thank R. Janik for the suggestion of adding more constraints to resolve the problem. We also thank N. Beisert for information about one way to find these constraints. The idea is to adjoin the $\alg{sl}(2)$ outer automorphisms to the original algebra, and construct the Yangian of the resulting bigger algebra. The Serre relations for such a Yangian are then naturally subdivided into the original Serre relations, those with mixed generators (original, and adjoined automorphisms), and those purely for the adjoined generators. While the first and third set of Serre relations independently guarantee the consistency of the procedure, the mixed one apparently produce a set of constraints {\it purely for the original generators}, as the adjoined ones would drop out of the mixed relations. Such extra constraints exactly rule out the representation which does {\it not} admit an $R$-matrix in our treatment. While this looks extremely comforting and promising, one is to face the problem that the resulting Yangian is hard to treat (for instance, it is difficult to obtain the corresponding Drinfeld's second realization). In any case, it is the personal opinion of the author that this route is at the moment the most interesting one to pursue in the quest for a universal R-matrix.}.

As we already pointed out, a
different Yangian representation, for which an S-matrix does
indeed exists, can be induced on the space of long
representations. {\rm This Yangian representation is obtained {\it via} the decomposition of long representations into
short ones, and it is therefore built upon the Yangian representations that have
already been constructed for short representations}. This induced
representation is quite different from the one described above
(cf. (\ref{eval})), and, in particular, it is not related to
(\ref{eval}) {\it via} any similarity transformation combined with
a redefinition of the spectral parameters.

Imposing invariance under the
symmetry algebra turns out not to be enough to completely fix the S-matrix. One coefficient function $\mathscr{X}(P,p$) remains undetermined, and can be
fixed by imposing that the S-matrix solves the Yang-Baxter
equation:
\begin{eqnarray}
\S_{12}(P,p_2)\S_{13}(P,p_3)\S_{23}(p_2,p_3) = \S_{23}(p_2,p_3)\S_{13}(P,p_3)\S_{12}(P,p_2).
\end{eqnarray}
By projecting on specific states, one obtains two quadratic equations for $\mathscr{X}$ of the
form
\begin{eqnarray}
\label{quadro}
A + B \mathscr{X}(P,p_2) + C \mathscr{X}(P,p_3) + D\mathscr{X}(P,p_2)\mathscr{X}(P,p_3) =0,
\end{eqnarray}
where $A,B,C,D$ are functions of $P,p_2,p_3$. It is easily seen
that there are {\it two} different solutions to these equations.
This means that we find {\it two} S-matrices. These two solutions are not related to each other by a similarity transformation. The solutions for
$\mathscr{X}$ appear rather complicated, and we refrain
from giving their explicit expressions. It can be checked
that {\it both} S-matrices satisfy the following conditions:
\begin{description}
    \item[\it Unitarity:] $\qquad \, \, \, \, \, \, \, \, \, \, \, \, \S_{12}\S_{21}=\mathbbm{1}$.
    \item[\it Hermiticity:] $\, \, \, \, \, \, \qquad \S_{12}(z_L,z)\S_{12}(z_L^*,z^*)^{\dag} = \mathbbm{1}$.
    \item[\it CPT Invariance:] $\, \, \, \, \S_{12}=\S_{12}^t$.
    \item[\it Yang-Baxter:] $\, \, \, \qquad \S_{12}\S_{13}\S_{23}=\S_{23}\S_{13}\S_{12}$.
\end{description}
Although the two solutions differ for the value of just one function $\mathscr{X}$, the way this function appears in the various matrix entries is non-trivial. Different values of $\mathscr{X}$ can determine whether certain entries ultimately vanish or not, resulting in a quite different form of the matrices for the two solutions. 

Consider the tensor product of two short representations labeled
by momenta $(p_1,p_2)$,
\begin{eqnarray}
V(p_1)\otimes V(p_2).
\end{eqnarray}
This vector space naturally carries a representation of $\alg{psl}(2|2)_c$ via the (opposite) coproduct. {\it I.e.}, for
any generator $\mathbb{J}$, we can consider
\begin{eqnarray}
\label{delt}
\mathbb{J}_{V(p_1)\otimes V(p_2)} = \Delta\mathbb{J}.
\end{eqnarray}
By considering the central charges on this space
we see that we are dealing with a long representation. To be precise, we
find
\begin{align}
(2q)^2 = \Delta\mathbb{H}^2 - 4
\Delta\mathbb{C}\Delta\mathbb{C}^{\dag} = [E(p_1)+E(p_2)]^2 -
E(p_1+p_2)^2 + 1,
\end{align}
where the energy $E(p)$ is given by
\begin{eqnarray}
\label{sq} E(p)^2= 1+4g^2 \sin^2 \frac{p}{2}.
\end{eqnarray}
The momentum of the long representation is found to be
\begin{eqnarray}
P= p_1+p_2.
\end{eqnarray}
One has therefore
\begin{eqnarray}
V(p_1)\otimes V(p_2) \cong V(P,q)
\end{eqnarray}
with
\begin{align}\label{eqn;Paramters:LongviaShort}
&P=p_1+p_2, & ~&~ q =
\frac{E(p_1)+E(p_2)}{\sqrt{[E(p_1)+E(p_2)]^2}}\frac{\sqrt{[E(p_1)+E(p_2)]^2
- E(p_1+p_2)^2 + 1}}{2}.
\end{align}
The dispersion relation (\ref{sq}) has two branches, corresponding
to particles and anti-particles. Fixing the momentum $p$ and
choosing a branch specifies the fundamental representation
completely. The tensor product of two such representations is
identified with a unique 16-dimensional long representation with
momentum $P$ and the central charge $q$ specified in (\ref{eqn;Paramters:LongviaShort}).

Suppose now that we are given a
long representation $(P,q)$ and we want to factorize it into the
tensor product of two fundamental representations. It is
convenient to label the representation space corresponding to
particles as $V_+$ and the one corresponding to antiparticles as
$V_{-}$. The module of the long representation can be
identified with one of the following four spaces:
\begin{enumerate}
\item[{ L1}:] \, \, $V_+\otimes V_+$,
\item[{L2}:] \, \, $V_+\otimes V_-$,
\item[{ L3}:] \, \, $V_-\otimes V_+$,
\item[{ L4}:] \, \, $V_-\otimes V_-$.
\end{enumerate}
Let us order the momenta
$p_1$ and $p_2$ such as, say,  $p_1\prec p_2$\footnote{The details of the
ordering are irrelevant, since the sole scope of the ordering is to choose a
unique representative between the couple $(p_1,p_2)$ and its permuted
couple $(p_2,p_1)$.}. Assuming for simplicity that $q$ is
real, we find that, to a given long representation $V(P,q)$, one can
associate two solutions in terms of `short' parameters. For
instance, for $q$ positive, the two solutions are both associated
with the space L1, or one of the solutions is from L1 and the second
is from L2. Analogous situation takes place for $q$ negative.
Thus, a given long representation can be written as a tensor product
of two {\it different} short representations. 

One can explicitly find \cite{Arutyunov:2009pw} the similarity
transformation $V_{\Delta}$ that relates the long algebra generators (constructed along the lines of \cite{zhang-2005-46} using the procedure of section \ref{2s}) to the
ones that arise from the coproduct (\ref{delt}).

The coproduct on the triple tensor product of short representations is given by
$(\Delta\otimes\mathbbm{1})\Delta$ (or, which is the same because of the {\it coassociativity} property of Hopf algebras, by $(\mathbbm{1}\otimes \Delta )\Delta$). It is easily seen that
\begin{align}
R_{13}R_{23} (\Delta\otimes\mathbbm{1})\Delta\mathbb{J}
&= R_{13}R_{23} (\Delta\mathbb{J}\otimes\mathbb{U}^{[[\mathbb{J}]]}+\mathbbm{1}_L\otimes\mathbb{J}) \nonumber \\
&= R_{13}R_{23}
(\mathbb{J}\otimes\mathbb{U}^{[[\mathbb{J}]]}\otimes\mathbb{U}^{[[\mathbb{J}]]}+\mathbbm{1}\otimes\mathbb{J}\otimes\mathbb{U}^{[[\mathbb{J}]]}+\mathbbm{1}\otimes\mathbbm{1}\otimes\mathbb{J})
\nonumber \\
&=
(\mathbb{J}\otimes\mathbb{U}^{[[\mathbb{J}]]}\otimes\mathbbm{1}+\mathbbm{1}\otimes\mathbb{J}\otimes\mathbbm{1}+\mathbb{U}^{[[\mathbb{J}]]}\otimes\mathbb{U}^{[[\mathbb{J}]]}\otimes\mathbb{J})
R_{13}R_{23}\nonumber\\
&= (\Delta\mathbb{J}\otimes\mathbbm{1} +
{\mathbb{U}_L}^{[[\mathbb{J}]]}\otimes\mathbb{J})R_{13}R_{23}.
\end{align}
Thus, we see that $R_{13}R_{23}$ intertwines the coproduct on the
tensor product of a long and a short representation. By the above
similarity transformation, we can interpret the S-matrix for $long \otimes short$ representations $\S$
as being built out of fundamental S-matrices:
\begin{eqnarray}
\S  = [V_{\Delta}\otimes\mathbbm{1}] \, \, R_{13}R_{23}
\, \, [V^{-1}_{\Delta}\otimes\mathbbm{1}].
\end{eqnarray}
The two different choices of short representations that give rise
to the long representation indeed gives two different
solutions for $\S$, which exactly coincide with the ones that are
found from the Yang-Baxter equation (cf. (\ref{quadro})).

As we announced, the fact that the
S-matrix in short representations possesses Yangian symmetries (in
evaluation representations) automatically induces, {\it via} the
above mentioned tensor product procedure, a Yangian representation
associated to the long representation. The generators are simply
given by
\begin{eqnarray}
\label{Ysh}
\widehat{\mathbb{J}}_{V(p_1)\otimes V(p_2)} = \Delta ( \, \widehat{\mathbb{J}}\, ).
\end{eqnarray}
$\Delta$ is projected into $short \otimes short$ Yangian
representations, the latter being characterized by the known
(`short') spectral parameters $u_1$ and $u_2$ (on the first and
second factor of the tensor product, respectively). These `short'
spectral parameters are linked to the parameters of the two
corresponding short representations as in (\ref{u}). 

\section{Yangian in spacetime $n$-point functions}\label{ampl}

Recently, Yangian symmetry has emerged in AdS/CFT from yet a quite different angle, {\it i.e.} in the study of spacetime $n$-point functions\footnote{We specify the attribute {\it spacetime} in order to avoid confusion with the {\it worldsheet} $n$-point functions. The spacetime in question is the four-dimensional Minkowski spacetime where the ${\cal{N}}=4$ SYM theory lives, as opposed to the two-dimensional world-sheet sigma model field theory characterizing the string side of the correspondence.} and their symmetries \cite{Drummond:2009fd}. There exist by now a few reviews on this rapidly developing subject \cite{Beisert:2010jq,Drummond:2010ep,Bargheer:2011mm,Bartels:2011nz}, and we are by no means trying to provide here an account of such developments. We would only like to draw attention to the remarkable fact that Yangian symmetry seems to permeate a wide variety of aspects of ${\cal{N}}=4$ SYM, and it is reasonable to wonder whether there exists a unified origin and description of such diverse manifestations. One striking example is the recent discovery of the secret symmetry of section \ref{segreto}, {\it mutatis mutandis}, in spacetime $n$-point amplitudes \cite{Beisert:2011pn}.

The way Yangian shows up in this new context is through the observation that tree-level spacetime $n$-point functions are annihilated by level-zero and -one generators of a $\alg{psu}(2,2|4)$ Yangian algebra, obtained by commuting the so-called {\it original} and {\it dual} superconformal symmetries. This Yangian algebra acts on the external legs of spacetime $n$-point functions as if they were periodic spin chains, much like the action we described in section \ref{spi}. This is possible because the spacetime $n$-point functions are reduced to their cyclically-invariant core before the symmetry can act. The well-definiteness of the Yangian charges on the cyclic-invariant ``chains" is ensured by the vanishing of the dual Coxeter number of the level-zero Lie superalgebra. One is likely to be dealing with a highly reducible singlet representation of the Yangian, and the natural question would be to investigate if non-singlet representations have a role to play in this analysis. On the other hand, a similar Yangian in non-singlet representations is intimately connected with the tree-level spectral problem, as we saw in section \ref{spi}.

The realization which looks most reminiscent of the spectral problem is given in terms of super-variables $Z^A$, with $A=1,...,8$. These variables are bosonic for $A=1,...,4$, and fermionic otherwise. One has, for $N$ external legs, 
\<
\label{charges}
&&x^{(0)}_{ba} = \sum_{m=1}^n \, \, Z_{m,b} \frac{\partial}{\partial Z_{m,a}},\nonumber\\
&&x^{(1)}_{ba} = \sum_{m<n=1}^N \, \, \sum_{c=1}^8 \, Z_{m,b} \frac{\partial}{\partial Z_{m,c}} Z_{n,c} \frac{\partial}{\partial Z_{n,a}},  
\>
and the ${\cal{Y}}(\alg{gl}(4|4))$-type of relations
\<
\label{comm}
&&[x^{(0)}_{ba}, x^{(0)}_{dc}] = \delta_{ad} x^{(0)}_{bc}
- (-)^{(deg(a)+deg(b))(deg(c)+deg(d))} \, \delta_{bc} x^{(0)}_{da},\nonumber\\
&&[x^{(0)}_{ba}, x^{(1)}_{dc}] = \delta_{ad} x^{(1)}_{bc}
- (-)^{(deg(a)+deg(b))(deg(c)+deg(d))} \, \delta_{bc} x^{(1)}_{da},
\>
where the supercommutator is defined as
\<
\label{superco}
&&[x^{(0)}_{ba}, x^{(0)}_{dc}] \equiv  x^{(0)}_{ba} x^{(0)}_{dc}
- (-)^{(deg(a)+deg(b))(deg(c)+deg(d))} \, x^{(0)}_{dc} x^{(0)}_{ba},\nonumber\\
&&[x^{(0)}_{ba}, x^{(1)}_{dc}] \equiv  x^{(0)}_{ba} x^{(1)}_{dc}
- (-)^{(deg(a)+deg(b))(deg(c)+deg(d))} \, x^{(1)}_{dc} x^{(0)}_{ba}.
\>
As always, we have used the rules
\<
&&[A \otimes B][C \otimes D] = (-)^{deg(B) deg(C)} AC \otimes BD,\nonumber\\
&&E_{ij} E_{kl} =\delta_{jk} E_{il}.
\>
One has $deg(E_{ij}) = deg(i)+deg(j)$ (modulo 2), with $deg(i)=0$ or $1$ if $i$ is a bosonic or fermionic index, respectively.

\section{Conclusions}\label{concl}
We have tried to give an account of the quantum group symmetry underlying the integrability of the AdS/CFT spectral problem. The main theme is the aim of finding all the hidden (local and non-local) symmetries of the problem, and unite them into a consistent algebraic framework\footnote{It would be very interesting to explore the question of how far this unifying program can be pushed, see for instance \cite{Fioravanti:1998ha,Fioravanti:1999th}, and also \cite{Fioravanti:2007un}.}.
One of the main lessons learned is that the infinite-dimensional algebra emerging from this analysis is apparently very close to the standard Yangian, at least in atypical representations, but the few crucial differences manage to be a serious challenge to the complete characterization of its mathematical structure. What we have been calling {\it Yangian} all the time, for historical reasons and for its closeness to the true Yangian as understood by mathematicians and mathematical physicists, may well be a completely different and new beast.

We would like to list some future directions of investigation, which may eventually lead to overcome these difficulties. 

The first thing that is necessary to obtain is a character formula for typical representations, and derive from first principles the T-system for the corresponding transfer matrices. In the case of Lie superalgebras, a uniform character formula does not exist even for the finite-dimensional simple case, and also for $\alg{gl}(m|n)$ no general expression is available\footnote{P. Papi, {\it Denominator identities for Lie superalgebras}, Algebra Seminar, University of York, UK, September 2010.}. Understanding how to deal with the two solutions for the `long-short' S-matrix we have described in section \ref{2s} should be instrumental to progress in this direction.

It should be quite instructive to study further the so called {\it near-flat space} limit \cite{Maldacena:2006rv,Klose:2007rz}, which simplifies the $R$-matrix and the algebra generators while maintaining the central extension. Preliminary results point towards the persistence of the secret symmetry, and actually of a whole Yangian tower of generators with the same signature $diag(1,1,-1,-1)$ and in the suitable evaluation representation \cite{unp}. Together with the recent discovery of \cite{Beisert:2011pn} that the secret symmetry is present in spacetime $n$-point functions, it has become highly relevant to understand the deep nature of this generator and if and how it is realized in the original string and gauge theory pictures. We would like to notice that the secret symmetry generator seems to play a special role in the $q$-deformed quantum algebra of \cite{Beisert:2011wq} (particularly with respect to its Yangian limit). Moreover \cite{Vidas}, the secret symmetry is found in the twisted boundary Yangian relevant to $D5$-branes, where one obtains supercharges of the new type discussed below (\ref{secretsymmetry}) also and directly {\it via} a {\it coideal} subalgebra construction \cite{Delius:2001yi}\footnote{Interestingly, in the so-called $Y=0$ system for $D3$-branes, only these type of supercharges are found {\it via} the coideal procedure, and not the secret symmetry directly.}. We remind that these new supercharges have a different spectral parameter in the `boson-fermion' with respect to the `fermion-boson' block. 

Another idea, relevant to the quantization of the classical $r$-matrix discussed in sections \ref{ssec:clpsu}, \ref{sec:univ}, would be to use the notion of `closure' of a universal enveloping algebra. Inside such a closure, denominators of central elements, like those appearing in  (\ref{eqn;Rmat}), can be given an abstract meaning. Unfortunately, it is very hard to equip such closures with a Hopf algebra structure in general\footnote{We thank E. Ragoucy for suggesting the idea and for discussions about this point.}.

\smallskip

A possible scenario is that the quantum universal R-matrix simply does not exist. It is perhaps not a surprise that one can write down seemingly universal formulas valid for atypical representations, since atypical representations are in some sense special. Typical representations are indeed traditionally more complicated\footnote{We thank P. Sorba for a discussion about this point.}, and a formula that encompasses both the typical and the atypical case may not exist. We would also like to notice that the results for long representations we presented in section \ref{lunghe} set quite rigid constraints on the possible extensions of the minimal Yangian that might admit a universal R-matrix. Basically, any extension of the minimal Yangian has to produce some Serre-type relations that rule out the evaluation representation. On the other hand, this fact may turn into a virtue, precisely {\it because} it severely restricts the allowed extensions\footnote{We thank N. Beisert for explanations about this point, see also footnote \ref{NiklasSerre}.}. Considering alternative physical setups, and/or restrictions of the algebra, for example of the type contained in  \cite{MacKay:2010ey}, may also turn out to be very fruitful.

\smallskip

Finally, non-planar corrections may organize themselves into powerful algebraic structures which might directly connect with and generalize the ones we have been describing in this review for the planar case, still providing a way of describing the spectrum exactly (see for instance \cite{Ramgoolam:2008yr} and the recent \cite{DeComarmond:2010ie}).   

\bigskip

\bigskip

\hfill{\it ``Before I came here I was confused about this subject. }

\hfill{\it Having listened to your lecture I am still confused. But on a higher level."} 

\smallskip

\hfill{\footnotesize(E. Fermi)}

\section{Acknowledgments}\label{ackn}
First, I would like to thank my collaborators on this topic, for the most enjoyable and fruitful time, and the endless list of things they have taught me: G. Arutyunov, M. de Leeuw, I. Heckenberger, T. Matsumoto, S. Moriyama, J. Plefka, C. Sieg, F. Spill, R. Suzuki and H. Yamane. 

I then would like to thank N. Beisert, for numerous discussions, crucial questions, and critical reading of several preprints; C. Ahn, Z. Bajnok, J. Balog, L. Palla and R. Nepomechie, for sharing their expertise in the theory of integrable systems (with infinite and finite size) and $R$-matrices, and for suggestions; F. Alday, C. Kristjansen, S. de Haro, D. Diaz, H. Dorn, N. Drukker, G. Jorjadze, P. Koroteev, S. Ramgoolam, M. Salizzoni, H. S. Yang and S. Zieme, for insightful discussions on wrapping diagrams and on integrability; I. Aniceto, M. Abbott, A. Jevicky, C. Kalousios, G. Papathanasiou, M. Spradlin and A. Volovich, for their interest in my work and for nice discussions; D.-S. Bak, N. Dorey, M. Magro, S.-J. Rey, S. Sch\"afer-Nameki, B. Vicedo, K. Zoubos, B. Zwiebel, for beautiful discussions on quantum groups and integrable structures, and most useful suggestions; A. Babichenko, for sharing insights on $R$-matrices and crossing symmetry, and for a collaboration which was unfortunately never finalized; A. Bassetto, G. De Pol, V. Forini, L. Griguolo, F. Lizzi, S. Pasquetti and D. Seminara, for discussions on supersymmetric field theories and quantum groups; I. Cherednik, for suggesting to interpret the centrally-extended Hopf algebra as a mini-Yangian, an idea which I keep promising myself I will soon pursue; G. Delius, for a very early discussion on quantum groups and integrable systems; B. de Wit, for thoroughly restructuring my writing, and for many inputs which eventually convened in the Introduction to this review; V. Dobrev, for very useful discussions; A. Doikou, for an insightful discussion about the universal R-matrix and the Yangian; P. Dorey, for interesting discussions and many stimulating questions; J. Drummons and L. Ferro, for an email exchange on Yangians in spacetime $n$-point functions and Drinfeld's second realization; P. Etingof, for guiding me through the computations related to the classical $r$-matrix, and for the constant interest he showed and advice he provided during my time at MIT; G. Ferretti, for pointing out that one may have integrable structures also in the absence of conformal symmetry (see for example \cite{Belitsky:2006av}); D. Fioravanti and F. Ravanini, for their constant support, and for numerous exchanges on Yangians, quantum deformations, algebraic structures in integrable systems, the Hubbard model, and the nature of the Beisert-Staudacher equations (in particular, their re-derivation in \cite{Martins:2007hb}); G. Feverati, for discussions about the distinction between the $R$-matrix of the inverse problem and the $R$-matrix associated to the scattering; V. Fomin, for very interesting discussions about Lie superalgebras, Yangians and the Hubbard model; L. Frappat, for a brief discussion on bilinear forms in Lie superalgebras; S. Frolov, for numerous discussions, and for complaining that the secret symmetry may be a product of the original supertraceless generators\footnote{Although missing so far, it is hard to exclude that such a rewriting might exist. We have tried to motivate in section \ref{classr} why one expects to have an independent status for the `secret' generator. In any case, such a rewriting may work in some (atypical) representations but not in others, much like writing the $2\times 2$ identity matrix as a product of $\alg{sl}(2)$ matrices (as pointed out by P. Sorba). We also remark that it is the coproduct which is a symmetry, and the coproduct is harder to factorize.}; N. Geer, for his interest in this subject from a purely mathematical point of view, and for a discussion; D. Grumiller, for several discussions and suggestions; B. Hoare, A. Rej, F. Spill and A. Tseytlin, for a discussion on classical $r$-matrices and weaker versions of the CYBE equation, where the r.h.s. is non-zero and proportional to a Casimir element; R. Jackiw, for discussions and penetrating questions that forced me to study more; R. Janik, for numerous discussions and suggestions, and an always formidably insightful email response; V. Kac, for useful comments and advise; M. Karowski and E. Kirchberg, for very early discussions on exact S-matrices and Hopf algebras; T. Klose, J. Minahan and T. McLoughlin, for numerous discussions, and for a stimulating collaboration and preliminary results on the Yangian and classical $r$-matrix in the near flat-space limit; N. MacKay and V. Regelskis, for sharing many insights and beautiful explanations, in particular on Yangians in the presence of boundaries; J. Maldacena, for a discussion on $2+1$ dimensional relativistic models in relation to the centrally-extended superalgebra; L. Mannelli, for interesting discussions on Hopf algebras; M. Martins, for discussions about bound state S- and transfer matrices, and for his interest in their factorizing twists; A. Odessky, for patience in reading my rustic exposition of the problem with the eye of a mathematician, and providing advise nevertheless; P. Orland, for discussions and email exchanges on the techniques used to derive scalar factors of integrable S-matrices; J. Plefka and J. M. Henn, for a stimulating discussion about Yangian symmetry in spacetime $n$-point amplitudes at the 2010 meeting in Stockholm; T. Quella, for discussions, and for pointing out the subtleties one has to beware of when using Young tableaux for Lie superalgebras; E. Ragoucy and P. Sorba, for stimulating discussions, and for starting a collaboration on the centrally-extended algebra and its completions; V. Regelskis, for very nice discussions and for kindly providing me with a copy of \cite{Vidas}; A. Rej, for discussions about the universal R-matrix for Yangians and its properties, and for pointing out how the crossing map is a coupling-constant effect, and becomes singular in one-loop gauge theory; E. Sklyanin, for sharing his illuminating insight on some of the challenging problems in the theory of integrable systems; B. Stefanski, for many interesting discussions, and for a beautiful collaboration on Landau-Lifshitz models, whose results were unfortunately never published; A. Turbiner, for a discussion about Belavin-Drinfeld's classification, and for explanations of his work; J. van de Leur, for interesting discussions on Lie algebras; S. van Tongeren, for interesting discussions about determinant operators, D-branes and open spin-chains; C. Young, for numerous enlightening discussions on Yangians and quantum algebras; M. Zabzine, for a discussion that originated my interest in classical $r$-matrices.  I would like to thank the referees of Journal of Physics A for very valuable comments and suggestions. I also would like to thank, and to apologize to for not mentioning explicitly, all the people I had conversations with about this topic, who formed my views with their input. This work has been supported by EPSRC through the grant no. EP/H000054/1.

I dedicate the effort of writing this review to my parents, to my brother and his wife, to Joe and Sunny, to Rob and Nina and their son Grant, and to my amazing wife, Jenni. 

\bibliographystyle{JHEP}
\bibliography{LitRmat}

\providecommand{\href}[2]{#2}\begingroup\raggedright\begin{thebibliography}{10%
0}

\bibitem{Lipatov:1993yb}
L.~N. Lipatov, {\it {High-energy asymptotics of multicolor QCD and exactly
  solvable lattice models}},
  \href{http://xxx.lanl.gov/abs/hep-th/9311037}{{\tt hep-th/9311037}}.

\bibitem{Minahan:2002ve}
J.~A. Minahan and K.~Zarembo, {\it The \textrm{B}ethe-ansatz for $\mathcal{N} =
  4$ super \textrm{Y}ang-\textrm{M}ills},  {\em JHEP} {\bf 03} (2003) 013,
  [\href{http://xxx.lanl.gov/abs/hep-th/0212208}{{\tt hep-th/0212208}}].

\bibitem{Maldacena:1997re}
J.~M. Maldacena, {\it {The large N limit of superconformal field theories and
  supergravity}},  {\em Adv. Theor. Math. Phys.} {\bf 2} (1998) 231--252,
  [\href{http://xxx.lanl.gov/abs/hep-th/9711200}{{\tt hep-th/9711200}}].

\bibitem{Gubser:1998bc}
S.~S. Gubser, I.~R. Klebanov, and A.~M. Polyakov, {\it {Gauge theory
  correlators from non-critical string theory}},  {\em Phys. Lett.} {\bf B428}
  (1998) 105--114, [\href{http://xxx.lanl.gov/abs/hep-th/9802109}{{\tt
  hep-th/9802109}}].

\bibitem{Witten:1998qj}
E.~Witten, {\it {Anti-de Sitter space and holography}},  {\em Adv. Theor. Math.
  Phys.} {\bf 2} (1998) 253--291,
  [\href{http://xxx.lanl.gov/abs/hep-th/9802150}{{\tt hep-th/9802150}}].

\bibitem{Aharony:1999ti}
O.~Aharony, S.~S. Gubser, J.~M. Maldacena, H.~Ooguri, and Y.~Oz, {\it {Large N
  field theories, string theory and gravity}},  {\em Phys. Rept.} {\bf 323}
  (2000) 183--386, [\href{http://xxx.lanl.gov/abs/hep-th/9905111}{{\tt
  hep-th/9905111}}].

\bibitem{D'Hoker:2002aw}
E.~D'Hoker and D.~Z. Freedman, {\it {Supersymmetric gauge theories and the
  AdS/CFT correspondence}},  \href{http://xxx.lanl.gov/abs/hep-th/0201253}{{\tt
  hep-th/0201253}}.

\bibitem{Tseytlin:2010jv}
A.~Tseytlin, {\it {Review of AdS/CFT Integrability, Chapter II.1: Classical
  AdS5xS5 string solutions}},  \href{http://xxx.lanl.gov/abs/1012.3986}{{\tt
  arXiv:1012.3986}}.

\bibitem{Beisert:2010jr}
N.~Beisert, C.~Ahn, L.~F. Alday, Z.~Bajnok, J.~M. Drummond, {\em et.~al.}, {\it
  {Review of AdS/CFT Integrability: An Overview}},
  \href{http://xxx.lanl.gov/abs/1012.3982}{{\tt arXiv:1012.3982}}.

\bibitem{Faddeev:1987ph}
L.~D. Faddeev and L.~A. Takhtajan, {\it {Hamiltonian Methods in the Theory of
  Solitons}},  {\em Springer Series in Soviet Mathematics, Berlin, Germany}
  (1987).

\bibitem{Sklyanin:1980ij}
E.~K. Sklyanin, {\it {Quantum version of the method of inverse scattering
  problem}},  {\em J. Sov. Math.} {\bf 19} (1982) 1546.

\bibitem{Evgeny}
E.~K. Sklyanin, {\it {Quantization of the Continuous Heisenberg Ferromagnet}},
  {\em Letters in Mathematical Physics} {\bf 15} (1988) 357.

\bibitem{GlebLect}
G.~Arutyunov, {\it {Student Seminar: Classical and Quantum Integrable
  Systems}},  {\em {\rm [http://www.phys.uu.nl/$\sim
  $arutyuno/StudentSeminar.pdf]}} (2006).

\bibitem{Douglas}
M.~R. Douglas, {\it {Report on the Status of the Yang-Mills Millenium Prize
  Problem}},  {\em \rm
  [http://claymath.org/millennium/Yang-Mills$\_$Theory/ym2.pdf]} (2004).

\bibitem{Staudacher:2004tk}
M.~Staudacher, {\it The factorized $\mathit{S}$-matrix of \textrm{CFT/AdS}},
  {\em JHEP} {\bf 05} (2005) 054,
  [\href{http://xxx.lanl.gov/abs/hep-th/0412188}{{\tt hep-th/0412188}}].

\bibitem{Beisert:2005tm}
N.~Beisert, {\it {The $su(2|2)$ dynamic S-matrix}},  {\em Adv. Theor. Math.
  Phys.} {\bf 12} (2008) 945,
  [\href{http://xxx.lanl.gov/abs/hep-th/0511082}{{\tt hep-th/0511082}}].

\bibitem{Arutyunov:2009ga}
G.~Arutyunov and S.~Frolov, {\it {Foundations of the $AdS_5 x S^5$ Superstring.
  Part I}},  {\em J. Phys.} {\bf A42} (2009) 254003,
  [\href{http://xxx.lanl.gov/abs/0901.4937}{{\tt arXiv:0901.4937}}].

\bibitem{1751-8121-44-12-124001}
D.~Serban, {\it Integrability and the ads/cft correspondence},  {\em Journal of
  Physics A: Mathematical and Theoretical} {\bf 44} (2011), no.~12 124001.

\bibitem{1751-8121-42-25-254001}
N.~Dorey, {\it Notes on integrability in gauge theory and string theory},  {\em
  Journal of Physics A: Mathematical and Theoretical} {\bf 42} (2009), no.~25
  254001.

\bibitem{Puletti:2010ge}
V.~G.~M. Puletti, {\it {On string integrability. A journey through the two-
  dimensional hidden symmetries in the AdS/CFT dualities}},  {\em Adv. High
  Energy Phys.} {\bf 2010} (2010) 471238,
  [\href{http://xxx.lanl.gov/abs/1006.3494}{{\tt arXiv:1006.3494}}].

\bibitem{Beisert:2004ry}
N.~Beisert, {\it {The dilatation operator of N = 4 super Yang-Mills theory and
  integrability}},  {\em Phys. Rept.} {\bf 405} (2005) 1--202,
  [\href{http://xxx.lanl.gov/abs/hep-th/0407277}{{\tt hep-th/0407277}}].

\bibitem{Sieg:2005kd}
C.~Sieg and A.~Torrielli, {\it {Wrapping interactions and the genus expansion
  of the 2- point function of composite operators}},  {\em Nucl. Phys.} {\bf
  B723} (2005) 3, [\href{http://xxx.lanl.gov/abs/hep-th/0505071}{{\tt
  hep-th/0505071}}].

\bibitem{Sieg:2010jt}
C.~Sieg, {\it {Review of AdS/CFT Integrability, Chapter I.2: The spectrum from
  perturbative gauge theory}},  \href{http://xxx.lanl.gov/abs/1012.3984}{{\tt
  arXiv:1012.3984}}.

\bibitem{Fiamberti:2007rj}
F.~Fiamberti, A.~Santambrogio, C.~Sieg, and D.~Zanon, {\it {Wrapping at four
  loops in N=4 SYM}},  {\em Phys. Lett.} {\bf B666} (2008) 100,
  [\href{http://xxx.lanl.gov/abs/0712.3522}{{\tt arXiv:0712.3522}}].

\bibitem{Bajnok:2008bm}
Z.~Bajnok and R.~A. Janik, {\it {Four-loop perturbative Konishi from strings
  and finite size effects for multiparticle states}},  {\em Nucl. Phys.} {\bf
  B807} (2009) 625, [\href{http://xxx.lanl.gov/abs/0807.0399}{{\tt
  arXiv:0807.0399}}].

\bibitem{1751-8121-44-12-124003}
D.~Volin, {\it Quantum integrability and functional equations: applications to
  the spectral problem of ads/cft and two-dimensional sigma models},  {\em
  Journal of Physics A: Mathematical and Theoretical} {\bf 44} (2011), no.~12
  124003.

\bibitem{Arutyunov:2007tc}
G.~Arutyunov and S.~Frolov, {\it {On String S-matrix, Bound States and TBA}},
  {\em JHEP} {\bf 12} (2007) 024,
  [\href{http://xxx.lanl.gov/abs/0710.1568}{{\tt arXiv:0710.1568}}].

\bibitem{Beccaria:2009eq}
M.~Beccaria, V.~Forini, T.~Lukowski, and S.~Zieme, {\it {Twist-three at five
  loops, Bethe Ansatz and wrapping}},  {\em JHEP} {\bf 03} (2009) 129,
  [\href{http://xxx.lanl.gov/abs/0901.4864}{{\tt arXiv:0901.4864}}].

\bibitem{Bajnok:2009vm}
Z.~Bajnok, A.~Hegedus, R.~A. Janik, and T.~Lukowski, {\it {Five loop Konishi
  from AdS/CFT}},  {\em Nucl.Phys.} {\bf B827} (2010) 426,
  [\href{http://xxx.lanl.gov/abs/0906.4062}{{\tt arXiv:0906.4062}}].

\bibitem{Fiamberti:2009jw}
F.~Fiamberti, A.~Santambrogio, and C.~Sieg, {\it {Five-loop anomalous dimension
  at critical wrapping order in N=4 SYM}},  {\em JHEP} {\bf 03} (2010) 103,
  [\href{http://xxx.lanl.gov/abs/0908.0234}{{\tt arXiv:0908.0234}}].

\bibitem{Lukowski:2009ce}
T.~Lukowski, A.~Rej, and V.~N. Velizhanin, {\it {Five-Loop Anomalous Dimension
  of Twist-Two Operators}},  {\em Nucl. Phys.} {\bf B831} (2010) 105,
  [\href{http://xxx.lanl.gov/abs/0912.1624}{{\tt arXiv:0912.1624}}].

\bibitem{Arutyunov:2010gb}
G.~Arutyunov, S.~Frolov, and R.~Suzuki, {\it {Five-loop Konishi from the Mirror
  TBA}},  {\em JHEP} {\bf 04} (2010) 069,
  [\href{http://xxx.lanl.gov/abs/1002.1711}{{\tt arXiv:1002.1711}}].

\bibitem{Gromov:2010vb}
N.~Gromov, V.~Kazakov, and Z.~Tsuboi, {\it {$PSU(2,2|4)$ Character of
  Quasiclassical AdS/CFT}},  {\em JHEP} {\bf 1007} (2010) 097,
  [\href{http://xxx.lanl.gov/abs/{arXiv:1002.3981}}{{\tt {arXiv:1002.3981}}}].

\bibitem{Velizhanin:2010cm}
V.~Velizhanin, {\it {Six-Loop Anomalous Dimension of Twist-Three Operators in
  N=4 SYM}},  {\em JHEP} {\bf 1011} (2010) 129,
  [\href{http://xxx.lanl.gov/abs/1003.4717}{{\tt arXiv:1003.4717}}].

\bibitem{Fioravanti:2010ge}
D.~Fioravanti and M.~Rossi, {\it {The high spin expansion of twist sector
  dimensions: the planar N=4 super Yang-Mills theory}},  {\em Adv.High Energy
  Phys.} {\bf 2010} (2010) 61413,
  [\href{http://xxx.lanl.gov/abs/{arXiv:1004.1081}}{{\tt {arXiv:1004.1081}}}].

\bibitem{Frolov:2010wt}
S.~Frolov, {\it {Konishi operator at intermediate coupling}},  {\em J.Phys.A}
  {\bf A44} (2011) 065401, [\href{http://xxx.lanl.gov/abs/1006.5032}{{\tt
  arXiv:1006.5032}}].

\bibitem{Gromov:2010km}
N.~Gromov, V.~Kazakov, S.~Leurent, and Z.~Tsuboi, {\it {Wronskian Solution for
  AdS/CFT Y-system}},  {\em JHEP} {\bf 1101} (2011) 155,
  [\href{http://xxx.lanl.gov/abs/1010.2720}{{\tt arXiv:1010.2720}}].

\bibitem{Roiban:2009aa}
R.~Roiban and A.~A. Tseytlin, {\it {Quantum strings in AdS(5) x S**5:
  Strong-coupling corrections to dimension of Konishi operator}},  {\em JHEP}
  {\bf 0911} (2009) 013, [\href{http://xxx.lanl.gov/abs/0906.4294}{{\tt
  arXiv:0906.4294}}].

\bibitem{Gromov:2011de}
N.~Gromov, D.~Serban, I.~Shenderovich, D.~Volin, and D.~Volin, {\it {Quantum
  folded string and integrability: From finite size effects to Konishi
  dimension}},  \href{http://xxx.lanl.gov/abs/1102.1040}{{\tt
  arXiv:1102.1040}}.

\bibitem{Roiban:2011fe}
R.~Roiban and A.~Tseytlin, {\it {Semiclassical string computation of
  strong-coupling corrections to dimensions of operators in Konishi
  multiplet}},  \href{http://xxx.lanl.gov/abs/1102.1209}{{\tt
  arXiv:1102.1209}}.

\bibitem{Vallilo:2011fj}
B.~C. Vallilo and L.~Mazzucato, {\it {The Konishi multiplet at strong
  coupling}},  \href{http://xxx.lanl.gov/abs/1102.1219}{{\tt arXiv:1102.1219}}.

\bibitem{Faddeev:1996iy}
L.~D. Faddeev, {\it How algebraic \textrm{B}ethe ansatz works for integrable
  model},  \href{http://xxx.lanl.gov/abs/hep-th/9605187}{{\tt hep-th/9605187}}.

\bibitem{Gromov:2009tv}
N.~Gromov, V.~Kazakov, and P.~Vieira, {\it {Exact Spectrum of Anomalous
  Dimensions of Planar N=4 Supersymmetric Yang-Mills Theory}},  {\em
  Phys.Rev.Lett.} {\bf 103} (2009) 131601,
  [\href{http://xxx.lanl.gov/abs/0901.3753}{{\tt arXiv:0901.3753}}].

\bibitem{Bombardelli:2009ns}
D.~Bombardelli, D.~Fioravanti, and R.~Tateo, {\it {Thermodynamic Bethe Ansatz
  for planar AdS/CFT: A Proposal}},  {\em J.Phys.A} {\bf A42} (2009) 375401,
  [\href{http://xxx.lanl.gov/abs/0902.3930}{{\tt arXiv:0902.3930}}].

\bibitem{Arutyunov:2009ur}
G.~Arutyunov and S.~Frolov, {\it {Thermodynamic Bethe Ansatz for the AdS5 x S5
  Mirror Model}},  {\em JHEP} {\bf 05} (2009) 068,
  [\href{http://xxx.lanl.gov/abs/0903.0141}{{\tt arXiv:0903.0141}}].

\bibitem{Puletti:2007hq}
V.~Giangreco Marotta~Puletti, T.~Klose, and O.~Ohlsson~Sax, {\it {Factorized
  world-sheet scattering in near-flat $AdS_5 x S^5$}},  {\em Nucl. Phys.} {\bf
  B792} (2008) 228, [\href{http://xxx.lanl.gov/abs/0707.2082}{{\tt
  arXiv:0707.2082}}].

\bibitem{Bargheer:2008jt}
T.~Bargheer, N.~Beisert, and F.~Loebbert, {\it {Boosting Nearest-Neighbour to
  Long-Range Integrable Spin Chains}},  {\em J. Stat. Mech.} {\bf 0811} (2008)
  L11001, [\href{http://xxx.lanl.gov/abs/0807.5081}{{\tt arXiv:0807.5081}}].

\bibitem{LeClair:1991cf}
A.~LeClair and F.~A. Smirnov, {\it {Infinite quantum group symmetry of fields
  in massive 2-D quantum field theory}},  {\em Int. J. Mod. Phys.} {\bf A7}
  (1992) 2997, [\href{http://xxx.lanl.gov/abs/hep-th/9108007}{{\tt
  hep-th/9108007}}].

\bibitem{Karowski:1978vz}
M.~Karowski and P.~Weisz, {\it {Exact Form-Factors in (1+1)-Dimensional Field
  Theoretic Models with Soliton Behavior}},  {\em Nucl. Phys.} {\bf B139}
  (1978) 455.

\bibitem{Smirnov}
F.~A. Smirnov, {\it {Form factors in completely integrable models of quantum
  field theory }},  {\em Advanced series in mathematical physics} (1992).

\bibitem{Drummond:2009fd}
J.~M. Drummond, J.~M. Henn, and J.~Plefka, {\it {Yangian symmetry of scattering
  amplitudes in N=4 super Yang-Mills theory}},  {\em JHEP} {\bf 05} (2009) 046,
  [\href{http://xxx.lanl.gov/abs/0902.2987}{{\tt arXiv:0902.2987}}].

\bibitem{Faddeev:1980zy}
L.~D. Faddeev, {\it Quantum completely integrable models of field theory},
  {\em Sov. Sci. Rev.} {\bf C1} (1980) 107--155.

\bibitem{Abe}
E.~Abe, {\it {Hopf Algebras}},  {\em Cambridge, UK: Univ. Press} (1980).

\bibitem{Drin}
V.~G. Drinfeld, {\it {Quantum groups}},  {\em Proc. of the International
  Congress of Mathematicians, Berkeley, 1986, American Mathematical Society}
  (1987) 798.

\bibitem{jimbo}
M.~Jimbo, {\it {Topics from representations of $U_q({\alg{g}})$ – An
  introductory guide to physicists}},  {\em Nankai Lecture Notes Series, {\rm
  ed.} M. L. Ge, (World Scientific, Singapore, 1992)}.

\bibitem{Chari}
V.~Chari and A.~Pressley, {\it {A Guide To Quantum Groups}},  {\em Cambridge,
  UK: Univ. Press} (1994).

\bibitem{Kassel}
C.~Kassel, {\it {Quantum Groups}},  {\em Graduate texts in Mathematics,
  Springer-Verlag New York, Inc.} (1995).

\bibitem{Dorey:1996gd}
P.~Dorey, {\it {Exact S matrices}},
  \href{http://xxx.lanl.gov/abs/hep-th/9810026}{{\tt hep-th/9810026}}.

\bibitem{Delius:1995tc}
G.~Delius, {\it {Exact S matrices with affine quantum group symmetry}},  {\em
  Nucl.Phys.} {\bf B451} (1995) 445,
  [\href{http://xxx.lanl.gov/abs/hep-th/9503079}{{\tt hep-th/9503079}}].

\bibitem{Fioravanti:1996rz}
D.~Fioravanti, A.~Mariottini, E.~Quattrini, and F.~Ravanini, {\it {Excited
  state Destri-De Vega equation for sine-Gordon and restricted sine-Gordon
  models}},  {\em Phys. Lett.} {\bf B390} (1997) 243,
  [\href{http://xxx.lanl.gov/abs/hep-th/9608091}{{\tt hep-th/9608091}}].

\bibitem{Arutyunov:2009mi}
G.~Arutyunov, M.~de~Leeuw, and A.~Torrielli, {\it {The Bound State S-Matrix for
  AdS5 x S5 Superstring}},  {\em Nucl. Phys.} {\bf B819} (2009) 319,
  [\href{http://xxx.lanl.gov/abs/0902.0183}{{\tt arXiv:0902.0183}}].

\bibitem{deLeeuw:2008ye}
M.~de~Leeuw, {\it {The Bethe Ansatz for $\mathit{AdS}_{5}\times \mathit{S}^5$
  Bound States}},  {\em JHEP} {\bf 01} (2009) 005,
  [\href{http://xxx.lanl.gov/abs/0809.0783}{{\tt arXiv:0809.0783}}].

\bibitem{Arutyunov:2009iq}
G.~Arutyunov, M.~de~Leeuw, R.~Suzuki, and A.~Torrielli, {\it {Bound State
  Transfer Matrix for AdS5 x S5 Superstring}},  {\em JHEP} {\bf 10} (2009) 025,
  [\href{http://xxx.lanl.gov/abs/0906.4783}{{\tt arXiv:0906.4783}}].

\bibitem{Beisert:2006qh}
N.~Beisert, {\it {The Analytic Bethe Ansatz for a Chain with Centrally Extended
  $su(2|2)$ Symmetry}},  {\em J. Stat. Mech.} {\bf 0701} (2007) P017,
  [\href{http://xxx.lanl.gov/abs/nlin/0610017}{{\tt nlin/0610017}}].

\bibitem{Bajnok:2010ke}
Z.~Bajnok, {\it {Review of AdS/CFT Integrability, Chapter III.6: Thermodynamic
  Bethe Ansatz}},  \href{http://xxx.lanl.gov/abs/1012.3995}{{\tt
  arXiv:1012.3995}}.

\bibitem{Bazhanov:2010ts}
V.~V. Bazhanov, T.~Lukowski, C.~Meneghelli, and M.~Staudacher, {\it {A Shortcut
  to the Q-Operator}},  {\em J.Stat.Mech.} {\bf 1011} (2010) P11002,
  [\href{http://xxx.lanl.gov/abs/1005.3261}{{\tt arXiv:1005.3261}}].

\bibitem{Arutyunov:2008zt}
G.~Arutyunov and S.~Frolov, {\it {The S-matrix of String Bound States}},  {\em
  Nucl. Phys.} {\bf B804} (2008) 90,
  [\href{http://xxx.lanl.gov/abs/0803.4323}{{\tt arXiv:0803.4323}}].

\bibitem{Beisert:2011pn}
N.~Beisert and B.~U. Schwab, {\it {Bonus Yangian Symmetry for the Planar
  S-Matrix of N=4 Super Yang-Mills}},
  \href{http://xxx.lanl.gov/abs/1103.0646}{{\tt arXiv:1103.0646}}.

\bibitem{Beisert:2010jq}
N.~Beisert, {\it {On Yangian Symmetry in Planar N=4 SYM}},
  \href{http://xxx.lanl.gov/abs/1004.5423}{{\tt arXiv:1004.5423}}.

\bibitem{Drummond:2010ep}
J.~M. Drummond, {\it {Hidden Simplicity of Gauge Theory Amplitudes}},  {\em
  Class. Quant. Grav.} {\bf 27} (2010) 214001,
  [\href{http://xxx.lanl.gov/abs/1010.2418}{{\tt arXiv:1010.2418}}].

\bibitem{Torrielli:2010kq}
A.~Torrielli, {\it {Review of AdS/CFT Integrability, Chapter VI.2: Yangian
  Algebra}},  \href{http://xxx.lanl.gov/abs/1012.4005}{{\tt arXiv:1012.4005}}.

\bibitem{Etingof}
P.~Etingof and O.~Schiffman, {\it {Lectures on Quantum Groups}},  {\em Lectures
  in Mathematical Physics, International Press, Boston} (1998).

\bibitem{MacKay:2004tc}
N.~J. MacKay, {\it {Introduction to Yangian symmetry in integrable field
  theory}},  {\em Int. J. Mod. Phys.} {\bf A20} (2005) 7189,
  [\href{http://xxx.lanl.gov/abs/hep-th/0409183}{{\tt hep-th/0409183}}].

\bibitem{Molev}
A.~Molev, {\it {Yangians and Classical Lie Algebras}},  {\em Mathematical
  Surveys and Monographs 143, American Mathematical Society, Providence, RI}
  (2007).

\bibitem{Curtright:1992kp}
T.~Curtright and C.~K. Zachos, {\it {Supersymmetry and the nonlocal Yangian
  deformation symmetry}},  {\em Nucl.Phys.} {\bf B402} (1993) 604--612,
  [\href{http://xxx.lanl.gov/abs/hep-th/9210060}{{\tt hep-th/9210060}}].
  Contribution to Prof. Biedenharn's Festschrift.

\bibitem{gz2}
Y.-Z. Zhang and M.~D. Gould, {\it {Quasi-Hopf superalgebras and elliptic
  quantum supergroups}},  {\em J. Math. Phys.} {\bf 40} (1999) 5264,
  [\href{http://xxx.lanl.gov/abs/math/9809156}{{\tt math/9809156}}].

\bibitem{2003CMaPh.240...31A}
D.~{Arnaudon}, N.~{Cramp{\'e}}, L.~{Frappat}, and E.~{Ragoucy}, {\it {Super
  Yangian $Y(osp(1|2))$ and the Universal R-matrix of Its Quantum Double}},
  {\em Commun. Math. Phys.} {\bf 240} (2003) 31,
  [\href{http://xxx.lanl.gov/abs/math/0209167}{{\tt math/0209167}}].

\bibitem{stuko}
V.~Stukopin, {\it {Yangians of classical Lie superalgebras: Basic
  constructions, quantum double and universal R-matrix}},  {\em Proceedings of
  the Institute of Mathematics of NAS of Ukraine} {\bf 50} (2004) 1195.

\bibitem{Gow}
L.~Gow, {\it {Gauss Decomposition of the Yangian $Y(\mathfrak{gl}(m|n))$}},
  {\em Commun. Math. Phys.} {\bf 276} (2007) 799.

\bibitem{Gowthesi}
L.~Gow, {\it {Yangians of Lie Superalgebras}},  {\em PhD thesis, {\rm 2007}}.

\bibitem{Creutzig:2010hr}
T.~Creutzig, {\it {Yangian Superalgebras in Conformal Field Theory}},
  \href{http://xxx.lanl.gov/abs/1011.6424}{{\tt arXiv:1011.6424}}.

\bibitem{Bernard:1992ya}
D.~Bernard, {\it {An Introduction to Yangian Symmetries}},  {\em Int. J. Mod.
  Phys.} {\bf B7} (1993) 3517,
  [\href{http://xxx.lanl.gov/abs/hep-th/9211133}{{\tt hep-th/9211133}}].

\bibitem{Dsecond}
V.~G. Drinfeld, {\it {A new realization of Yangians and quantum affine
  algebras}},  {\em Soviet Math. Dokl.} {\bf 36} (1988) 212.

\bibitem{Khoroshkin:1994uk}
S.~M. Khoroshkin and V.~N. Tolstoy, {\it {Yangian double}},  {\em Lett. Math.
  Phys.} {\bf 36} (1996) 373,
  [\href{http://xxx.lanl.gov/abs/hep-th/9406194}{{\tt hep-th/9406194}}].

\bibitem{Faddeev:1987ih}
L.~D. Faddeev, N.~Y. Reshetikhin, and L.~A. Takhtajan, {\it {Quantization of
  Lie Groups and Lie Algebras}},  {\em Leningrad Math. J.} {\bf 1} (1990) 193.

\bibitem{KR1}
A.~N. Kirillov and N.~Y. Reshetikhin, {\it {The Yangians, Bethe Ansatz and
  Combinatorics}},  {\em Lett. Math. Phys.} {\bf 12} (1986) 199.

\bibitem{KR2}
A.~N. Kirillov and N.~Y. Reshetikhin, {\it {Representations of Yangians and
  multiplicities of occurrence of the irreducible components of the tensor
  product of representations of simple Lie algebras}},  {\em Journal of
  mathematical sciences} {\bf 52} (1990) 3156.

\bibitem{CPr}
V.~Chari and A.~Pressley, {\it {Fundamental representations of Yangians and
  singularities of R-matrices}},  {\em Journal fuer die reine und angewandte
  Mathematik} {\bf 417} (1991) 87.

\bibitem{Nazarov}
M.~L. Nazarov, {\it {Quantum Berezinian and the Classical Capelli Identity}},
  {\em Lett. Math. Phys.} {\bf 21} (1991) 123.

\bibitem{GowBer}
L.~Gow, {\it {On the Yangian $Y(gl(m|n))$ and its quantum Berezinian}},  {\em
  Czechoslovak Journal of Physics} {\bf 55} (2005) 1415.

\bibitem{Kulish:1981bi}
P.~Kulish and E.~Sklyanin, {\it {Quantum Spectral Transform Method. Recent
  Developments}},  {\em Lect. Notes Phys.} {\bf 151} (1982) 61.

\bibitem{Kac}
V.~G. Kac, {\it {Lie superalgebras}},  {\em Advances in mathematics} {\bf 26}
  (1977) 8.

\bibitem{Frappat:1996pb}
L.~Frappat, P.~Sorba, and A.~Sciarrino, {\it {Dictionary on Lie
  superalgebras}},  \href{http://xxx.lanl.gov/abs/hep-th/9607161}{{\tt
  hep-th/9607161}}.

\bibitem{Goddard:1986bp}
P.~Goddard and D.~I. Olive, {\it {Kac-Moody and Virasoro Algebras in Relation
  to Quantum Physics}},  {\em Int. J. Mod. Phys.} {\bf A1} (1986) 303.

\bibitem{kt2}
S.~M. Khoroshkin and V.~N. Tolstoy, {\it {On Drinfeld's realization of quantum
  affine algebras}},  {\em J. Geom. Phys.} {\bf 11} (1993) 445.

\bibitem{DingFrenkel}
J.~Ding and I.~B. Frenkel, {\it {Isomorphism of two realizations of quantum
  affine algebra $U_q (\widehat{\mathfrak{gl} (n)} )$}},  {\em Commun. Math.
  Phys.} {\bf 156} (1993) 277.

\bibitem{Khoroshkin:1994uj}
S.~M. Khoroshkin and V.~N. Tolstoy, {\it {Twisting of quantum (super)algebras:
  Connection of Drinfeld's and Cartan-Weyl realizations for quantum affine
  algebras}},  \href{http://xxx.lanl.gov/abs/hep-th/9404036}{{\tt
  hep-th/9404036}}.

\bibitem{jing}
N.~Jing, {\it {On Drinfeld realization of quantum affine algebras}},  {\em
  Proceedings of Conf. on Lie Alg. at Ohio State Univ., May 1996; {\rm in}
  Monster and Lie Algebras, {\rm ed.} J. Ferrar and K. Harada, OSU Math Res
  Inst Publ. {\bf 7}, de Gruyter, Berlin, {\rm (1998) 195}}
  [\href{http://xxx.lanl.gov/abs/q-alg/9610035}{{\tt q-alg/9610035}}].

\bibitem{HaMi}
N.~Hayaishi and K.~Miki, {\it {L operators and Drinfeld's generators}},  {\em
  J. Math. Phys.} {\bf 39} (1998) 1623,
  [\href{http://xxx.lanl.gov/abs/q-alg/9705018}{{\tt q-alg/9705018}}].

\bibitem{Dolan:2003uh}
L.~Dolan, C.~R. Nappi, and E.~Witten, {\it {A relation between approaches to
  integrability in superconformal Yang-Mills theory}},  {\em JHEP} {\bf 10}
  (2003) 017, [\href{http://xxx.lanl.gov/abs/hep-th/0308089}{{\tt
  hep-th/0308089}}].

\bibitem{Dolan:2004ps}
L.~Dolan, C.~R. Nappi, and E.~Witten, {\it {Yangian symmetry in D = 4
  superconformal Yang-Mills theory}},
  \href{http://xxx.lanl.gov/abs/hep-th/0401243}{{\tt hep-th/0401243}}.

\bibitem{Dolan:2004ys}
L.~Dolan and C.~R. Nappi, {\it {Spin models and superconformal Yang-Mills
  theory}},  {\em Nucl. Phys.} {\bf B717} (2005) 361,
  [\href{http://xxx.lanl.gov/abs/hep-th/0411020}{{\tt hep-th/0411020}}].

\bibitem{Serban:2004jf}
D.~Serban and M.~Staudacher, {\it {Planar N = 4 gauge theory and the Inozemtsev
  long range spin chain}},  {\em JHEP} {\bf 06} (2004) 001,
  [\href{http://xxx.lanl.gov/abs/hep-th/0401057}{{\tt hep-th/0401057}}].

\bibitem{Agarwal:2004sz}
A.~Agarwal and S.~G. Rajeev, {\it {Yangian symmetries of matrix models and spin
  chains: The dilatation operator of N = 4 SYM}},  {\em Int. J. Mod. Phys.}
  {\bf A20} (2005) 5453, [\href{http://xxx.lanl.gov/abs/hep-th/0409180}{{\tt
  hep-th/0409180}}].

\bibitem{Agarwal:2005ed}
A.~Agarwal, {\it {Comments on higher loop integrability in the $su(1|1)$ sector
  of N = 4 SYM: Lessons from the su(2) sector}},
  \href{http://xxx.lanl.gov/abs/hep-th/0506095}{{\tt hep-th/0506095}}.

\bibitem{Zwiebel:2006cb}
B.~I. Zwiebel, {\it {Yangian symmetry at two-loops for the $su(2|1)$ sector of
  N = 4 SYM}},  {\em J. Phys.} {\bf A40} (2007) 1141,
  [\href{http://xxx.lanl.gov/abs/hep-th/0610283}{{\tt hep-th/0610283}}].

\bibitem{Beisert:2007sk}
N.~Beisert and B.~I. Zwiebel, {\it {On Symmetry Enhancement in the $psu(1,1|2)$
  Sector of N=4 SYM}},  {\em JHEP} {\bf 10} (2007) 031,
  [\href{http://xxx.lanl.gov/abs/0707.1031}{{\tt arXiv:0707.1031}}].

\bibitem{Beisert:2003tq}
N.~Beisert, C.~Kristjansen, and M.~Staudacher, {\it {The dilatation operator of
  N = 4 super Yang-Mills theory}},  {\em Nucl. Phys.} {\bf B664} (2003) 131,
  [\href{http://xxx.lanl.gov/abs/hep-th/0303060}{{\tt hep-th/0303060}}].

\bibitem{Beisert:2005wv}
N.~Beisert and T.~Klose, {\it {Long-range gl(n) integrable spin chains and
  plane-wave matrix theory}},  {\em J. Stat. Mech.} {\bf 0607} (2006) P006,
  [\href{http://xxx.lanl.gov/abs/hep-th/0510124}{{\tt hep-th/0510124}}].

\bibitem{Beisert:2007jv}
N.~Beisert and D.~Erkal, {\it {Yangian Symmetry of Long-Range gl(N) Integrable
  Spin Chains}},  {\em J. Stat. Mech.} {\bf 0803} (2008) P03001,
  [\href{http://xxx.lanl.gov/abs/0711.4813}{{\tt arXiv:0711.4813}}].

\bibitem{Zwiebel:2008gr}
B.~I. Zwiebel, {\it {Iterative Structure of the N=4 SYM Spin Chain}},  {\em
  JHEP} {\bf 07} (2008) 114, [\href{http://xxx.lanl.gov/abs/0806.1786}{{\tt
  arXiv:0806.1786}}].

\bibitem{Agarwal:2004cb}
A.~Agarwal and S.~G. Rajeev, {\it {The dilatation operator of N = 4 SYM and
  classical limits of spin chains and matrix models}},  {\em Mod. Phys. Lett.}
  {\bf A19} (2004) 2549, [\href{http://xxx.lanl.gov/abs/hep-th/0405116}{{\tt
  hep-th/0405116}}].

\bibitem{Mikhailov:2004ca}
A.~Mikhailov, {\it {Anomalous dimension and local charges}},  {\em JHEP} {\bf
  12} (2005) 009, [\href{http://xxx.lanl.gov/abs/hep-th/0411178}{{\tt
  hep-th/0411178}}].

\bibitem{Agarwal:2006nv}
A.~Agarwal and A.~P. Polychronakos, {\it {BPS operators in N = 4 SYM: Calogero
  models and 2D fermions}},  {\em JHEP} {\bf 08} (2006) 034,
  [\href{http://xxx.lanl.gov/abs/hep-th/0602049}{{\tt hep-th/0602049}}].

\bibitem{Ihry:2008gm}
J.~N. Ihry, {\it {Yangians in Deformed Super Yang-Mills Theories}},  {\em JHEP}
  {\bf 04} (2008) 051, [\href{http://xxx.lanl.gov/abs/0802.3644}{{\tt
  arXiv:0802.3644}}].

\bibitem{Mansson:2008xv}
T.~Mansson and K.~Zoubos, {\it {Quantum Symmetries and Marginal Deformations}},
   {\em JHEP} {\bf 1010} (2010) 043,
  [\href{http://xxx.lanl.gov/abs/0811.3755}{{\tt arXiv:0811.3755}}].

\bibitem{Zwiebel:2009vb}
B.~I. Zwiebel, {\it {Two-loop Integrability of Planar N=6 Superconformal
  Chern-Simons Theory}},  {\em J.Phys.A} {\bf A42} (2009) 495402,
  [\href{http://xxx.lanl.gov/abs/0901.0411}{{\tt arXiv:0901.0411}}].

\bibitem{Zoubos:2010kh}
K.~Zoubos, {\it {Review of AdS/CFT Integrability, Chapter IV.2: Deformations,
  Orbifolds and Open Boundaries}},
  \href{http://xxx.lanl.gov/abs/1012.3998}{{\tt arXiv:1012.3998}}.

\bibitem{Luscher:1977rq}
M.~Luscher and K.~Pohlmeyer, {\it {Scattering of Massless Lumps and Nonlocal
  Charges in the Two-Dimensional Classical Nonlinear Sigma Model}},  {\em Nucl.
  Phys.} {\bf B137} (1978) 46.

\bibitem{Brezin:1979am}
E.~Brezin, C.~Itzykson, J.~Zinn-Justin, and J.~B. Zuber, {\it {Remarks About
  the Existence of Nonlocal Charges in Two- Dimensional Models}},  {\em Phys.
  Lett.} {\bf B82} (1979) 442.

\bibitem{Curtright:1979am}
T.~L. Curtright and C.~K. Zachos, {\it {Nonlocal currents for supersymmetric
  nonlinear models}},  {\em Phys. Rev.} {\bf D21} (1980) 411.

\bibitem{Ridout:2011wx}
D.~Ridout and J.~Teschner, {\it {Integrability of a family of quantum field
  theories related to sigma models}},
  \href{http://xxx.lanl.gov/abs/1102.5716}{{\tt arXiv:1102.5716}}.

\bibitem{Luscher:1977uq}
M.~Luscher, {\it {Quantum Nonlocal Charges and Absence of Particle Production
  in the Two-Dimensional Nonlinear Sigma Model}},  {\em Nucl. Phys.} {\bf B135}
  (1978) 1.

\bibitem{Tarasov:1983cj}
V.~O. Tarasov, L.~A. Takhtajan, and L.~D. Faddeev, {\it {Local Hamiltonians for
  integrable quantum models on a lattice}},  {\em Theor. Math. Phys.} {\bf 57}
  (1983) 1059.

\bibitem{Evans:1999mj}
J.~M. Evans, M.~Hassan, N.~J. MacKay, and A.~J. Mountain, {\it {Local conserved
  charges in principal chiral models}},  {\em Nucl. Phys.} {\bf B561} (1999)
  385, [\href{http://xxx.lanl.gov/abs/hep-th/9902008}{{\tt hep-th/9902008}}].

\bibitem{KR}
A.~N. Kirillov and N.~Y. Reshetikhin, {\it {The Yangians, Bethe Ansatz and
  combinatorics}},  {\em Letters in Mathematical Physics} {\bf 12} (1986) 199.

\bibitem{Bena:2003wd}
I.~Bena, J.~Polchinski, and R.~Roiban, {\it Hidden symmetries of the
  $\mathit{AdS}_{5}\times \mathit{S}^5$ superstring},  {\em Phys. Rev.} {\bf
  D69} (2004) 046002, [\href{http://xxx.lanl.gov/abs/hep-th/0305116}{{\tt
  hep-th/0305116}}].

\bibitem{Mandal:2002fs}
G.~Mandal, N.~V. Suryanarayana, and S.~R. Wadia, {\it {Aspects of semiclassical
  strings in AdS(5)}},  {\em Phys. Lett.} {\bf B543} (2002) 81,
  [\href{http://xxx.lanl.gov/abs/hep-th/0206103}{{\tt hep-th/0206103}}].

\bibitem{Alday:2003zb}
L.~F. Alday, {\it {Non-local charges on AdS(5) x S**5 and pp-waves}},  {\em
  JHEP} {\bf 12} (2003) 033,
  [\href{http://xxx.lanl.gov/abs/hep-th/0310146}{{\tt hep-th/0310146}}].

\bibitem{Arutyunov:2003rg}
G.~Arutyunov and M.~Staudacher, {\it {Matching higher conserved charges for
  strings and spins}},  {\em JHEP} {\bf 03} (2004) 004,
  [\href{http://xxx.lanl.gov/abs/hep-th/0310182}{{\tt hep-th/0310182}}].

\bibitem{Hou:2004ru}
B.-Y. Hou, D.-T. Peng, C.-H. Xiong, and R.-H. Yue, {\it {The affine hidden
  symmetry and integrability of type IIB superstring in AdS(5) x S**5}},
  \href{http://xxx.lanl.gov/abs/hep-th/0406239}{{\tt hep-th/0406239}}.

\bibitem{Hatsuda:2004it}
M.~Hatsuda and K.~Yoshida, {\it {Classical integrability and super Yangian of
  superstring on AdS(5) x S**5}},  {\em Adv. Theor. Math. Phys.} {\bf 9} (2005)
  703, [\href{http://xxx.lanl.gov/abs/hep-th/0407044}{{\tt hep-th/0407044}}].

\bibitem{Das:2004hy}
A.~K. Das, J.~Maharana, A.~Melikyan, and M.~Sato, {\it {The algebra of
  transition matrices for the AdS(5) x S**5 superstring}},  {\em JHEP} {\bf 12}
  (2004) 055, [\href{http://xxx.lanl.gov/abs/hep-th/0411200}{{\tt
  hep-th/0411200}}].

\bibitem{Alday:2005gi}
L.~F. Alday, G.~Arutyunov, and A.~A. Tseytlin, {\it {On integrability of
  classical superstrings in AdS(5) x S**5}},  {\em JHEP} {\bf 07} (2005) 002,
  [\href{http://xxx.lanl.gov/abs/hep-th/0502240}{{\tt hep-th/0502240}}].

\bibitem{Frolov:2005dj}
S.~Frolov, {\it {Lax pair for strings in Lunin-Maldacena background}},  {\em
  JHEP} {\bf 05} (2005) 069,
  [\href{http://xxx.lanl.gov/abs/hep-th/0503201}{{\tt hep-th/0503201}}].

\bibitem{Das:2005hp}
A.~K. Das, A.~Melikyan, and M.~Sato, {\it {The algebra of flat currents for the
  string on AdS(5) x S**5 in the light-cone gauge}},  {\em JHEP} {\bf 11}
  (2005) 015, [\href{http://xxx.lanl.gov/abs/hep-th/0508183}{{\tt
  hep-th/0508183}}].

\bibitem{Arutyunov:2008if}
G.~Arutyunov and S.~Frolov, {\it {Superstrings on $AdS_4 \times CP^3$ as a
  Coset Sigma-model}},  {\em JHEP} {\bf 09} (2008) 129,
  [\href{http://xxx.lanl.gov/abs/0806.4940}{{\tt arXiv:0806.4940}}].

\bibitem{Stefanski:2008ik}
B.~Stefanski~jr., {\it {Green-Schwarz action for Type IIA strings on
  $AdS_4\times CP^3$}},  {\em Nucl. Phys.} {\bf B808} (2009) 80,
  [\href{http://xxx.lanl.gov/abs/0806.4948}{{\tt arXiv:0806.4948}}].

\bibitem{Bernard:1992mu}
D.~Bernard and A.~Leclair, {\it {The Quantum double in integrable quantum field
  theory}},  {\em Nucl. Phys.} {\bf B399} (1993) 709,
  [\href{http://xxx.lanl.gov/abs/hep-th/9205064}{{\tt hep-th/9205064}}].

\bibitem{MacKay:1992rc}
N.~J. MacKay, {\it {On the classical origins of Yangian symmetry in integrable
  field theory}},  {\em Phys. Lett.} {\bf B281} (1992) 90.

\bibitem{Zakharov:1978wc}
V.~E. Zakharov and A.~V. Mikhailov, {\it {Example of nontrivial interaction of
  solitons in two-dimensional classical field theory}},  {\em Pisma Zh. Eksp.
  Teor. Fiz.} {\bf 27} (1978) 47.

\bibitem{Arutyunov:2006ak}
G.~Arutyunov, S.~Frolov, J.~Plefka, and M.~Zamaklar, {\it The off-shell
  symmetry algebra of the light-cone $\mathit{AdS}_{5}\times \mathit{S}^5$
  superstring},  {\em J. Phys.} {\bf A40} (2007) 3583,
  [\href{http://xxx.lanl.gov/abs/hep-th/0609157}{{\tt hep-th/0609157}}].

\bibitem{Kac:1977qb}
V.~G. Kac, {\it {A Sketch of Lie Superalgebra Theory}},  {\em Commun. Math.
  Phys.} {\bf 53} (1977) 31.

\bibitem{Evans:1990qq}
J.~Evans and T.~J. Hollowood, {\it {Supersymmetric Toda field theories}},  {\em
  Nucl. Phys.} {\bf B352} (1991) 723.

\bibitem{SchaferNameki:2009xr}
S.~Schafer-Nameki, M.~Yamazaki, and K.~Yoshida, {\it {Coset Construction for
  Duals of Non-relativistic CFTs}},  {\em JHEP} {\bf 05} (2009) 038,
  [\href{http://xxx.lanl.gov/abs/0903.4245}{{\tt arXiv:0903.4245}}].

\bibitem{Babichenko:2009dk}
A.~Babichenko, B.~Stefanski, Jr., and K.~Zarembo, {\it {Integrability and the
  AdS(3)/CFT(2) correspondence}},  {\em JHEP} {\bf 03} (2010) 058,
  [\href{http://xxx.lanl.gov/abs/0912.1723}{{\tt arXiv:0912.1723}}].

\bibitem{FabianThesis}
F.~Spill, {\it {Hopf Algebras in the AdS/CFT Correspondence}},  {\em Diploma
  Thesis, Humboldt University of Berlin} (2007).

\bibitem{IoharaKoga}
K.~Iohara and Y.~Koga, {\it {Central extensions of Lie superalgebras}},  {\em
  Comment. Math. Helv.} {\bf 76} (2001) 110.

\bibitem{Hofman:2006xt}
D.~M. Hofman and J.~M. Maldacena, {\it {Giant magnons}},  {\em J. Phys.} {\bf
  A39} (2006) 13095, [\href{http://xxx.lanl.gov/abs/hep-th/0604135}{{\tt
  hep-th/0604135}}].

\bibitem{Gomez:2006va}
C.~Gomez and R.~Hernandez, {\it {The magnon kinematics of the AdS/CFT
  correspondence}},  {\em JHEP} {\bf 11} (2006) 021,
  [\href{http://xxx.lanl.gov/abs/hep-th/0608029}{{\tt hep-th/0608029}}].

\bibitem{Plefka:2006ze}
J.~Plefka, F.~Spill, and A.~Torrielli, {\it {On the Hopf algebra structure of
  the AdS/CFT S-matrix}},  {\em Phys. Rev.} {\bf D74} (2006) 066008,
  [\href{http://xxx.lanl.gov/abs/hep-th/0608038}{{\tt hep-th/0608038}}].

\bibitem{Arutyunov:2006yd}
G.~Arutyunov, S.~Frolov, and M.~Zamaklar, {\it {The Zamolodchikov-Faddeev
  algebra for $\mathit{AdS}_{5}\times \mathit{S}^5$ superstring}},  {\em JHEP}
  {\bf 04} (2007) 002, [\href{http://xxx.lanl.gov/abs/hep-th/0612229}{{\tt
  hep-th/0612229}}].

\bibitem{Janik:2006dc}
R.~A. Janik, {\it The $\mathit{AdS}_{5}\times \mathit{S}^5$ superstring
  worldsheet $\mathit{S}$-matrix and crossing symmetry},  {\em Phys. Rev.} {\bf
  D73} (2006) 086006, [\href{http://xxx.lanl.gov/abs/hep-th/0603038}{{\tt
  hep-th/0603038}}].

\bibitem{Torrielli:2007mc}
A.~Torrielli, {\it {Classical r-matrix of the $su(2|2)$ SYM spin-chain}},  {\em
  Phys. Rev.} {\bf D75} (2007) 105020,
  [\href{http://xxx.lanl.gov/abs/hep-th/0701281}{{\tt hep-th/0701281}}].

\bibitem{Serganova}
V.~V. Serganova, {\it {Automorphisms of Simple Lie Superalgebras}},  {\em Math.
  USSR Izv.} {\bf 24} (1985) 539.

\bibitem{Klose:2006zd}
T.~Klose, T.~McLoughlin, R.~Roiban, and K.~Zarembo, {\it Worldsheet scattering
  in $\mathit{AdS}_{5}\times \mathit{S}^5$},  {\em JHEP} {\bf 03} (2007) 094,
  [\href{http://xxx.lanl.gov/abs/hep-th/0611169}{{\tt hep-th/0611169}}].

\bibitem{Arutyunov:2011uz}
G.~Arutyunov and S.~Frolov, {\it {Comments on the Mirror TBA}},
  \href{http://xxx.lanl.gov/abs/1103.2708}{{\tt arXiv:1103.2708}}.

\bibitem{Staudacher:2010jz}
M.~Staudacher, {\it {Review of AdS/CFT Integrability, Chapter III.1: Bethe
  Ansaetze and the R-Matrix Formalism}},
  \href{http://xxx.lanl.gov/abs/1012.3990}{{\tt arXiv:1012.3990}}.

\bibitem{Ahn:2010ka}
C.~Ahn and R.~I. Nepomechie, {\it {Review of AdS/CFT Integrability, Chapter
  III.2: Exact world-sheet S-matrix}},
  \href{http://xxx.lanl.gov/abs/1012.3991}{{\tt arXiv:1012.3991}}.

\bibitem{Beisert:2005fw}
N.~Beisert and M.~Staudacher, {\it Long-range $\mathfrak{psu}(2,2|4)$
  \textrm{B}ethe ansaetze for gauge theory and strings},  {\em Nucl. Phys.}
  {\bf B727} (2005) 1, [\href{http://xxx.lanl.gov/abs/hep-th/0504190}{{\tt
  hep-th/0504190}}].

\bibitem{Balog:2011nm}
J.~Balog and A.~Hegedus, {\it {$AdS_5\times S^5$ mirror TBA equations from
  Y-system and discontinuity relations}},
  \href{http://xxx.lanl.gov/abs/1104.4054}{{\tt arXiv:1104.4054}}.

\bibitem{Torrielli:2011zz}
A.~Torrielli, {\it {The Hopf superalgebra of AdS/CFT}},  {\em J.Geom.Phys.}
  {\bf 61} (2011) 230--236.

\bibitem{Gomez:2007zr}
C.~Gomez and R.~Hernandez, {\it {Quantum deformed magnon kinematics}},  {\em
  JHEP} {\bf 03} (2007) 108,
  [\href{http://xxx.lanl.gov/abs/hep-th/0701200}{{\tt hep-th/0701200}}].

\bibitem{Young:2007wd}
C.~A.~S. Young, {\it {q-Deformed Supersymmetry and Dynamic Magnon
  Representations}},  {\em J. Phys.} {\bf A40} (2007) 9165,
  [\href{http://xxx.lanl.gov/abs/0704.2069}{{\tt arXiv:0704.2069}}].

\bibitem{Beisert:2008tw}
N.~Beisert and P.~Koroteev, {\it {Quantum Deformations of the One-Dimensional
  Hubbard Model}},  {\em J. Phys.} {\bf A41} (2008) 255204,
  [\href{http://xxx.lanl.gov/abs/0802.0777}{{\tt arXiv:0802.0777}}].

\bibitem{Beisert:2011wq}
N.~Beisert, W.~Galleas, and T.~Matsumoto, {\it {A Quantum Affine Algebra for
  the Deformed Hubbard Chain}},  \href{http://xxx.lanl.gov/abs/1102.5700}{{\tt
  arXiv:1102.5700}}.

\bibitem{Martins:2007hb}
M.~J. Martins and C.~S. Melo, {\it {The Bethe ansatz approach for factorizable
  centrally extended S-matrices}},  {\em Nucl. Phys.} {\bf B785} (2007) 246,
  [\href{http://xxx.lanl.gov/abs/hep-th/0703086}{{\tt hep-th/0703086}}].

\bibitem{Shastry:1986zz}
B.~S. Shastry, {\it {Exact Integrability of the One-Dimensional Hubbard
  Model}},  {\em Phys. Rev. Lett.} {\bf 56} (1986) 2453.

\bibitem{Rej:2005qt}
A.~Rej, D.~Serban, and M.~Staudacher, {\it {Planar N = 4 gauge theory and the
  Hubbard model}},  {\em JHEP} {\bf 03} (2006) 018,
  [\href{http://xxx.lanl.gov/abs/hep-th/0512077}{{\tt hep-th/0512077}}].

\bibitem{Feverati:2006hh}
G.~Feverati, D.~Fioravanti, P.~Grinza, and M.~Rossi, {\it {Hubbard's adventures
  in N = 4 SYM-land? Some non- perturbative considerations on finite length
  operators}},  {\em J. Stat. Mech.} {\bf 0702} (2007) P001,
  [\href{http://xxx.lanl.gov/abs/hep-th/0611186}{{\tt hep-th/0611186}}].

\bibitem{Beisert:2007ds}
N.~Beisert, {\it {The S-Matrix of AdS/CFT and Yangian Symmetry}},  {\em PoS}
  {\bf SOLVAY} (2006) 002, [\href{http://xxx.lanl.gov/abs/0704.0400}{{\tt
  arXiv:0704.0400}}].

\bibitem{Spill:2008zz}
F.~Spill, {\it {Symmetries of the AdS/CFT S-matrix}},  {\em Acta Phys. Polon.}
  {\bf B39} (2008) 3135.

\bibitem{BazhanovTalk}
V.~Bazhanov, {\it {Talk at the Conference on ``Integrability in Gauge and
  String Theory", Max Planck Institute for Gravitational Physics
  (Albert-Einstein Institute), Potsdam, {\rm 29 June - 3 July 2009}}},  {\em
  {\rm [http://int09.aei.mpg.de/]}}.

\bibitem{Freidel:1992cd}
L.~Freidel and J.~M. Maillet, {\it {The Universal R matrix and its associated
  quantum algebra as functionals of the classical r matrix: The sl(2) case}},
  {\em Phys. Lett.} {\bf B296} (1992) 353,
  [\href{http://xxx.lanl.gov/abs/hep-th/9210039}{{\tt hep-th/9210039}}].

\bibitem{BD1}
A.~A. Belavin and V.~G. Drinfeld, {\it {Solutions of the classical Yang-Baxter
  equation for simple Lie algebras}},  {\em Funct. Anal. Appl} {\bf 16} (1982)
  159.

\bibitem{BD2}
A.~A. Belavin and V.~G. Drinfeld, {\it {Triangle equation for simple Lie
  algebras}},  {\em Mathematical Physics Reviews (ed. Novikov et al.), Harwood,
  New York} {\bf 96} (1984).

\bibitem{Leites:1984pt}
D.~A. Leites and V.~V. Serganova, {\it {Solutions of the Classical Yang-Baxter
  Equation for Simple Superalgebras}},  {\em Theor. Math. Phys.} {\bf 58}
  (1984) 16.

\bibitem{Zhang:1990du}
R.-B. Zhang, M.~D. Gould, and A.~J. Bracken, {\it {Solutions of the graded
  classical Yang-Baxter equation and integrable models}},  {\em J. Phys.} {\bf
  A24} (1991) 1185.

\bibitem{Karaali1}
G.~Karaali, {\it {Constructing r-matrices on simple Lie superalgebras}},  {\em
  J. Algebra} {\bf 282} (2004) 83.

\bibitem{Karaali2}
G.~Karaali, {\it {A New Lie Bialgebra Structure on sl(2,1)}},  {\em Contemp.
  Math.} {\bf 413} (2006) 101.

\bibitem{Yang:1967bm}
C.-N. Yang, {\it {Some exact results for the many body problems in one
  dimension with repulsive delta function interaction}},  {\em Phys. Rev.
  Lett.} {\bf 19} (1967) 1312.

\bibitem{KT}
S.~M. Khoroshkin and V.~N. Tolstoy, {\it {Universal R-matrix for quantized
  (super)algebras}},  {\em Commun. Math. Phys.} {\bf 276} (1991) 599.

\bibitem{BD3}
A.~A. Belavin and V.~G. Drinfeld, {\it {Classical Yang-Baxter equation for
  simple Lie algebras}},  {\em Funct. Anal. Appl.} {\bf 17} (1983) 220.

\bibitem{Arutyunov:2006iu}
G.~Arutyunov and S.~Frolov, {\it On $\mathit{AdS}_{5}\times \mathit{S}^5$
  string $\mathit{S}$-matrix},  {\em Phys. Lett.} {\bf B639} (2006) 378,
  [\href{http://xxx.lanl.gov/abs/hep-th/0604043}{{\tt hep-th/0604043}}].

\bibitem{Cai:q-alg9709038}
J.~Cai, S.~Wang, K.~Wu, and C.~Xiong, {\it {Universal R-matrix Of The Super
  Yangian Double $DY(gl(1|1))$}},  {\em Comm. Theor. Phys.} {\bf 29} (1998)
  173, [\href{http://xxx.lanl.gov/abs/q-alg/9709038}{{\tt q-alg/9709038}}].

\bibitem{Beisert:2005wm}
N.~Beisert, {\it {An $SU(1|1)$-invariant S-matrix with dynamic
  representations}},  {\em Bulg. J. Phys.} {\bf 33S1} (2006) 371,
  [\href{http://xxx.lanl.gov/abs/hep-th/0511013}{{\tt hep-th/0511013}}].

\bibitem{Arutyunov:2009ce}
G.~Arutyunov, M.~de~Leeuw, and A.~Torrielli, {\it {Universal blocks of the
  AdS/CFT Scattering Matrix}},  {\em JHEP} {\bf 05} (2009) 086,
  [\href{http://xxx.lanl.gov/abs/0903.1833}{{\tt arXiv:0903.1833}}].

\bibitem{Rej:2010mu}
A.~Rej and F.~Spill, {\it {The Yangian of $sl(n|m)$ and the universal
  R-matrix}},  \href{http://xxx.lanl.gov/abs/1008.0872}{{\tt arXiv:1008.0872}}.

\bibitem{Moriyama:2007jt}
S.~Moriyama and A.~Torrielli, {\it {A Yangian Double for the AdS/CFT Classical
  r-matrix}},  {\em JHEP} {\bf 06} (2007) 083,
  [\href{http://xxx.lanl.gov/abs/0706.0884}{{\tt 0706.0884}}].

\bibitem{deLeeuw:2010nd}
M.~de~Leeuw, {\it {The S-matrix of the $AdS_5 x S^5$ superstring, {\rm Based on
  PhD thesis}}},  \href{http://xxx.lanl.gov/abs/{arXiv:1007.4931}}{{\tt
  {arXiv:1007.4931}}}.

\bibitem{Beisert:2007ty}
N.~Beisert and F.~Spill, {\it {The Classical r-matrix of AdS/CFT and its Lie
  Bialgebra Structure}},  {\em Commun. Math. Phys.} {\bf 285} (2009) 537,
  [\href{http://xxx.lanl.gov/abs/0708.1762}{{\tt arXiv:0708.1762}}].

\bibitem{deLeeuw:2008dp}
M.~de~Leeuw, {\it {Bound States, Yangian Symmetry and Classical r-matrix for
  the $\mathit{AdS}_{5}\times \mathit{S}^5$ Superstring}},  {\em JHEP} {\bf 06}
  (2008) 085, [\href{http://xxx.lanl.gov/abs/0804.1047}{{\tt
  arXiv:0804.1047}}].

\bibitem{Dorey:2006mx}
N.~Dorey and B.~Vicedo, {\it {A symplectic structure for string theory on
  integrable backgrounds}},  {\em JHEP} {\bf 03} (2007) 045,
  [\href{http://xxx.lanl.gov/abs/hep-th/0606287}{{\tt hep-th/0606287}}].

\bibitem{Aoyama:2007tz}
S.~Aoyama, {\it {Classical Exchange Algebra of the Superstring on $S^5$ with
  the AdS-time}},  \href{http://xxx.lanl.gov/abs/0709.3911}{{\tt
  arXiv:0709.3911}}.

\bibitem{Mikhailov:2007eg}
A.~Mikhailov and S.~Schafer-Nameki, {\it {Algebra of transfer-matrices and
  Yang-Baxter equations on the string worldsheet in AdS(5) x S(5)}},  {\em
  Nucl. Phys.} {\bf B802} (2008) 1,
  [\href{http://xxx.lanl.gov/abs/0712.4278}{{\tt arXiv:0712.4278}}].

\bibitem{Vicedo:2008jy}
B.~Vicedo, {\it {Semiclassical Quantisation of Finite-Gap Strings}},  {\em
  JHEP} {\bf 06} (2008) 086, [\href{http://xxx.lanl.gov/abs/0803.1605}{{\tt
  arXiv:0803.1605}}].

\bibitem{Vicedo:2008jk}
B.~Vicedo, {\it {Finite-g Strings}},  {\em J.Phys.A} {\bf 44} (2011) 124002,
  [\href{http://xxx.lanl.gov/abs/0810.3402}{{\tt arXiv:0810.3402}}].

\bibitem{Magro:2008dv}
M.~Magro, {\it {The Classical Exchange Algebra of AdS5 x S5}},  {\em JHEP} {\bf
  01} (2009) 021, [\href{http://xxx.lanl.gov/abs/0810.4136}{{\tt
  arXiv:0810.4136}}].

\bibitem{Vicedo:2010qd}
B.~Vicedo, {\it {The classical R-matrix of AdS/CFT and its Lie dialgebra
  structure}},  {\em Lett.Math.Phys.} {\bf 95} (2011) 249--274,
  [\href{http://xxx.lanl.gov/abs/1003.1192}{{\tt arXiv:1003.1192}}].

\bibitem{Magro:2010jx}
M.~Magro, {\it {Review of AdS/CFT Integrability, Chapter II.3: Sigma Model,
  Gauge Fixing}},  \href{http://xxx.lanl.gov/abs/1012.3988}{{\tt
  arXiv:1012.3988}}.

\bibitem{Beisert:2010kk}
N.~Beisert, {\it {The Classical Trigonometric r-Matrix for the Quantum-Deformed
  Hubbard Chain}},  \href{http://xxx.lanl.gov/abs/1002.1097}{{\tt
  arXiv:1002.1097}}.

\bibitem{Matsumoto:2007rh}
T.~Matsumoto, S.~Moriyama, and A.~Torrielli, {\it {A Secret Symmetry of the
  AdS/CFT S-matrix}},  {\em JHEP} {\bf 09} (2007) 099,
  [\href{http://xxx.lanl.gov/abs/0708.1285}{{\tt arXiv:0708.1285}}].

\bibitem{Spill:2008tp}
F.~Spill and A.~Torrielli, {\it {On Drinfeld's second realization of the
  AdS/CFT $su(2|2)$ Yangian}},  {\em J. Geom. Phys.} {\bf 59} (2009) 489,
  [\href{http://xxx.lanl.gov/abs/0803.3194}{{\tt arXiv:0803.3194}}].

\bibitem{Dobrev:2009bb}
V.~K. Dobrev, {\it {Note on Centrally Extended $su(2|2)$ and Serre Relations}},
   {\em Fortsch. Phys.} {\bf 57} (2009) 542,
  [\href{http://xxx.lanl.gov/abs/0903.0511}{{\tt arXiv:0903.0511}}].

\bibitem{Cornwell}
J.~F. Cornwell, {\it {Group theory in physics, volume III: Supersymmetries and
  infinite-dimensional algebras}},  {\em Academic Press, New York} (1989).

\bibitem{Torrielli:2008wi}
A.~Torrielli, {\it {Structure of the string R-matrix}},  {\em J. Phys.} {\bf
  A42} (2009) 055204, [\href{http://xxx.lanl.gov/abs/0806.1299}{{\tt
  arXiv:0806.1299}}].

\bibitem{Spill:2008yr}
F.~Spill, {\it {Weakly coupled N=4 Super Yang-Mills and N=6 Chern-Simons
  theories from $u(2|2)$ Yangian symmetry}},  {\em JHEP} {\bf 03} (2009) 014,
  [\href{http://xxx.lanl.gov/abs/0810.3897}{{\tt arXiv:0810.3897}}].

\bibitem{Heckenberger:2007ry}
I.~Heckenberger, F.~Spill, A.~Torrielli, and H.~Yamane, {\it {Drinfeld second
  realization of the quantum affine superalgebras of $D^{(1)}(2,1:x)$ via the
  Weyl groupoid}},  {\em RIMS Kokyuroku Bessatsu} {\bf B8} (2008) 171,
  [\href{http://xxx.lanl.gov/abs/0705.1071}{{\tt arXiv:0705.1071}}].

\bibitem{Matsumoto:2008ww}
T.~Matsumoto and S.~Moriyama, {\it {An Exceptional Algebraic Origin of the
  AdS/CFT Yangian Symmetry}},  {\em JHEP} {\bf 04} (2008) 022,
  [\href{http://xxx.lanl.gov/abs/0803.1212}{{\tt arXiv:0803.1212}}].

\bibitem{Matsumoto:2009rf}
T.~Matsumoto and S.~Moriyama, {\it {Serre Relation and Higher Grade Generators
  of the AdS/CFT Yangian Symmetry}},  {\em JHEP} {\bf 0909} (2009) 097,
  [\href{http://xxx.lanl.gov/abs/0902.3299}{{\tt arXiv:0902.3299}}].

\bibitem{Zamolodchikov:1989cf}
A.~B. Zamolodchikov, {\it {Thermodynamic Bethe Ansatz in Relativistic Models.
  Scaling Three State Potts and Lee-Yang Models}},  {\em Nucl. Phys.} {\bf
  B342} (1990) 695.

\bibitem{Ambjorn:2005wa}
J.~Ambjorn, R.~A. Janik, and C.~Kristjansen, {\it {Wrapping interactions and a
  new source of corrections to the spin-chain / string duality}},  {\em Nucl.
  Phys.} {\bf B736} (2006) 288,
  [\href{http://xxx.lanl.gov/abs/hep-th/0510171}{{\tt hep-th/0510171}}].

\bibitem{Zamolodchikov:1978xm}
A.~B. Zamolodchikov and A.~B. Zamolodchikov, {\it Factorized
  $\mathit{S}$-matrices in two dimensions as the exact solutions of certain
  relativistic quantum field models},  {\em Annals Phys.} {\bf 120} (1979) 253.

\bibitem{Kulish:1981gi}
P.~P. Kulish, N.~Y. Reshetikhin, and E.~K. Sklyanin, {\it {Yang-Baxter Equation
  and Representation Theory. 1}},  {\em Lett. Math. Phys.} {\bf 5} (1981) 393.

\bibitem{Arutyunov:2009kf}
G.~Arutyunov and S.~Frolov, {\it {The Dressing Factor and Crossing Equations}},
   {\em J.Phys.A} {\bf A42} (2009) 425401,
  [\href{http://xxx.lanl.gov/abs/0904.4575}{{\tt arXiv:0904.4575}}].

\bibitem{Ahn:2010xa}
C.~Ahn and R.~I. Nepomechie, {\it {Yangian symmetry and bound states in AdS/CFT
  boundary scattering}},  {\em JHEP} {\bf 05} (2010) 016,
  [\href{http://xxx.lanl.gov/abs/1003.3361}{{\tt arXiv:1003.3361}}].

\bibitem{MacKay:2010zb}
N.~MacKay and V.~Regelskis, {\it {On the reflection of magnon bound states}},
  {\em JHEP} {\bf 08} (2010) 055,
  [\href{http://xxx.lanl.gov/abs/1006.4102}{{\tt arXiv:1006.4102}}].

\bibitem{Palla:2011eu}
L.~Palla, {\it {Yangian symmetry of boundary scattering in AdS/CFT and the
  explicit form of bound state reflection matrices}},  {\em JHEP} {\bf 1103}
  (2011) 110, [\href{http://xxx.lanl.gov/abs/1102.0122}{{\tt
  arXiv:1102.0122}}].

\bibitem{Chen:2006gq}
H.-Y. Chen, N.~Dorey, and K.~Okamura, {\it {On the scattering of magnon
  boundstates}},  {\em JHEP} {\bf 11} (2006) 035,
  [\href{http://xxx.lanl.gov/abs/hep-th/0608047}{{\tt hep-th/0608047}}].

\bibitem{Roiban:2006gs}
R.~Roiban, {\it {Magnon bound-state scattering in gauge and string theory}},
  {\em JHEP} {\bf 04} (2007) 048,
  [\href{http://xxx.lanl.gov/abs/hep-th/0608049}{{\tt hep-th/0608049}}].

\bibitem{Beisert:2006ib}
N.~Beisert, R.~Hernandez, and E.~Lopez, {\it A crossing-symmetric phase for
  $\mathit{AdS}_{5}\times \mathit{S}^5$ strings},  {\em JHEP} {\bf 11} (2006)
  070, [\href{http://xxx.lanl.gov/abs/hep-th/0609044}{{\tt hep-th/0609044}}].

\bibitem{Beisert:2006ez}
N.~Beisert, B.~Eden, and M.~Staudacher, {\it Transcendentality and crossing},
  {\em J. Stat. Mech.} {\bf 0701} (2007) P021,
  [\href{http://xxx.lanl.gov/abs/hep-th/0610251}{{\tt hep-th/0610251}}].

\bibitem{Coleman:1978kk}
S.~R. Coleman and H.~J. Thun, {\it {On the prosaic origin of the double poles
  in the Sine-Gordon S matrix}},  {\em Commun. Math. Phys.} {\bf 61} (1978) 31.

\bibitem{twi}
V.~G. Drinfeld, {\it {Quasi-Hopf algebras}},  {\em Leningrad Math. J.} {\bf 1}
  (1990) 1419.

\bibitem{pfeiffer}
H.~Pfeiffer, {\it {Factorizing twists and the universal R-matrix of the Yangian
  $Y(sl(2))$}},  {\em J. Phys. A} {\bf 33} (2000) 8929,
  [\href{http://xxx.lanl.gov/abs/math-ph/0006032}{{\tt math-ph/0006032}}].

\bibitem{Gromov:2010kf}
N.~Gromov and V.~Kazakov, {\it {Review of AdS/CFT Integrability, Chapter III.7:
  Hirota Dynamics for Quantum Integrability}},
  \href{http://xxx.lanl.gov/abs/1012.3996}{{\tt arXiv:1012.3996}}.

\bibitem{Kuniba:2010ir}
A.~Kuniba, T.~Nakanishi, and J.~Suzuki, {\it {T-systems and Y-systems in
  integrable systems}},  {\em J. Phys.} {\bf A44} (2011) 103001,
  [\href{http://xxx.lanl.gov/abs/1010.1344}{{\tt arXiv:1010.1344}}].

\bibitem{Korepin}
F.~H.~L. Essler, H.~Frahm, F.~Goehmann, A.~Kluemper, and V.~E. Korepin, {\it
  {The One-Dimensional Hubbard Model}},  {\em Cambridge University Press}
  (2005).

\bibitem{Arutyunov:2009zu}
G.~Arutyunov and S.~Frolov, {\it {String hypothesis for the AdS5 x S5 mirror}},
   {\em JHEP} {\bf 03} (2009) 152,
  [\href{http://xxx.lanl.gov/abs/0901.1417}{{\tt arXiv:0901.1417}}].

\bibitem{Krichever:1996qd}
I.~Krichever, O.~Lipan, P.~Wiegmann, and A.~Zabrodin, {\it {Quantum integrable
  models and discrete classical Hirota equations}},  {\em Commun. Math. Phys.}
  {\bf 188} (1997) 267--304,
  [\href{http://xxx.lanl.gov/abs/hep-th/9604080}{{\tt hep-th/9604080}}].

\bibitem{Kazakov:2007fy}
V.~Kazakov, A.~Sorin, and A.~Zabrodin, {\it {Supersymmetric Bethe ansatz and
  Baxter equations from discrete Hirota dynamics}},  {\em Nucl. Phys.} {\bf
  B790} (2008) 345--413, [\href{http://xxx.lanl.gov/abs/hep-th/0703147}{{\tt
  hep-th/0703147}}].

\bibitem{Arutyunov:2009pw}
G.~Arutyunov, M.~de~Leeuw, and A.~Torrielli, {\it {On Yangian and Long
  Representations of the Centrally Extended $su(2|2)$ Superalgebra}},  {\em
  JHEP} {\bf 06} (2010) 033, [\href{http://xxx.lanl.gov/abs/0912.0209}{{\tt
  arXiv:0912.0209}}].

\bibitem{zhang-2005-46}
Y.-Z. Zhang and M.~D. Gould, {\it A unified and complete construction of all
  finite dimensional irreducible representations of $gl(2|2)$},  {\em Journal
  of Mathematical Physics} {\bf 46} (2005) 013505.

\bibitem{Kamupingene:1989wj}
A.~H. Kamupingene, N.~A. Ky, and T.~D. Palev, {\it {Finite-dimensional
  representations of the Lie superalgebra $gl(2|2)$ in a $gl(2)\oplus gl(2)$
  basis. I. Typical representations}},  {\em J. Math. Phys.} {\bf 30} (1989)
  553.

\bibitem{Palev:1990wm}
T.~D. Palev and N.~I. Soilova, {\it {Finite dimensional representations of the
  Lie superalgebra $gl(2|2)$ in a $gl(2) \times gl(2)$ basis. 2. Nontypical
  representations}},  {\em J. Math. Phys.} {\bf 31} (1990) 953.

\bibitem{Gotz:2005ka}
G.~Gotz, T.~Quella, and V.~Schomerus, {\it {Tensor products of $psl(2|2)$
  representations}},  \href{http://xxx.lanl.gov/abs/hep-th/0506072}{{\tt
  hep-th/0506072}}.

\bibitem{Gotz:2005jz}
G.~Gotz, T.~Quella, and V.~Schomerus, {\it {Representation theory of
  $sl(2|1)$}},  {\em J. Algebra} {\bf 312} (2007) 829,
  [\href{http://xxx.lanl.gov/abs/hep-th/0504234}{{\tt hep-th/0504234}}].

\bibitem{Kazakov:2007na}
V.~Kazakov and P.~Vieira, {\it {From Characters to Quantum (Super)Spin Chains
  via Fusion}},  {\em JHEP} (2008) 050,
  [\href{http://xxx.lanl.gov/abs/0711.2470}{{\tt arXiv:0711.2470}}].

\bibitem{BahaBalantekin:1980qy}
A.~Baha~Balantekin and I.~Bars, {\it {Dimension and Character Formulas for Lie
  Supergroups}},  {\em J. Math. Phys.} {\bf 22} (1981) 1149.

\bibitem{Bargheer:2011mm}
T.~Bargheer, N.~Beisert, and F.~Loebbert, {\it {Exact Superconformal and
  Yangian Symmetry of Scattering Amplitudes}},
  \href{http://xxx.lanl.gov/abs/1104.0700}{{\tt arXiv:1104.0700}}.

\bibitem{Bartels:2011nz}
J.~Bartels, L.~Lipatov, and A.~Prygarin, {\it {Integrable spin chains and
  scattering amplitudes}},  \href{http://xxx.lanl.gov/abs/1104.0816}{{\tt
  arXiv:1104.0816}}.

\bibitem{Fioravanti:1998ha}
D.~Fioravanti and M.~Stanishkov, {\it {On the null-vectors in the spectra of
  the 2D integrable hierarchies}},  {\em Phys. Lett.} {\bf B430} (1998)
  109--119, [\href{http://xxx.lanl.gov/abs/hep-th/9806090}{{\tt
  hep-th/9806090}}].

\bibitem{Fioravanti:1999th}
D.~Fioravanti and M.~Stanishkov, {\it {Hidden local, quasi-local and non-local
  symmetries in integrable systems}},  {\em Nucl. Phys.} {\bf B577} (2000)
  500--528, [\href{http://xxx.lanl.gov/abs/hep-th/0001151}{{\tt
  hep-th/0001151}}].

\bibitem{Fioravanti:2007un}
D.~Fioravanti and M.~Rossi, {\it {On the commuting charges for the highest
  dimension SU(2) operator in planar ${\cal N}=4$ SYM}},  {\em JHEP} {\bf 08}
  (2007) 089, [\href{http://xxx.lanl.gov/abs/0706.3936}{{\tt
  arXiv:0706.3936}}].

\bibitem{Maldacena:2006rv}
J.~M. Maldacena and I.~Swanson, {\it {Connecting giant magnons to the pp-wave:
  An interpolating limit of $AdS_5 \times S^5$}},  {\em Phys. Rev.} {\bf D76}
  (2007) 026002, [\href{http://xxx.lanl.gov/abs/hep-th/0612079}{{\tt
  hep-th/0612079}}].

\bibitem{Klose:2007rz}
T.~Klose, T.~McLoughlin, J.~A. Minahan, and K.~Zarembo, {\it {World-sheet
  scattering in AdS(5) x S**5 at two loops}},  {\em JHEP} {\bf 08} (2007) 051,
  [\href{http://xxx.lanl.gov/abs/0704.3891}{{\tt arXiv:0704.3891}}].

\bibitem{unp}
T.~Klose, T.~McLoughlin, J.~Minahan, and A.~Torrielli {\em {\rm
  (unpublished)}}.

\bibitem{Vidas}
V.~Regelskis, {\it {The secret symmetries of the AdS/CFT reflection matrices,
  {\rm to appear}}},  {\em {\rm 2011}}.

\bibitem{Delius:2001yi}
G.~Delius, N.~MacKay, and B.~Short, {\it {Boundary remnant of Yangian symmetry
  and the structure of rational reflection matrices}},  {\em Phys.Lett.} {\bf
  B522} (2001) 335--344, [\href{http://xxx.lanl.gov/abs/hep-th/0109115}{{\tt
  hep-th/0109115}}].

\bibitem{MacKay:2010ey}
N.~MacKay and V.~Regelskis, {\it {Yangian symmetry of the Y=0 maximal giant
  graviton}},  {\em JHEP} {\bf 1012} (2010) 076,
  [\href{http://xxx.lanl.gov/abs/1010.3761}{{\tt arXiv:1010.3761}}].

\bibitem{Ramgoolam:2008yr}
S.~Ramgoolam, {\it {Schur-Weyl duality as an instrument of Gauge-String
  duality}},  {\em AIP Conf.Proc.} {\bf 1031} (2008) 255--265,
  [\href{http://xxx.lanl.gov/abs/0804.2764}{{\tt arXiv:0804.2764}}].

\bibitem{DeComarmond:2010ie}
V.~De~Comarmond, R.~de~Mello~Koch, and K.~Jefferies, {\it {Surprisingly Simple
  Spectra}},  {\em JHEP} {\bf 1102} (2011) 006,
  [\href{http://xxx.lanl.gov/abs/1012.3884}{{\tt arXiv:1012.3884}}].

\bibitem{Belitsky:2006av}
A.~V. Belitsky, G.~P. Korchemsky, and D.~Mueller, {\it {Towards Baxter equation
  in supersymmetric Yang-Mills theories}},  {\em Nucl. Phys.} {\bf B768} (2007)
  116, [\href{http://xxx.lanl.gov/abs/hep-th/0605291}{{\tt hep-th/0605291}}].

\end{thebibliography}\endgroup

\end{document}